\documentclass[aps,prd,nofootinbib,superscriptaddress,11pt]{revtex4}  
\usepackage{graphicx}
\usepackage{epstopdf}
\usepackage{amsmath}
\usepackage{amsfonts}
\usepackage{amssymb}
\usepackage{appendix}
\usepackage{comment}
\usepackage{color}
\usepackage{slashed}
\usepackage{subfigure}
\usepackage{setspace}
\usepackage{footnote}
\usepackage{multirow}
\usepackage{mathrsfs}
\usepackage[normalem]{ulem}

\newcommand{\be}{\begin{equation}}
\newcommand{\ee}{\end{equation}}
\newcommand{\ba}{\begin{array}}
\newcommand{\ea}{\end{array}}
\newcommand{\bea}{\begin{eqnarray}}
\newcommand{\eea}{\end{eqnarray}}
\newcommand{\balg}{\begin{align}}
\newcommand{\ealg}{\end{align}}
\newcommand{\bit}{\begin{itemize}}
\newcommand{\eit}{\end{itemize}}
\newcommand{\trm}[1]{\textrm{#1}}
\newcommand{\mbf}[1]{\mathbf{#1}}

\newcommand{\mcl}[1]{\mathcal{#1}}
\newcommand{\mbb}[1]{\mathbb{#1}}
\newcommand{\msc}[1]{\mathscr{#1}}

\newcommand{\ra}{\rightarrow}

\newcommand{\Mpc}{\trm{\Mpc}}
\newcommand{\yr}{\trm{\yr}}
\newcommand{\eV}{\trm{\eV}}

\newcommand{\nn}{\nonumber}

\newcommand{\vtw}{\vspace{.2cm}}

\usepackage[letterpaper, margin=0.5in]{geometry}

\begin{document}

\singlespacing

{\hfill NUHEP-TH/16-08}

\title{Standard model flavor from an $SU(2)$ symmetry}

\author{Jeffrey M. Berryman}
\affiliation{Northwestern University, Department of Physics and Astronomy, 2145 Sheridan Road, Evanston, IL 60208, USA}

\author{Daniel Hern\'andez}
\affiliation{Northwestern University, Department of Physics and Astronomy, 2145 Sheridan Road, Evanston, IL 60208, USA}


\begin{abstract}
We propose that the flavor structure of the standard model is based on a horizontal $SU(2)$ symmetry. It generically predicts (i) a parametrically small mass for the lightest charged fermions, (ii) small mixings in the quark sector, and (iii) suppression of flavor-changing neutral currents. Supplemented with the assumption of a strong hierarchy between the second- and third- generation masses, it also predicts (iv) a large $CP$-violating phase in the quark sector. Only Majorana neutrinos allow for large mixings in the lepton sector. In this case, this framework further predicts (v) near-maximal $\theta_{23}^l$, (vi) a normal hierarchy of neutrino masses, and (vii) large $CP$ violation in the lepton sector.
\end{abstract}

\maketitle


\section{Introduction}
\label{SecI}


The Standard Model (SM) matter fields organize into three generations of quarks and leptons. The spectrum of these generations, with the possible exception of the neutrinos, is distinctly hierarchical. Moreover, generations mix with large angles in the lepton sector while relatively small angles are observed for the quarks. In the SM, the information about these physical parameters is encoded in the Yukawa couplings. Unobservable in themselves, an explanation for the structure in the Yukawas that gives rise to the seemingly whimsical masses and mixings has remained elusive. This is the so-called Flavor Puzzle.

Ever since the SM was proposed, several ideas have been put forward to resolve the Flavor Puzzle. Of note is the proposal of Froggatt and Nielsen that the origin of the fermionic mass hierarchies is dynamical \cite{Froggatt:1978nt}. This was achieved by positing the existence of a $U(1)$ symmetry under which fermions and a new scalar field were charged. Upon symmetry breaking, masses appeared proportional to powers of the vacuum expectation value of the $U(1)$ scalar.

However, typical Froggatt-Nielsen models struggle to explain the suppression of flavor-changing neutral currents (FCNCs) that the SM elegantly accounts for, a cappella, via the GIM mechanism \cite{Glashow:1970gm}. In the past decades, experimental measurements have been pushing the limits of FCNCs measurement. Generic arguments now indicate  that for nonstandard FCNCs to exist, new physics related to flavor should appear at least at the PeV scale. In view of this, a different approach to flavor, the Minimal Flavor Violation (MFV) ansatz, has been formulated \cite{Chivukula:1987py,D'Ambrosio:2002ex,Buras:2003jf,Cirigliano:2005ck,Agashe:2005hk,Kagan:2009bn,Gavela:2009cd,Alonso:2011yg,Alonso:2012fy}.

Within MFV, a prominent role is played by the flavor symmetry the SM would have if the Yukawa couplings were removed. MFV hypothesizes that this flavor symmetry is only broken by the Yukawa matrices at low energies. The SM can then be rephrased as a flavor-invariant theory if one introduces a formal transformation rule for the Yukawas under this flavor group. MFV goes on to posit that any nonrenormalizable operator made of SM fields should be flavor-invariant as well. In particular, the coefficients of flavorful operators, possibly contributing to exotic processes, must be functions of the Yukawa matrices such that the flavor charges of the fields composing the operator are cancelled.

In this way, MFV has two main consequences. First and foremost, it provides a way out of the ever-looming FCNC problem. The structure forced by the SM Yukawa couplings onto the coefficients of the nonrenormalizable operators is enough to lower the smallest possible scale of new flavor physics down to a few TeV. Secondly, it provides predictability, to some extent, since the same Yukawa couplings link the SM masses and mixings with the rates for exotic flavor processes.

In contrast, MFV does not explain, nor is it designed to explain, how the SM Yukawa structure comes about. In this regard, an old idea of Cabibbo \cite{Cabibbo:1970rza} has been resurrected recently. The proposal is to take the MFV hypothesis seriously and promote the Yukawa couplings to flavor-charged scalar fields. It is now possible to try to reproduce the SM observables by extremizing a flavor-invariant Yukawa potential. This approach has achieved partial success. In particular, it naturally produces no mixing in the quark sector while in the lepton sector, by invoking the Majorana character of neutrinos, it can explain at least one large angle. On the other hand, other features pertaining to the flavor puzzle are harder to account for, such as the hierarchy of masses and the observed values of the mixing angles, both in the quark and in the lepton sector \cite{Alonso:2011yg,Alonso:2012fy,Alonso:2011jd,Alonso:2013mca,Alonso:2013nca}.

In this paper, we put forward an alternative hypothesis to MFV. We keep the assumption that the SM is formally invariant under some flavor symmetry, but we abandon the requirement that the Yukawas are fundamental fields under it. We focus on a scenario in which the flavor symmetry of the SM is a single $SU(2)$ group, which we dub Flavorspin, that is the same for all fermions. Continuous flavor symmetries have been previously discussed in, for instance, Refs.~\cite{Terazawa:1976xx,Terazawa:1977ee,Maehara:1978ts,Wilczek:1978xi,Yanagida:1979gs,Chikashige:1980ht,Yanagida:1980wd,Terazawa:2011ci,Aulakh:2013kha,Aulakh:2014wsa,Terazawa:2015bsa}. Under Flavorspin, quarks and leptons transform as triplets and Yukawa matrices are upgraded to composite spurions, formed by linear combinations of fundamental ones that transform as symmetric or antisymmetric real matrices under flavor $SU(2)$. The ansatz proposed in this paper shares with MFV the capacity to suppress FCNCs. At the same time, it can account for several features that a solution to the Flavor Puzzle should target. Moreover, as we shall discuss in detail below, this simple case provides a way to link the flavor features of the quark and lepton sectors, by using the same fundamental spurions everywhere.

The paper is organized as follows. In the first three sections, our framework is presented in detail and theoretical and analytical results are described. In the later sections, we perform a complete numerical exploration of the framework and delve into phenomenological features such as the absence of FCNCs. In the final section, we discuss the results and comment on several ways this work could be extended.


\section{Flavorspin}
\label{flavorspin}
\setcounter{equation}{0}


\newcommand{\Gfl}{\mcl{G}_{fl}}
\newcommand{\veps}{\varepsilon}
\newcommand{\bveps}{\overline{\varepsilon}}

We consider a theory $\msc{L}$ that can generically be written as:
\be
\msc{L} = \msc{L}_{SM} + \msc{L}_{\nu} + \msc{L}_{NR} \label{mscL}  \,.
\ee
where $\msc{L}_{SM}$ is the SM Lagrangian, $\msc{L}_{\nu}$ are renormalizable terms that account for neutrino masses and $\msc{L}_{NR}$ are possible nonrenormalizable operators composed of SM fields. The SM piece $\msc{L}_{SM}$ can be split into flavorful and flavorless terms as
\be
\msc{L}_{SM} = \msc{L}_{0} + \msc{L}_{Yuk} \,,
\ee
where $\msc{L}_0$ contains the standard kinetic terms, Higgs potential and gauge interactions. Here, we are mostly interested in the flavorful Yukawa terms, contained in $\msc{L}_{Yuk}$. These have the form
\be
\label{YukLag}
-\msc{L}_{Yuk} = \overline{Q_L} Y_u U_R \cdot \widetilde{H} + \overline{Q_L} Y_d D_R \cdot H +  \overline{L_L} Y_l E_R \cdot H + \trm{h.c.} \,.
\ee

It is well known that in order to account for neutrino masses, the SM has to be extended. There are many possibilities for doing so consistently; in this work, we will focus on three of them, namely, Dirac neutrinos and the type I and type II seesaw Majorana neutrinos.
\begin{itemize}
\item \emph{Dirac neutrinos:} This possibility involves the introduction of a set of right-handed neutrinos $N_R$. The SM neutrinos acquire a mass in a way analogous to the rest of the fermions,
  \be
  \msc{L}_{\nu} = -\overline{L_L} Y_\nu N_R \cdot \widetilde{H} + \trm{h.c.} \,. \label{Dirac-nus}
  \ee
  As is well known, in the purely Dirac neutrino scenario, the SM preserves Lepton Number (LN) symmetry, the $U(1)$ global symmetry under which both $L_L$ and $N_R$ have charge +1. 
\item \emph{Type I Seesaw:} Right-handed neutrinos are introduced, in this case with a heavy Majorana mass, profiting from the fact that they are SM singlets,
  \be
  \msc{L}_{\nu} = -\overline{L_L} Y_\nu N_R \cdot \widetilde{H} + M\overline{N_R} N_R^c + \trm{h.c.} \,. \label{TypeI-nus}
  \ee
  This Lagrangian violates LN. With the charge assignment above, the Yukawa term preserves LN; only the Majorana mass breaks it. In this work, it will be assumed for simplicity that $M$ is proportional to the identity,
  \be
  M \propto \mbb{I} \,,
  \ee
though the results of this work do not depend strongly on this assumption. In addition, the type I seesaw is able to explain the low scale of the neutrino masses. Below the electroweak symmetry breaking (EWSB) scale and after integrating out $N_R$, the light neutrinos acquire a Majorana mass,
  \be
  \frac{v^2Y_\nu Y_\nu^T}{M}\overline{\nu_L}\nu_L^c \,.
  \ee
  which yields the right order of magnitude for the neutrino masses if $M/Y_\nu^2 \sim 10^{15}$ GeV.
  
\item \emph{Type II Seesaw:} The SM is augmented with an $SU(2)_W$ triplet $\Delta$ that couples to the leptons and the Higgs boson as
  \be
  \msc{L}_{\nu} =  -\mu_\Delta HH\Delta -  Y_\Delta \overline{L_L} L_L^c \Delta + \trm{h.c.} + \dots
  \ee
  where $\mu_\Delta$ has energy dimensions and  $Y_\Delta$ is flavor-charged. In this case, after integrating out the triplet and EWSB, the Majorana mass term for active neutrinos appears again, given by
  \be
  \frac{\mu_\Delta v^2 Y_\Delta}{M_\Delta^2}\overline{\nu_L}\nu_L^c \,.
  \ee
\end{itemize}
In all of the above, summation over flavor indices is implicit.

The nonrenormalizable term in Eq.~\eqref{mscL}, $\msc{L}_{NR}$, consists of all the gauge-invariant operators of dimension higher than 4 that can be constructed out of SM fields \cite{Buchmuller:1985jz,Manohar:1996cq,Burgess:2007pt},
\be
\msc{L}_{NR} = \sum_{d, \alpha} \frac{c_\alpha^{(d)}}{\Lambda^{d-4}}\mcl{O}^{(d)}_\alpha  \,,\quad d \in 4 + \mbb{N}_+,\; \alpha\in \mbb{N}_+
\ee
where $d$ is the energy dimension of the operator and $\alpha$ runs over all operators of a given dimension. We will consider here the phenomenologically-relevant $d=5$, 6 flavorful operators that include one or more fermionic bilinear so that we can write
\be
c_\alpha^{(d)} \mcl{O}^{(d)}_\alpha = c_{\alpha,}^{(d)}F_iF_j\cdot\mcl{Q}_\alpha^{(d)} \,.
\ee
where $F$ stands for any fermion or antifermion and $i$, $j$ are flavor indices. The Lorentz structure of such operators is not relevant here.

To define our scenario, the Yukawa couplings are upgraded to spurions, i.e., couplings that formally transform under a flavor symmetry $\mcl{G}_{fl}$. The Lagrangian $\msc{L}$ in Eq.~\eqref{mscL} must be $\Gfl$-invariant under simultaneous transformations of the spurions and the SM fields. The departure from MFV comes in the choice of the flavor group. In standard MFV, one factor of $SU(3)$ is introduced for each type of fermion. We hypothesize instead that the flavor group is
\be
\mcl{G}_{fl} = SU(2) \,,
\ee
under which fermions transform as triplets,
\be
F \rightarrow OF \,,\quad F = Q_L,\,L_L,\,D_R,\,U_R,\,E_R,\,N_R \,.
\ee
Here, $O$ is an orthogonal $3\times 3$ matrix and it is the same for all fermionic fields. This group is the only flavor symmetry we impose. In particular, the SM global $SU(3)^5$ flavor symmetry, apparent when the Yukawa couplings are set to zero, is understood to be mostly accidental. We refer to this flavor $SU(2)$ as Flavorspin.\footnote{We will use the name ``Flavorspin" to refer to the $SU(2)$ of flavor proposed here and, more generally, to the framework constructed using this group; its meaning will be clear from context.}

Demanding that the Yukawa terms are Flavorspin-invariant restricts the possible transformation laws for the Yukawa couplings. In this case, they  must formally belong in the
\be
\mbf{3}\times\mbf{3} = \mbf{5}\oplus\mbf{3}\oplus\mbf{1}
\ee
representations of $\Gfl$. Out of the above, the singlet term is flavorless, corresponding to a Yukawa matrix proportional to the identity. The possible fundamental Yukawa spurions with nontrivial flavor structure can therefore be represented by a $3\times 3$, traceless, symmetric, real Yukawa tensor, corresponding to the $\mathbf{5}$, and a $3\times 3$ antisymmetric real one, corresponding to the $\mathbf{3}$. 

The main hypothesis of this work is that all of the SM flavor can be understood from a minimalistic set of $SU(2)$ spurions. Specifically, we assume flavor is determined by two unique spurions in the $\mbf{3}$ and $\mbf{5}$ representations of $SU(2)$. These are denoted by $Y_3$ and $Y_5$ respectively. Under $\mcl{G}_{fl}$, $Y_3$ and $Y_5$ transform as
\be
Y_3 \rightarrow OY_3O^T \,,\quad Y_5 \rightarrow OY_5O^T \,.
\ee
where $O$ is an orthogonal $3\times 3$ matrix. This rule guarantees that the Lagrangian $\msc{L}$ is $\mcl{G}_{fl}$-invariant as long as $Y_X$, $Y_\Delta$ and the $c_\alpha^{(d)}$ are polynomial functions of $Y_3$, $Y_5$. Thus, a first approximation to the SM flavor structure is given by
\be
Y_X \equiv Y_X(Y_3,\,Y_5) + \bveps_X\mbb{I} \,,\quad Y_\Delta \equiv Y_\Delta(Y_3,\,Y_5) + \bveps_\Delta\mbb{I} \,,\quad  c_{\alpha,\,ij}^{(d)} \equiv c_{\alpha,\,ij}^{(d)}(Y_3,\,Y_5) + \bveps_\alpha^{(d)}\mbb{I} \,. \label{first-approx}
\ee
with $X = u,d,l,\nu$ and where the $\bveps$ coefficients are arbitrary complex numbers.

However, as it stands, Eq.~\eqref{first-approx} does not include an evident parameter on which to perform a perturbative expansion. Indeed, one can explicitly check that masses and mixing angles derived from it can be arbitrarily large. On the other hand, several SM observables pertaining flavor are parametrically small. These include the mixing angles in the quark sector and the masses of the first and second generations relative to the third. The idea then is to restrict the parameter space allowed by Eq.~\eqref{first-approx} by making some of the couplings above perturbative. In particular, we will assume a hierarchy between the contributions from the symmetric and antisymmetric spurions to flavor. More specifically, we demand
\be
Y_X \equiv Y_{X}(Y_3,\,\veps_X Y_5) + \bveps_X\mbb{I} \,,\quad \quad c_{\alpha,\,ij}^{(6)} \equiv c_{\alpha,\,ij}^{(6)}\left( Y_3,\, \veps_\alpha^{(6)}Y_5\right) + \bveps_\alpha^{(6)} \mbb{I} \,. \label{final-form} 
\ee
where $|\veps_X|$, $|\bveps_X|$, $|\veps_\alpha^{(6)}|$, $|\bveps_\alpha^{(6)}| \ll 1$. Note that no such assumptions are made for $Y_\Delta$, nor for the coefficient of the $d=5$ Weinberg operator. We will provide a possible argument to justify this apparently arbitrary distinction in a later section based on the fact that these operators violate $B-L$.

\vtw
In the remainder of this section, we analyze the features of flavor to be expected at zeroth order from Eq.~\eqref{final-form}. Consider the LN-conserving Yukawa coefficients $Y_X$. Explicitly, Eq.~\eqref{final-form} amounts to the Yukawa matrices taking the form
\be
\label{YX}
Y_X =  \mu_X \big(Y_X^0 + Y_{\veps X}  \big)
\ee 
with
\be
Y_X^0 = \mu_X \left(iY_3 + A_X e^{i \alpha_X} \cdot Y_3^2 \right) \,, \quad Y_{\veps X}= \veps_X Y_{5} + \overline{\veps}_X \cdot \mbb{I} + \dots \,. \label{Ydefs} 
\ee
The normalization factor $\mu_X$ sets the overall mass scale of each fermion type and it is fixed so that
\be
\frac{1}{2}\trm{Tr}[Y_3Y_3^T] = 1 \,.
\ee
The real antisymmetric flavor spurion $Y_3$ is assumed to be $\mcl{O}(1)$ and it is universal. That is, the zeroth-order terms in $Y_X^0$ are formed by linear combinations of the same $Y_3$ and $Y_3^2$ for all fermion types. The real constants $A_X$ and $\alpha_X$ specify the relative phase and weight of the two terms composing $Y_X^0$. Notice the relative factor of $i$ in Eq.~\eqref{Ydefs}; this amounts to a phase redefinition of the quark fields and is a useful convention, as we will make clear.

By means of $\mcl{G}_{fl}$ transformations, it is always possible to choose a basis in which the $Y_3$ spurion takes the form
\be
\label{Y3canonical}
Y_3 = \left( \ba{ccc} 0 & 0 & 0 \\ 0 & 0 & 1 \\ 0 & -1 & 0 \ea \right).
\ee
In this basis, it is evident why the truncation of the series at the quadratic order in $Y_3$ in Eq.~\eqref{Ydefs} is justified. Higher powers of $Y_3$ need not be introduced since $Y_3^3 = - Y_3$, as can be readily verified. For the remainder of this work, we work in this basis.

The term $Y_{\veps X}$ represents the perturbation term and it is formed by a linear combination of the universal, real, symmetric and traceless spurion $Y_5$,
\be
\label{Y5}
Y_5 = \left( \begin{array}{ccc}
  \;y_{11}\;  & \;y_{12}\; & y_{13} \\
  y_{12 }& y_{22} & y_{23} \\
  y_{13} & y_{23} & -(y_{11} + y_{22})
  \end{array} \right),
\ee
and the singlet term, see Eq.~\eqref{Ydefs}. It is assumed that $|\veps_X|, |\overline{\veps}_X| \ll 1$.

Any $Y_X^0$ of the form in Eq.~\eqref{Ydefs} has one null eigenvalue. Setting aside the neutrinos for the time being -- the possibility of Majorana masses changes this picture -- it is clear that, in the unperturbed setup, the lightest charged fermions have vanishing masses. Hence, our scenario automatically leads to a spectrum in which the first generation is much lighter than the other two.

Let us introduce the parameter
\be
\xi_X \equiv 1 - A_Xe^{i\alpha_X}
\ee
It can be easily shown that the two remaining eigenvalues are generically nonzero, their values given by
\begin{align}
  m_1^2 & = 0 \nn \\
  m_2^2 & = \mu_Xv^2|\xi_X|^2 \label{masses}  \\
  m_3^2 & =  \mu_Xv^2|2-\xi_X|^2
\end{align}
From Eq.~\eqref{masses}, it is possible for $m_2$ to vanish as well, if the relation
\be
\label{Hcond}
\xi_X =  0 \,
\ee
is satisfied. It follows that a large hierarchy between the second- and third-generation masses is obtained for $|\xi_X|\ll 1$. Since such a hierarchy is observed for both quarks and charged leptons, we adopt the final assumption
\be
|\xi_X| \ll 1 \,.
\ee
In this context, this assumption is equivalent to assuming that there is a strong hierarchy between $m_2^2$ and $m_3^2$. In other words, in order to explain the SM spectrum in Flavorspin, aside from $\{\veps_X, \bveps_X\}$, another set of perturbative parameters, the $\xi_X$, must exist.

The zeroth-order quark mixing can be quickly computed as well. The Yukawa matrices are generically diagonalized by biunitary transformations,
\be
\tilde{Y}_X = V_{XL}Y_XV_{XR}^\dagger \,, \quad\quad \tilde{Y}_X = \trm{diag}\{y_{X1},\, y_{X2},\, y_{X3}\} \,, \label{defineVX}
\ee
and it is apparent that at zeroth order in $\{\veps_X, \bveps_X\}$, the equality
\be
V_{uL} = V_{dL} = V_{\ell L} = V_{\nu L} =  V_{uR} = V_{dR} = V_{\ell R} = V_{\nu R} \label{Vequals}
\ee
holds. This is because $Y_3$, being fully antisymmetric, is diagonalized by a similarity transformation $V^0$. Thus, we have
\be
\tilde{Y}_X^0 = V^0Y_X^0V^{0\dagger}
\ee
where $\tilde{Y}_X^0$ is diagonal. $V^0$ is found to be
\be
V^0 \equiv V_{XL}(\veps_X = 0) = V_{XR}(\veps_X = 0) = \left( \begin{array}{ccc}
  1 & 0 & 0 \\
  0 & \frac{1}{\sqrt{2}} & \frac{1}{\sqrt{2}} \\
  0 & -\frac{1}{\sqrt{2}} & \frac{1}{\sqrt{2}} \\
\end{array} \right) \cdot \left( \begin{array}{ccc}
  1 && \\ & -i & \\ && 1
  \end{array} \right) \,. \label{V0}
\ee

The quark mixing matrix $V_{CKM}$ is defined as
\be
V_{CKM} = V_{uL}^\dagger V_{dL} \,. \label{CKM}
\ee
Hence, from Eq.~\eqref{V0}, it follows that there is no mixing in the quark sector at zeroth order in $\{\veps_X, \bveps_X\}$,
\be
V_{CKM} = \mbb{I} + \mcl{O}(\veps) \,.
\ee

In the spirit of the ansatz proposed above, the coefficients $c_\alpha^{(d)}$ accompanying LN-conserving operators are also assumed to be linear combinations of $Y_3$, $Y_5$ and $\mbb{I}$. With respect to perturbativity, however, the LN-violating couplings $Y_\Delta$ of the type II seesaw and $c_W^{(5)}$ are treated differently. In particular, no hierarchy is assumed between the coefficients of the linear combination of fundamental spurions from which $Y_\Delta$ is formed. We have, for instance,
\be
Y_\Delta = \eta_{33}Y_3Y_3^T + \eta_5Y_5 + \eta_1\mbb{I} + \cdots
\ee
with all the coefficients being, in principle, of $\mcl{O}(1)$.

Summarizing, we have introduced a framework that posits an $SU(2)$ horizontal flavor group, Flavorspin, under which SM fermions transform as triplets. Based on phenomenological considerations, in this paper we will focus on a specific scenario in which the following hypotheses hold:
\begin{enumerate}
\item The Lagrangian $\msc{L}$ including the SM and possible flavor-charged higher-dimensional operators is invariant under $\Gfl$.
\item Only two spurions, $Y_3$ and $Y_5$, in the $\mbf{3}$ and $\mbf{5}$ representations of $\Gfl$, respectively, are introduced.
\item The symmetric contribution to the Yukawa couplings, represented by $Y_5$ and the singlet term, is small compared to that of $Y_3$ for $B-L$-conserving, flavor charged operators. That is, $\veps_X$, $\bveps_X$ in Eq.~\eqref{final-form} satisfy $|\veps_X|,\, |\bveps_X| \ll 1$.
\item The parameters $\xi_X$ parametrizing the hierarchy between the second and third generation satisfy $|\xi_X| \ll 1$.
\end{enumerate}
No hierarchy is assumed between the perturbative parameters $\veps_X$ and $\xi_X$.

Although the large mass difference between the second and third generations of fermions is imposed by hand, $|\xi_X| \ll 1$, note that it can only appear intrinsically connected to the large relative phase between the $Y_3$ and $Y_3^2$ contributions to $Y_X$.
In particular, the $CP$-invariant possibility $\alpha_X = \pi/2$\footnote{Technically, there is no $CP$ violation at this stage. All the phases in the Lagrangian can be reabsorbed by unitary redefinitions of the quark and lepton fields. However, once perturbations are introduced, the large phase diffrence between the two terms in $Y_X^0$ will indeed lead to large values for $CP$ violation.} would have led to phenomenologically unrealistic degeneracy of the masses of the second and third generations states.
Looking forward, in general, this will lead to large $CP$ violation once nonvanishing mixings emerge due to the $Y_5$ perturbations.


\section{Perturbations in the quark sector}
\label{Sec-quarks}
\setcounter{equation}{0}


There are two main effects of introducing the perturbation $Y_{\veps X}$ in Eq.~\eqref{YX}: (i) To lift the lightest quark masses from zero, and (ii) to give rise to small mixing angles.
Using $Y_5$ from Eq.~\eqref{Y5} we find
\be
Y_{\veps X} = \veps_X\left( \begin{array}{ccc}
  \;y_{11} + y_X\;  & \;y_{12}\; & y_{13} \\
  y_{12 }& y_{22} + y_X & y_{23} \\
  y_{13} & y_{23} & -(y_{11} + y_{22}) + y_X
  \end{array} \right) \,,\quad \quad \bveps_X = \veps_X y_X.
\ee
The perturbations induced by $Y_{\veps X}$ to Eq.~\eqref{masses} can be computed up to the most relevant order. We obtain the following expressions for the perturbed eigenvalues:
\begin{align}
  m_{1X}^2 & = \mu_X^2 v^2 \left(F_X - G_X  \right) \nn \\
  m_{2X}^2 & = \mu_X^2 v^2 \left( F_X + G_X \right) \label{masses-pert}\\
  m_{3X}^2 & = 2\mu_X^2v^2\left( 2 + y_{11}\veps_X - 2\bveps_X + 2\xi_X \right)\nn
\end{align}
where
\begin{align}
  F_X & = \frac{1}{8}\left[\left(5y_{11}^3 + 4y_{12}^2 + 4y_{13}^2\right) \veps_X^2 + 8\bveps_X^2 + 4\xi_X^2 + 4 y_{11}\left(\veps_X\bveps_X +\veps_X\xi_X\right) + 8\bveps_X\xi_X \right] \,,\nn \\
  G_X & = \frac{1}{8}\left( 4\bveps_X + 2\xi_X - y_{11}\veps_X\right)\sqrt{\left(9y_{11}^2 + 8y_{12}^2 + 8y_{13}^2 \right)\veps_X^2 - 12y_{11}\veps_X\xi_X + 4\xi_X^2} \,. \label{defineFG}
\end{align}
In Eq.~\eqref{masses-pert}, terms have been kept up to the lowest relevant order for the first two eigenvalues, $m_{1,2}^2$. The mass of the third eigenvalue is corrected to order $\mcl{O}(|\veps_X|, \, |\bveps_X|, \, |\xi_X|)$ and its order of magnitude is determined by the EWSB scale and by the dimensionless coupling $\mu_X$.  It is straightforward to check that for $|\veps_X|, \, |\bveps_X| \rightarrow 0$, the perturbed spectrum reduces to Eq.~\eqref{masses}. Thus, the simplest way to implement the hierarchy between the first- and second-generation masses without imposing any artificial tuning is to assume a hierarchy between the perturbative parameters,
\be
|\veps_X|,\, |\bveps_X| \lesssim |\xi_X| \ll 1 \,.
\ee
In this case, the second-to-third-generation mass ratio can be approximated by:
\be
\label{XiRatio}
\frac{m_{2X}^2}{m_{3X}^2} \sim |\xi_X|^2\,.
\ee
Replacing in all of what follows $X\rightarrow u, d$, we obtain $\xi_{u,d} \lesssim x_{u,d}\cdot10^{-2}$ where $x_{u,d}$ are $\mcl{O}(1)$.

A crude, yet useful estimation for the first-to-second-generation mass ratio can also be obtained by keeping only the highest order terms in $\xi_{u,d}$ in the ratio
\be
\frac{m_{1X}^2}{m_{2X}^2} \sim \frac{3\veps_X}{\xi_X}\cdot y_{11} \sim \veps_X y_{11} \cdot 10^2  \label{mass-est}
\ee
For the quarks we have: $m_u^2/m_c^2 \sim 10^{-6}$, $m_d^2/m_s^2 \sim 10^{-4}$. Thus, generically we obtain
\be
\frac{\veps_u}{\xi_u} < \frac{\veps_d}{\xi_d} \,. \label{ud-comparison1}
\ee
This result will be validated by our numerical analysis in Sec.~\ref{results}. The latter also shows that the mass ratios and the relatively large size of the Cabibbo angle cannot both be accounted simply by setting $|\veps_X/\xi_X|$ small. Thus, it is necessary that
\be
y_{11}, \, |\bveps_X| \ll 1 \,.
\ee
Finally note that the ratio between the $\mu_X$ determines the scales of the up and down sectors
\be
\frac{m_b}{m_t} \sim \frac{\mu_u}{\mu_d} \,.
\ee

The small $\veps$-parameters also give rise to small mixing angles. The sines of these mixing angles, to leading order in $\veps_X$, are given by
\begin{align}
  \sin \theta_{12}^q & \simeq \frac{1}{\sqrt{2}} \left|(y_{12} - iy_{13})\left(\frac{\veps_d}{\xi_d} - \frac{\veps_u}{\xi_u} \right)\right|  \label{th12} \\
  \sin \theta_{13}^q & \simeq \frac{1}{\sqrt{2}}\big|(y_{12} + iy_{13})(\veps_d - \veps_u) \big|  \label{th13} \\
  \sin \theta_{23}^q & \simeq \frac{1}{4}\big| (y_{11} + 2y_{22} + 2iy_{23})(\veps_d-\veps_u) \big|  \sim  \frac{1}{2}\big| (y_{22} + iy_{23})(\veps_d-\veps_u) \big|\label{th23}
\end{align}
Hence, approximately,
\be
\label{Ratio_13_23_Quark}
\frac{\sin \theta_{13}^q}{\sin\theta_{23}^q} \sim \left| \frac{y_{12} + iy_{13}}{y_{22} + iy_{23}} \right| \sim 10^{-1}.
\ee
Thus, a mild hierarchy between the values of $\{ y_{12}, \, y_{13} \}$ and those of $ \{y_{22},\, y_{23} \}$ is expected. Also, from Eq.~\eqref{ud-comparison1}, we can roughly approximate:
\be
\sin \theta_{12}^q = \frac{1}{\sqrt{2}} \left|(y_{12} - iy_{13})\frac{\veps_d}{\xi_d} \right| \sim  \frac{m_d}{\sqrt{2}m_s} \left|(y_{12} - iy_{13}) \right| \,.
\ee

Eqs.~(\ref{th12}-\ref{th23}) illustrate an important consequence of Flavorspin. That is, an enhancement of $\sin\theta_{12}^q$ with respect to $\sin\theta_{13}^q$ and $\sin \theta_{23}^q$. A rough order of magnitude estimate of this enhancement is
\be
\label{Ratio_12_13_Quark}
\frac{\sin\theta_{12}}{\sin\theta_{13}} \simeq \frac{1}{|\xi_d|} \sim \frac{m_3}{m_2}\sim 10^{2}.
\ee
We stress that this enhancement is a prediction that emerges, in this context, as a consequence of the $SU(2)$ structure coupled to the intergenerational mass hierarchy.

The $CP$-violating phase $\delta_{CP}^q$ in the quark sector can be computed via the Jarlskog invariant $J$:
\be
\label{defineJ}
J = \left( \sum_{m,n} \epsilon_{ikm} \epsilon_{jln} \right) \Im \left[V_{ij} V_{kl} V^*_{il} V^*_{kj}  \right]= c_{12} c_{13}^2 c_{23} s_{12} s_{13} s_{23} \sin \delta_{CP}^q,
\ee
defining $s_{ij} \equiv \sin \theta_{ij}^q$ and $c_{ij} \equiv \cos \theta_{ij}^q$. The leading-order contribution to $J$ is $\mcl{O}(|\veps_X|^3)$. In order to compute it, it is enough to consider the CKM matrix to $\mcl{O}(|\veps_X|)$, use $\{i,j,k,l\} = \{1,2,2,3\}$ in Eq.~\eqref{defineJ} and replace the values for the mixing angles found in Eqs.~\eqref{th12}-\eqref{th23}.


\section{Leptons}
\label{Sec-leptons}
\setcounter{equation}{0}


The formalism established in the last two sections for the quarks generalizes straightforwardly to the charged leptons. Neutrinos, however, are a different story since the character of the neutrino masses is not known. We consider both the Dirac and Majorana options.


\subsection*{Pure Dirac Masses}


 For purely Dirac neutrinos, the zeroth-order neutrino masses are given by Eq.~\eqref{masses} and the effects resulting from the perturbations are given by Eq.~\eqref{masses-pert}. The leptonic mixing angles are given by the same expressions as in Eqs.~\eqref{th12}-\eqref{th23} with the replacements $d\ra\ell$, $u\ra\nu$:
\begin{align}
  \sin \theta_{12}^l & = \frac{1}{\sqrt{2}} \left|(y_{12} - iy_{13})\left(\frac{\veps_\ell}{\xi_\ell} - \frac{\veps_\nu}{\xi_\nu} \right)\right|  \label{sin12lep}   \\
  \sin \theta_{13}^l & = \frac{1}{\sqrt{2}}\big|(y_{12} + iy_{13})(\veps_\ell - \veps_\nu) \big|   \label{sin13lep}  \\
  \sin \theta_{23}^l  & = \frac{1}{4}\big| (y_{11} + 2y_{22} + 2iy_{23})(\veps_\ell -\veps_\nu) \big|  \label{sin23lep}
\end{align}
Following the analysis performed in the previous section, it is clear that these formulae cannot reproduce the observed mixing properties of the leptons. In particular, in the lepton sector, all three observable angles are sizable. Hence, while the enhancement of $\theta_{12}^l$ shown in Eq.~\eqref{sin12lep} is still relevant and desirable, $\theta_{23}^l$ is still predicted to be perturbatively small in Eq.~\eqref{sin23lep}. This angle is known to be  close to maximal and cannot be explained in the perturbative framework we have introduced.


\subsection*{Majorana Masses}


\newcommand{\tveps}{\tilde{\veps}}
\newcommand{\Mnu}{\mcl{M}_\nu}
For the case of Majorana neutrinos, several possibilities can be investigated for the structure of the Yukawa matrices. Let us consider, then, a general Majorana mass term for the light neutrino states. It can be written as
\be
v\mu_\nu\overline{\nu} \mcl{M}_\nu \nu^c, \label{nu-mass}
\ee
where $\Mnu$ is a flavor-charged, symmetric matrix, $\mcl{M}_\nu\equiv \mcl{M}_\nu(Y_3, Y_5)$, and  $\mu$ is a possibly small, dimensionless parameter. This mass term is not invariant under the gauge symmetry of the SM and extra fields should be added in order to compensate the $SU(2)_W$ and hypercharge charges of the neutrino states. This leads to nonrenormalizable neutrino mass operators. The simplest such possibility is to add two Higgs fields to form the dimension-five Weinberg operator,
\be
\frac{c_W^{(5)}}{\Lambda_{LN}}\overline{L} L^c HH , \quad \quad \frac{v c_W^{(5)}}{\Lambda_{LN}} = \mu_\nu \mcl{M}_\nu.
\ee
As is well known, there are 3 ways to generate the Weinberg operator at tree level, the so-called type I, II and III seesaw mechanisms.

In the Flavorspin context, $\Mnu$ should be considered a polynomial function of the spurions $Y_3$ and $Y_5$ with complex coefficients. In what follows, we are interested in truncations of this polynomial inspired by the seesaw mechanisms. They can all be parametrized by an $\Mnu$ of the form:
\be
\mcl{M}_\nu = \eta_{33}Y_3Y_3^T + \eta_5 Y_5 + \eta_1 \mbb{I} + \eta_{35}(Y_3Y_5^T + Y_5Y_3^T) +  \eta_{55}Y_5Y_5^T + \eta_{335}(Y_3Y_3^TY_5^T + Y_5Y_3 Y_3^T) + \eta_{353} Y_3 Y_5 Y_3^T \,. \label{Mnu} 
\ee
where for simplicity we have kept terms of at most third-order in $Y_3, Y_5$ and that are at most second-order in $Y_5$.

A naive consideration and a driving idea of this work is the fact that in the limit
\be
\{ \eta_5,\,\eta_1,\, \eta_{35},\, \eta_{55},\, \eta_{335}, \, \eta_{353} \} \rightarrow 0,
\ee
one obtains
\be
\mcl{M}_\nu \ra \mcl{M}_\nu^0 \equiv \eta_{33}Y_3Y_3^T = \left( \begin{array}{ccc}
  0 & 0 & 0 \\
  0 & \eta_{33} & 0 \\
  0 & 0 & \eta_{33}
\end{array} \right) \,. \label{Mnu0}
\ee
That is, it appears that the mass matrix for Majorana neutrinos is already diagonal in the basis employed here. Eq.~\eqref{Mnu0} suggests an inverted hierarchy of neutrino masses with two nonzero eigenvalues and another one vanishingly small. Since the flavor and the mass bases for the neutrinos are the same in this limit, the leptonic mixing matrix, analogous to $V_{CKM}$ in Eq.~\eqref{CKM}, would then be given by:
\be
V_{PMNS} = V_{\ell L} = V^0 = \left( \begin{array}{ccc}
  1 & 0 & 0 \\
  0 & \frac{1}{\sqrt{2}} & \frac{1}{\sqrt{2}} \\
  0 & -\frac{1}{\sqrt{2}} & \frac{1}{\sqrt{2}} \\
\end{array} \right) \cdot \left( \begin{array}{ccc}
  1 && \\ & -i & \\ && 1
  \end{array} \right) \,. \label{PMNS0}
\ee
In particular, as opposed to the quark case, Flavorspin appears to automatically predicts one large mixing angle -- maximal, at zeroth order -- in the lepton sector with Majorana neutrinos.

On the other hand, it is wrong to associate the angle in Eq.~\eqref{PMNS0} with $\theta_{23}^l$. Since the favored pattern in this case is an inverted hierarchy and Eq.~\eqref{PMNS0} mixes the two eigenvalues different than zero, this mixing angle is more naturally identified with $\theta_{12}^l$. Therefore, presumably large deviations from the zeroth-order structure are required to generate close-to-maximal $\theta_{23}^l$. 

Nonetheless, the fact that, as opposed to the quark case, one mixing angle automatically comes out large is encouraging. Coming back to the general form in Eq.~\eqref{Mnu}, we explore the following cases in detail:
\begin{itemize}
\item \emph{Type I Seesaw:} $\Mnu$ is assumed to be given by
  \be
  \Mnu = Y_\nu Y_\nu^T \label{typeI}
  \ee
  with $Y_\nu$ defined in Eqs.~\eqref{YX} and \eqref{Ydefs}. More explicitly, the coefficients in Eq.~\eqref{Mnu} are given by:
  \be
  \eta_{33} = 1 + 2\bveps_\nu - (1-\xi_\nu)^2\,, \quad \eta_5 = \veps_\nu\bveps_\nu \,,\quad \eta_1 = \bveps_{\nu}^2 \label{etastypeI1}\,, 
  \ee
  \be
  \eta_{35} = \veps_\nu\,,\quad \eta_{55} = \veps_\nu^2\,,\quad \eta_{335} = (1-\xi_\nu)\veps_\nu \label{etastypeI2}, \quad \eta_{353} = 0,
  \ee
  where we have used that $(Y_3Y_3^T)^2 = -Y_3Y_3^T$. Importantly, in this case, the coefficients in $\Mnu$ are correlated and depend only on the complex numbers $\xi_\nu$, $\veps_\nu$ and $\bveps_\nu$, where $|\xi_\nu|, \, |\veps_\nu|, \, |\bveps_\nu| \ll 1$.
\item \emph{Type II Seesaw:} In this scenario, $\Mnu$ is identified with $Y_\nu + Y_\nu^T$. There is no term proportional to $Y_3$ because $\Mnu$ is symmetric and, moreover, we set
  \be
  \eta_{35} = \eta_{55} = \eta_{335} = \eta_{353} = 0 \,. \label{typeIIconds}
  \ee
  For the other parameters, only $|\eta_{33}|$, $|\eta_5|$,  $|\eta_1| \lesssim 1$ is assumed.
\end{itemize}
Although we are referring to these two scenarios as type I and type II seesaws, these naming conventions should not be taken too literally. In particular, there is no strong argument for the conditions imposed in Eq.~\eqref{typeIIconds} other than simplicity. We will briefly comment on the most general case, as defined in Eq.~\eqref{Mnu}, in Sec.~\ref{results}.


\subsubsection*{Type I Seesaw}


The explicit form of the type I seesaw mass matrix in Eq.~\eqref{typeI} is
\bea
\label{typeImatrix}
\Mnu &=& \frac{\mu_\nu^2 v^2}{\Lambda_{LN}} \left( \begin{array}{ccc}
0 & \veps_\nu \left( i y_{13} - y_{12} \right) & - \veps_\nu (i y_{12} + y_{13}) \\
\veps_\nu \left( i y_{13} - y_{12} \right) & -2(y_{22} \veps_\nu -i y_{23} \veps_\nu + \overline{\veps}_\nu + \xi_\nu) & -\veps_\nu (i y_{11} + 2i y_{22} + 2 y_{23}) \\
- \veps_\nu (i y_{12} + y_{13}) & -\veps_\nu (i y_{11} + 2i y_{22} + 2 y_{23}) & 2 (y_{11} \veps_\nu + y_{22} \veps_\nu - i y_{23} \veps_\nu - \overline{\veps}_\nu - \xi_\nu) 
\end{array} \right) \nn \\
& + & \mcl{O}\left( \{ |\xi_\nu|, |\veps_\nu|, |\overline{\veps}_\nu| \}^2 \right),
\eea
where $\Lambda_{LN}$ is the scale of LN-violating physics. We consider the masses in the limit $ y_{11}, \, y_{12}, \, y_{13}, \, |\xi_\nu|,  \, |\veps_\nu|, \, |\overline{\veps}_\nu|  \ll 1$; we will justify these limits of $y_{11}$, $y_{22}$ and $y_{23}$. The neutrino masses become
\bea
\label{typeImasses}
m_1^2 &=& \mcl{O}\left(  \{y_{12},\, y_{13}\}^2 \right), \\
m_2^2 &=& \frac{\mu_\nu^4 v^4}{\Lambda_{LN}^2} \big(F_\nu - G_\nu \big) + \mcl{O}\left( \{ y_{11}, \,y_{12},\, y_{13} \}^2;\; \{ |\xi_\nu|,\, |\veps_\nu|,\, |\overline{\veps}_\nu| \}^3 \right), \\
m_3^2 &=& \frac{\mu_\nu^4 v^4}{\Lambda_{LN}^2} \big(F_\nu + G_\nu \big) + \mcl{O}\left( \{ y_{11},\, y_{12},\, y_{13} \}^2;\; \{ |\xi_\nu|,\, |\veps_\nu|,\, |\overline{\veps}_\nu| \}^3 \right),
\eea
where we have introduced the quantities
\bea
\label{FGnu}
F_\nu & = & 8 |\veps_\nu|^2 (y_{22}^2 + y_{23}^2) + 4 |\xi_\nu + \overline{\veps}_\nu^*|^2, \\
G_\nu & = & 8 |\veps_\nu| \sqrt{\left( y_{22}^2 + y_{23}^2 \right) \big[ |\veps_\nu|^2 (y_{22}^2 + y_{23}^2) + |\xi_\nu + \overline{\veps}_\nu^*|^2 \big]}.
\eea
The ordering of the neutrino mass eigenstates here differs slightly from that used in analyses of neutrino oscillations.\footnote{We remind the reader that for neutrino oscillations, the two closest values of $m^2$ are defined to be $m_1^2$ and $m_2^2$, with $m_1^2$ being the lighter of the two. The third is then defined to be $m_3^2$, which may be heavier or lighter than the other two.} Here, the masses are strictly ordered from least to greatest: $m_1^2 < m_2^2 < m_3^2$. For the latter, however, the ordering depends on the hierarchy; for the normal hierarchy (NH), the ordering is the same as the one used here, but for the inverted hierarchy (IH) the masses are ordered $m_3^2 < m_1^2 < m_2^2$. It will be important to establish, for the rest of this work, the conditions for the NH and IH in the ordering scheme we employ:
\bea
m_3^2 - m_2^2 > m_2^2 - m_1^2 & \implies & \text{NH,} \nn \\
m_3^2 - m_2^2 < m_2^2 - m_1^2 & \implies & \text{IH.}
\label{hierarchy}
\eea
For the masses in Eq.~\eqref{typeImasses} to constitute a NH, the condition $3G_\nu > F_\nu$ must be satisfied; otherwise, neutrinos are organized in an IH. It is instructive to consider the neutrino mass spectrum in various limit. When $|\xi_\nu + \overline{\veps}_\nu^*| \ll |\veps_\nu|$, then $G_\nu \sim F_\nu$, and the masses may constitute a NH. In the opposite limit, 
\be
G_\nu/F_\nu \to 2 |\veps_\nu| \cdot (y_{22}^2 + y_{23}^2)/|\xi_\nu + \overline{\veps}_\nu^*| \ll 1,
\ee
and the masses organize in an IH.


\subsubsection*{Type II Seesaw}

The explicit form of the type II seesaw mass matrix, from Eqs.~\eqref{Mnu} and \eqref{typeIIconds}, is
\bea
\label{typeIImatrix}
\Mnu &=& \left( \begin{array}{ccc}
\eta_1+ \eta_5 y_{11} & \eta_5 y_{12} & \eta_5 y_{13} \\
\eta_5 y_{12} & \eta_1 + \eta_5 y_{22} + \eta_{33} & \eta_5 y_{23} \\
\eta_5 y_{13} & \eta_5 y_{23} & \eta_1 - \eta_5(y_{11} + y_{22}) + \eta_{33}
\end{array} \right).
\eea
For $ y_{11},\, y_{12},\, y_{13} \ll 1$, as well as $|\eta_{33}|,\, |\eta_5|,\, |\eta_1| \lesssim 1$, the neutrino masses become
\bea
\label{typeIImasses}
m_1^2 &=& \frac{\mu_\nu^4 v^4}{\Lambda_{LN}^2} \times \eta_1^2 + \mcl{O}\left(  y_{11}, y_{12}, y_{13}  \right), \\
m_2^2 &=& \frac{\mu_\nu^4 v^4}{\Lambda_{LN}^2} \big(H_\nu - K_\nu \big) + \mcl{O}\left(  y_{11},\, y_{12},\, y_{13};\; \{ |\eta_{33}|, \,|\eta_5|,\, |\eta_1| \}^3 \right), \\
m_3^2 &=& \frac{\mu_\nu^4 v^4}{\Lambda_{LN}^2} \big(H_\nu + K_\nu \big) + \mcl{O}\left(  y_{11},\, y_{12},\, y_{13}; \;\{ |\eta_{33}|,\, |\eta_5|,\, |\eta_1| \}^3 \right),
\eea
where we have introduced the quantities
\bea
\label{HKnu}
H_\nu & = & |\eta_1 + \eta_{33}^* |^2 + |\eta_5|^2 (y_{22}^2 + y_{23}^2), \\
K_\nu & = & \big| (\eta_5 + \eta_{33}) \eta_1^* + \eta_1 (\eta_5^* + \eta_{33}^*) \big| \sqrt{(y_{22}^2 + y_{23}^2)}.
\eea
These quantities must satisfy $3K_\nu > H_\nu - |\eta_1|^2$ for neutrinos to form a NH, else neutrinos are organized in an IH. Note the difference with the type I seesaw; here, the singlet term corresponding to the coefficient $\eta_1$ is not suppressed. An IH would follow if the singlet term, $\eta_1$, were subdominant to $\eta_{33}$ and $\eta_5$, since then $H_\nu$ would dominate $K_\nu$. If $|\eta_{33}|$, $|\eta_5|$ and $|\eta_{1}|$ are all comparable in magnitude, then it is possible that $H_\nu$ and $K_\nu$ are likewise comparable. In this case, a NH would be produced. 


\subsubsection*{Leptonic Mixing Matrix}


The matrix $V_\nu$ that diagonalizes the neutrino mass matrix is defined via
\be
\label{defineVnu}
V_\nu \cdot (Y_\nu Y_\nu^T) \cdot (Y^*_\nu Y_\nu^\dagger) \cdot V_\nu^\dagger = \mcl{P}^T \cdot \text{diag}(m_1^2, m_2^2, m_3^2) \cdot \mcl{P}.
\ee
Here, $\mcl{P}$ is a permutation matrix that reorders the neutrino masses according to the standard mass ordering conventions used in neutrino oscillations, following the discussion surrounding Eq.~\eqref{hierarchy}:
\bea
\label{defineP}
m_3^2 - m_2^2 > m_2^2 - m_1^2: \mcl{P} &=& \mbb{I}, \\
m_3^2 - m_2^2 < m_2^2 - m_1^2: \mcl{P} &=& \left( \begin{array}{ccc} 0 & 0 & 1 \\ 1 & 0 & 0 \\ 0 & 1 & 0 \end{array} \right).
\eea
For nontrivial $V_\nu$, the leptonic mixing matrix $V_{PMNS}$ is given by
\be
\label{leptonmixing}
V_{PMNS} = V_{\ell L} V_\nu^\dagger,
\ee
similar to Eq.~\eqref{PMNS0}. We can anticipate some of the numerical results of Sec.~\ref{results} by inspecting Eqs.~\eqref{typeImatrix} and \eqref{typeIImatrix}. For the type I seesaw, and assuming a NH, $\sin \theta_{12}^l$ is approximately given by 
\be
\sin \theta_{12}^l \sim \left| \frac{\Mnu^{12}}{\Mnu^{22}} \right| = \frac{1}{2} \frac{|\veps_\nu| \sqrt{y_{12}^2+y_{13}^2}}{ |(y_{22}+i y_{23}) \veps_\nu-\overline{\veps}_\nu - \xi_\nu|}.
\ee
In the limit $y_{12}, \, y_{13} \ll y_{22}, \, y_{23}$, this mixing angle is predicted to be small.\footnote{Recall that the charged-lepton contribution to this mixing angle given by Eq.~\eqref{sin12lep} with $\veps_\nu \to 0$. The charged-lepton and neutrino contributions to this angle are roughly comparable, as we will see in the next section, so this conclusion is robust.} A similar estimate for the type II seesaw, again assuming a NH, gives
\be
\sin \theta_{12}^l \sim \left| \frac{\Mnu^{12}}{\Mnu^{22}} \right| = \left| \frac{\eta_5 y_{12}}{ \eta_1 + \eta_5 y_{22} + \eta_{33}} \right| .
\ee
From this equation, assuming $|\eta_{33}|$, $|\eta_5|$ and $|\eta_1|$ are comparable in magnitude, if $|y_{22}|$ were approximately 1, the terms in the denominator could give a sizable cancellation. In the next section, we will show that, indeed, $|y_{22}|$ must be approximately 1. In this case, even with $y_{12} \ll 1$, a sizable neutrino contribution to $\sin \theta_{12}^l$ is possible.


\section{Results}
\setcounter{equation}{0}
\label{results}


In this section, we numerically explore the parameter space of Flavorspin and show that it provides a plausible description of all the SM flavor. We stress that the goal is not to show that there is a set of precise values for the Flavorspin parameters that exactly reproduces the low-energy observables of interest to us. Setting aside that the low-energy observables are known with finite precision, attempting to find such a solution is computationally expensive and ultimately unenlightening. We demonstrate instead that general agreement between the predictions of Flavorspin and the experimentally-determined values of low-energy observables can be obtained by looking at specific regions of parameter space that satisfy the constraints enumerated in Sec.~\ref{flavorspin}. 

The method used is as follows. First, random values are generated over specified ranges of the Flavorspin parameters, assuming a flat prior in these ranges, and the values for the relevant observables are calculated for all these points. These pseudodata are binned together in two-dimensional subspaces of the space of observables, and a likelihood $L$ is assigned to each bin, proportional to its population, $N$. A $\Delta \chi^2$ is calculated for each bin, via \cite{Agashe:2014kda}
\be
\label{definechi2}
\Delta \chi^2 = 2 \ln L_0/L \approx 2 \ln N_0/N,
\ee
where $N_0$ and $L_0$ are the population and likelihood, respectively, of the bin with the highest population; $\Delta \chi^2 = 0$ corresponds to the center of this bin. A smooth interpolation of the $\Delta \chi^2$ function is calculated, and the contours along which $\Delta \chi^2 = 2.30$, 5.99 and 9.21 are drawn, corresponding approximately to the 68.3\%, 95\% and 99\% confidence intervals (CI);\footnote{This correspondence is only exact in the limit of vanishing bin size and a large number of pseudodata points, but this yields a sufficiently precise approximation of the true confidence intervals for our purposes here.} in the figures that follow, these will be represented by dark, medium and light shadings of the color assigned to each scenario, respectively. The goal of this section is to show that there are ranges of the Flavorspin parameters such that these confidence intervals contain the experimentally-determined values of the observables of interest with a high degree of confidence.


\subsection*{Charged Fermions}


\begin{table}[]
\begin{center}
\begin{tabular}{|c||c|}\hline
Observable & $\overline{MS}$ Value, $\mu$ = 1 TeV \\ \hline \hline
$y_u$ & $6.3 \times 10^{-6}$ \\ \hline
$y_c$ & $3.104 \times 10^{-3}$ \\ \hline
$y_t$ & $0.8685$ \\ \hline \hline
$y_d$ & $1.364 \times 10^{-5}$ \\ \hline
$y_s$ & $2.74 \times 10^{-4}$ \\ \hline
$y_b$ & $1.388 \times 10^{-2}$ \\ \hline \hline
$y_e$ & $2.8482 \times 10^{-6}$ \\ \hline
$y_\mu$ & $6.0127 \times 10^{-4}$ \\ \hline
$y_\tau$ & $1.02213 \times 10^{-2}$ \\ \hline \hline
$\sin \theta_{12}^q$ & $0.2254$ \\ \hline
$\sin \theta_{13}^q$ & $3.770 \times 10^{-3}$ \\ \hline
$\sin \theta_{23}^q$ & $4.363 \times 10^{-2}$ \\ \hline
$\sin \delta_{CP}^q$ & $0.9349$ \\ \hline
\end{tabular}
\end{center}
\caption{The values for the low-energy observables used in our analysis. We use the $\overline{MS}$ values at $\mu$ = 1 TeV calculated in Ref.~\cite{Antusch:2013jca}; see text for details.}
\label{table:Observables}
\end{table}

The starting point is the determination of the approximate ranges in which the Flavorspin parameters must lie in order to reproduce the quark masses and mixing parameters. Table~\ref{table:Observables} shows the $\overline{MS}$ values of charged-fermion masses and quark mixing observables at the scale $\mu$ = 1 TeV, which we have taken from Ref.~\cite{Antusch:2013jca}. The ranges of the parameters that we consider are as follows:
\be
\label{ranges1}
|\xi_u| \in [6 \times 10^{-3}, 8 \times 10^{-3}]\,,\quad  |\veps_u| \in [1 \times 10^{-3}, 2 \times 10^{-3}]\,,\quad |\overline{\veps}_u| \in [1 \times 10^{-5}, 2 \times 10^{-5}],
\ee
\be
|\xi_d| \in [0.035, 0.037]\,,\quad |\veps_d| \in[0.06, 0.07]\,,\quad |\overline{\veps}_d| \in  [6 \times 10^{-4}, 7 \times 10^{-4}],
\ee
\begin{align}
\label{ranges2}
y_{11} & \in [-0.01, 0.01], \nn \\
y_{22} & = 1, \nn \\
y_{23} & \in \pm [0.88, 0.92], \\
\Phi \equiv \sqrt{y_{12}^2 + y_{13}^2} & \in [0.15, 0.16], \nn \\
\varphi \equiv \arctan \left[ \frac{y_{13}}{y_{12}} \right] & \in [-\pi, \pi]. \nn
\end{align}
We arrive at these ranges guided by the following considerations:
\begin{enumerate}
\item The ratios of the masses determine the $|\xi_X|$. For instance, $|\xi_X|$ is, to a good approximation, equal to the ratio of the second- and third-generation quark masses, as per Eq.~\eqref{XiRatio}.
\item The largest of the $y_{ij}$ parameters can be defined to be equal to unity using the freedom to rescale all the $\veps_X$ and $y_{ij}$  to ensure this. From Eqs.~\eqref{mass-est} and \eqref{Ratio_13_23_Quark}, we estimate that $y_{22}$, $y_{23} \sim \mcl{O}(1)$; $y_{12}$, $y_{13} \sim \mcl{O}(0.1)$; and $y_{11} \lesssim \mcl{O}(0.01)$.
\item The required sizes of the $|\veps_X|$ are estimated via their contribution to the mixing angles and to the first generation masses. For instance, since $m_d^2/m_b^2 \gg m_u^2/m_t^2$, we must have $|\veps_d| \gg |\veps_u|$. Therefore, $|\veps_d|$ dominates over $|\veps_u|$ in Eqs.~\eqref{th13} and \eqref{th23}. This allows for an estimate of $|\veps_d|$ by comparing these expressions with their best-fit values in Table~\ref{table:Observables}. 
\item In the leading approximation, the first-generation masses are given by $|y_{11} \veps_X + \overline{\veps}_X|$. Therefore, $y_{11} \lesssim \mcl{O}(0.01)$ must hold in order to suppress the down quark mass, given the value of $|\veps_d|$ in Eq.~\eqref{ranges1}. Furthermore, we require that $|\overline{\veps}_d|$ be at least a factor of $10^{-2}$ smaller than $|\veps_d|$; otherwise, a tuned cancellation would be required to get a small down-quark mass.
\item Given the range of $y_{11}$ and the value of the up-quark mass, ranges for $|\veps_u|$ and $|\overline{\veps}_u|$ are determined.
  \item  The phases on $\xi_u$, $\xi_d$, $\veps_u$, $\veps_d$, $\overline{\veps}_u$ and $\overline{\veps}_d$ are allowed to vary uniformly on $[-\pi,\pi]$.
\end{enumerate}
The take-home message is that it is possible to find regions of parameter space such that the quark observables are reproduced. The ranges for the parameter ranges are later refined by comparing calculations of the observables against the values of Table~\ref{table:Observables} until general agreement between the two is attained.

\begin{figure}[]
\begin{center}
\includegraphics[width=4in]{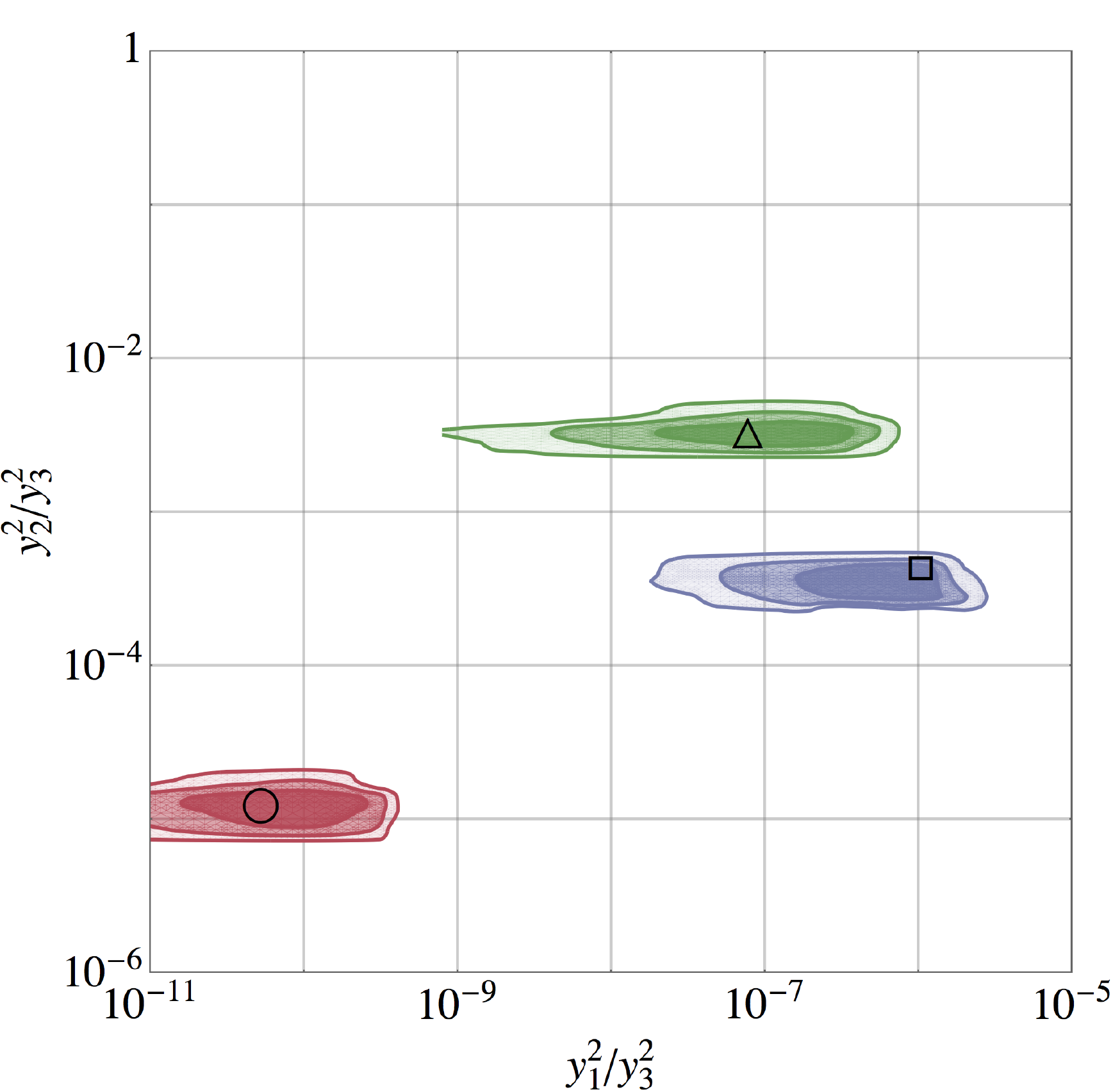}
\end{center}
\caption{The predicted ratio of Yukawa couplings for the up-type quarks (red), down-type quarks (blue) and charged leptons (green), calculated from the ranges in Eqs~\eqref{ranges1}, \eqref{ranges2} and \eqref{ranges3}. The phases on $\xi_u$, $\xi_d$, $\veps_u$, $\veps_d$, $\overline{\veps}_u$ and $\overline{\veps}_d$ are allowed to vary uniformly on $[-\pi,\pi]$. The 68.3\% (dark), 95\% (medium) and 99\% (light) confidence intervals for each sector are shown. The circle, square and triangle indicate the values of the Yukawa ratios for the up-type quarks, down-type quarks and charged leptons, respectively, as given in Table~\ref{table:Observables}.}
\label{fig:Masses}
\end{figure}

\begin{figure}[]
\begin{center}
\subfigure[]{\includegraphics[width=3.15in]{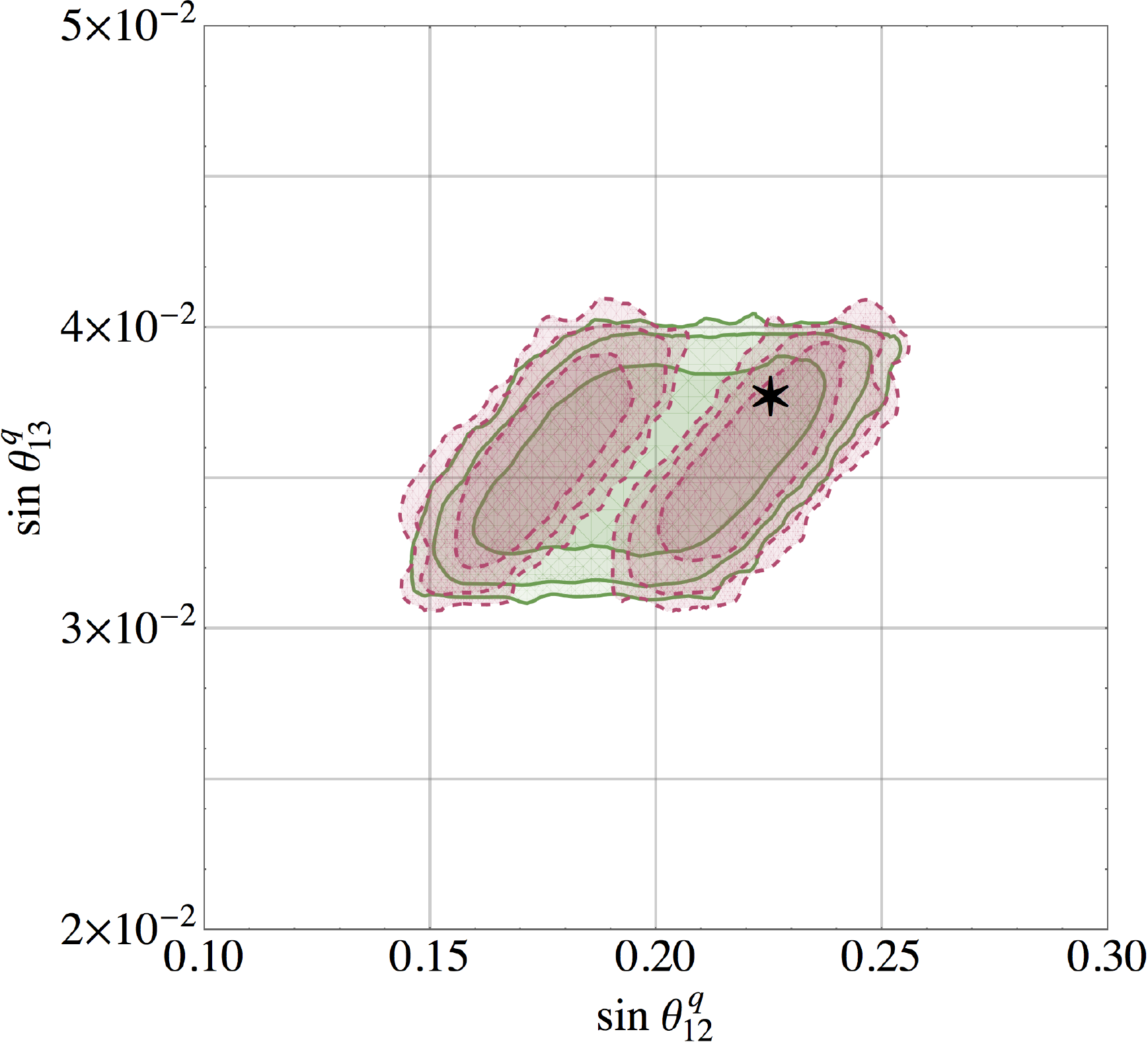}}
\subfigure[]{\includegraphics[width=3in]{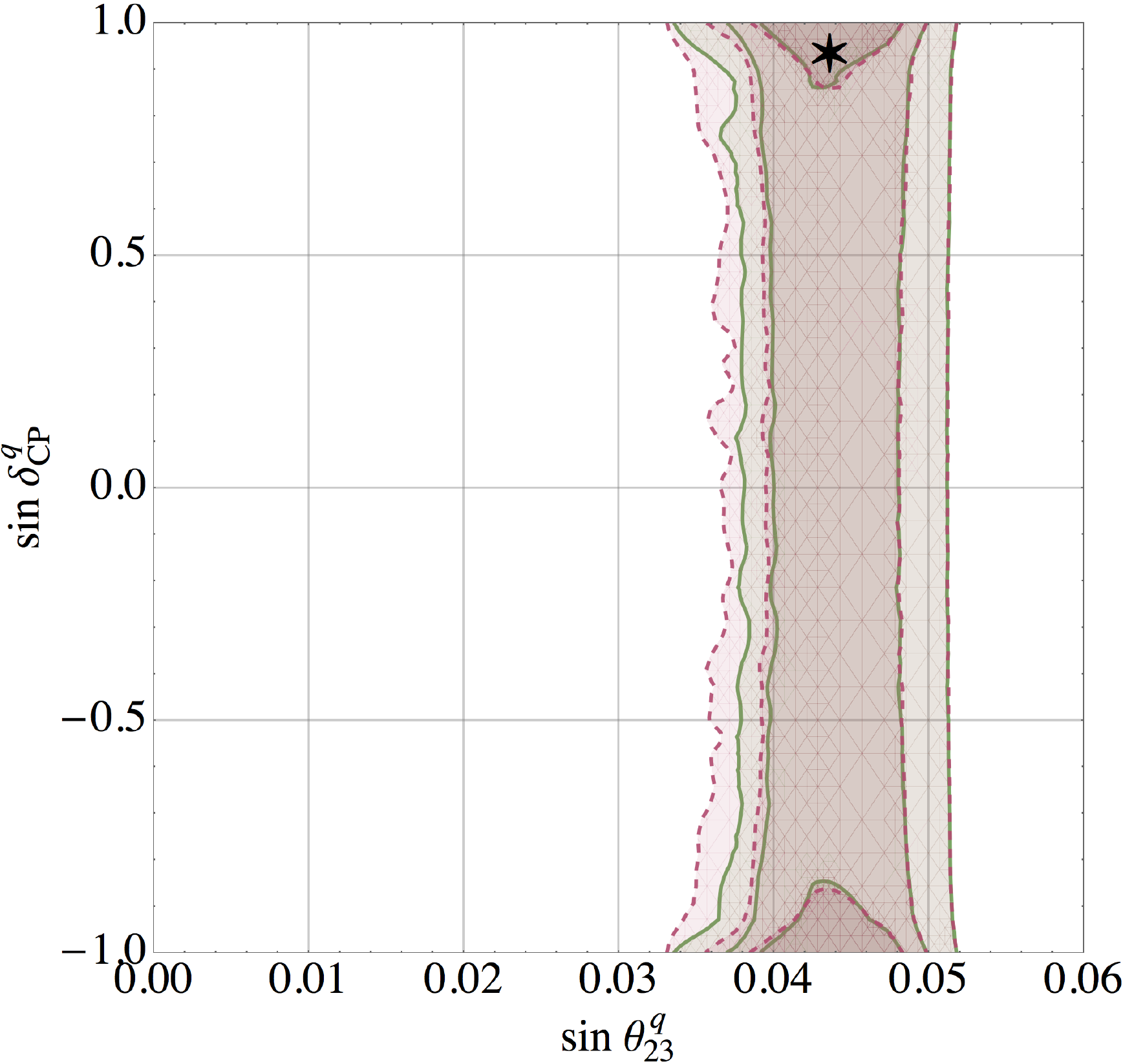}}
\end{center}
\caption{Quark mixing parameters, calculated from the ranges in Eq.~\eqref{ranges1} and \eqref{ranges2}. The six-pointed star in each figure represents the measured values of these observables from Table~\ref{table:Observables}. The red regions with dashed outlines are calculated assuming $\xi_X$, $\veps_X$ and $\overline{\veps}_X$, ($X = u, d$) are real, while the green regions with solid outlines are calculated allowing these quantities to be complex. Note that the former almost completely covers the latter in panel (b).}
\label{fig:QuarkAngles}
\end{figure}

Fig.~\ref{fig:Masses} shows the regions of $y_1^2/y_3^2$--$y_2^2/y_3^2$ space covered by the ranges for the parameters in Eqs.~\eqref{ranges1} and \eqref{ranges2} for the up-type (red) and down-type (blue) quark masses. Fig.~\ref{fig:QuarkAngles}(a) shows the regions of $\sin \theta^q_{12}$--$\sin \theta^q_{13}$ space covered by the same choices of parameter regions as in Eqs.~\eqref{ranges1} and \eqref{ranges2}, and Fig.~\ref{fig:QuarkAngles}(b) is the same in $\sin \theta^q_{23}$--$\sin \delta_{CP}^q$ space. The red regions with dashed outlines take $\xi_u$, $\xi_d$, $\veps_u$, $\veps_d$, $\overline{\veps}_u$ and $\overline{\veps}_d$ to be real, while the green regions with solid outlines allow these parameters to be complex with a phase on $[-\pi,\pi]$. Note that the red region almost completely covers the green region in Fig.~\ref{fig:QuarkAngles}(b). The six-pointed star in each panel represents the best-fit point from Table~\ref{table:Observables}. 

The phase parameters in the quark sector cannot be constrained by this analysis. Part of the reason for this is that $|\veps_d|$ ($|\veps_d/\xi_d|$) numerically dominates $|\veps_u|$ ($|\veps_u/\xi_u|$) in Eqs.~\eqref{th13} and \eqref{th23} (Eq.~\eqref{th12}), so the magnitude of their difference is largely insensitive to their relative phase. Moreover, the phase of $\xi_X$ is irrelevant in determining $m_2^2/m_3^2$, and the phases on $\veps_X$ and $\overline{\veps}_X$ do not dramatically alter the range of possible values for $y_1^2/y_3^2$.  This insensitivity to the phases is demonstrated in Fig.~\ref{fig:QuarkAngles}(b). Although there is a small preference for maximal $CP$ violation, all possible values of $\sin \delta^q_{CP}$ are contained in the 95\% CI for real-valued $\xi_X$, $\veps_X$ and $\overline{\veps}_X$. Letting these parameters be complex produces no appreciable changes. The $CP$ violation that arises when these parameters stems from the imaginary coefficient of $Y_3$ that appears in Eq.~\eqref{Ydefs}. This factor of $i$, coupled with the finite spread and sign indeterminacy of the ranges in Eq.~\eqref{ranges2}, is enough to populate the entire allowable range for $\sin \delta^q_{CP}$. Regarding the elements of $Y_5$, at this order, separate ranges for $y_{12}$ and $y_{13}$ are not specified in Eq.~\eqref{ranges2}. These parameters only appear in the particular combination $(y_{12}^2 + y_{13}^2)$ in Eqs.~\eqref{defineFG}, \eqref{th12} and \eqref{th13}. Therefore, we reparametrize these as
\bea
y_{12} =  \Phi \cos \varphi, & & y_{13} = \Phi \sin \varphi.
\eea
The quark masses and mixing observables inform the range of $\Phi$, but the angle $\varphi$ is completely undetermined.

Next, we use the parameter ranges for the $y_{ij}$ found for the quarks to compute the charged lepton masses. This system of equation still has enough freedom due to the new parameters  $\xi_\ell$, $\veps_\ell$ and $\overline{\veps}_\ell$ that determine the lepton spectrum. We obtain the following ranges for the latter:
\bea
\label{ranges3}
|\xi_\ell| \in [0.11, 0.12], & |\veps_\ell| \in[0.05, 0.06], & |\overline{\veps}_\ell| \in  [5 \times 10^{-4}, 6 \times 10^{-4}].
\eea
The region of $y_1^2/y_3^2$--$y_2^2/y_3^2$ space covered by these parameter ranges and those in Eq.~\eqref{ranges2} are shown in green in Fig.~\ref{fig:Masses}. The phases on $\xi_\ell$, $\veps_\ell$ and $\overline{\veps}_\ell$ are varied between $[-\pi, \pi]$. The triangle represents the observed ratios of charged-lepton masses in Table~\ref{table:Observables}. 


\subsection*{Neutrinos}


The parameter ranges obtained in Eqs.~\eqref{ranges2} and \eqref{ranges3} for $Y_5$ are now used to determine the neutrino masses and leptonic mixing observables. Table~\ref{table:Neutrinos} lists the current best-fit values for the neutrino mass-squared differences and leptonic mixing angles determined by the NuFIT collaboration \cite{Esteban:2016qun} both for a NH and for an IH of neutrino masses. While the calculation of the renormalization-group evolution of these observables has been calculated in, for instance, Ref.~\cite{Antusch:2005gp,Mei:2005qp,Ellis:2005dr,Xing:2007fb,Lin:2009sq,Ohlsson:2013xva}, we use the low-energy values in order to keep pace with current experimental observations and to avoid making model-dependent assumptions about the renormalization group flow.

\begin{table}[b]
\begin{center}
\begin{tabular}{|c||c|c|}\hline
Observable & Normal Hierarchy ($\Delta \chi^2 = 0$) & Inverted Hierarchy ($\Delta \chi^2 = 0.83$) \\ \hline \hline
$\Delta m^2_{12}$ & $7.50 \times 10^{-5}$ eV$^2$ & $7.50 \times 10^{-5}$ eV$^2$ \\ \hline
$\Delta m^2_{13}$ & $+2.524 \times 10^{-3}$ eV$^2$ & $-2.444 \times 10^{-3}$ eV$^2$ \\ \hline \hline
$\sin^2 \theta_{12}^l$ & 0.306 & 0.306 \\ \hline
$\sin^2 \theta_{13}^l$ & 0.02166 & 0.02179 \\ \hline
$\sin^2 \theta_{23}^l$ & 0.441 & 0.587 \\ \hline
$\sin \delta_{CP}^l$ & $-0.988$ & $-0.993$ \\ \hline
\end{tabular}
\end{center}
\caption{The values for the neutrino observables used in our analysis, from the NuFIT collaboration \cite{Esteban:2016qun}. Shown are the NH and IH fits to oscillation data. We do not consider the renormalizaton-group-evolved values of these observables as we did with the quark and charged-lepton observables.}
\label{table:Neutrinos}
\end{table}

We studied numerically the seesaw scenarios described in Sec.~\ref{Sec-leptons} using the same  method we used for the charged fermions. As before, the values of the neutrino-specific parameters, as well as the parameters of Eqs.~\eqref{ranges2} and \eqref{ranges3}, are scanned over specified ranges. The parameters $\{\xi_X, \, \veps_X, \, \overline{\veps}_X, \, \eta_X\}$ are all allowed to be complex with their phases on $[-\pi, \pi]$. For each set of parameters, the low-energy observables are calculated. These observables are the three leptonic mixing angles (via $\sin^2 \theta_{12}^l$, $\tan^2 \theta_{13}^l$ and $\sin^2 \theta_{23}^l$), the lone leptonic $CP$-violating phase ($\sin \delta_{CP}^l$) and the ratio $R_\nu$ of the neutrino mass-squared splittings,
\be
R_\nu = \left\{ \begin{array}{l} \dfrac{m_3^2 - m_1^2}{m_2^2 - m_1^2}, \quad m_3^2-m_2^2 > m_2^2 - m_1^2 \\ \\ \dfrac{m_1^2 - m_2^2}{m_3^2 - m_2^2}, \quad m_3^2-m_2^2 < m_2^2 - m_1^2 \end{array} \right. .
\ee
In this convention, $R_\nu$ is positive (negative) for the NH (IH), and its magnitude is strictly greater than two.

The pseudodata then are binned in $R_\nu$ and, using Eq.~\eqref{definechi2}, $\Delta \chi^2$ is calculated for each bin (with the most populous bin having $\Delta \chi^2 = 0$). A smooth interpolation of the $\Delta \chi^2$ is calculated, and the 68.3\%, 95\% and 99\% confidence levels (CL) are set at $\Delta \chi^2 = 1.00$, 3.84 and 6.63, respectively; in figures, these are respectively drawn as solid, dashed and dot-dashed black lines. A flat posterior is imposed on $R_\nu$, so that only pseudodata for which $30 < |R_\nu| < 35$ are kept, consistent with the measurements in Table~\ref{table:Neutrinos}. Separate pseudodata are generated for the NH and the IH. The pseudodata are binned in two-dimensional subspaces of the space of observables, and $\Delta \chi^2$ is calculated over each subspace, once again using Eq.~\eqref{definechi2}. The 68.3\%, 95\% and 99\% CI are drawn as the contours along which $\Delta \chi^2 = 2.30$, $5.99$ and $9.21$; as before, these contours are depicted as dark, medium and light shadings of the appropriate color, respectively, in the figures that follow. Finally, a one-dimensional $\Delta \chi^2$ function is produced for $\sin \delta_{CP}^l$ -- precisely as was done for $R_\nu$, above -- both for the NH and the IH.


\subsubsection*{Type I Neutrino Seesaw}


\begin{figure}[]
\begin{center}
\subfigure[]{\includegraphics[width=3in]{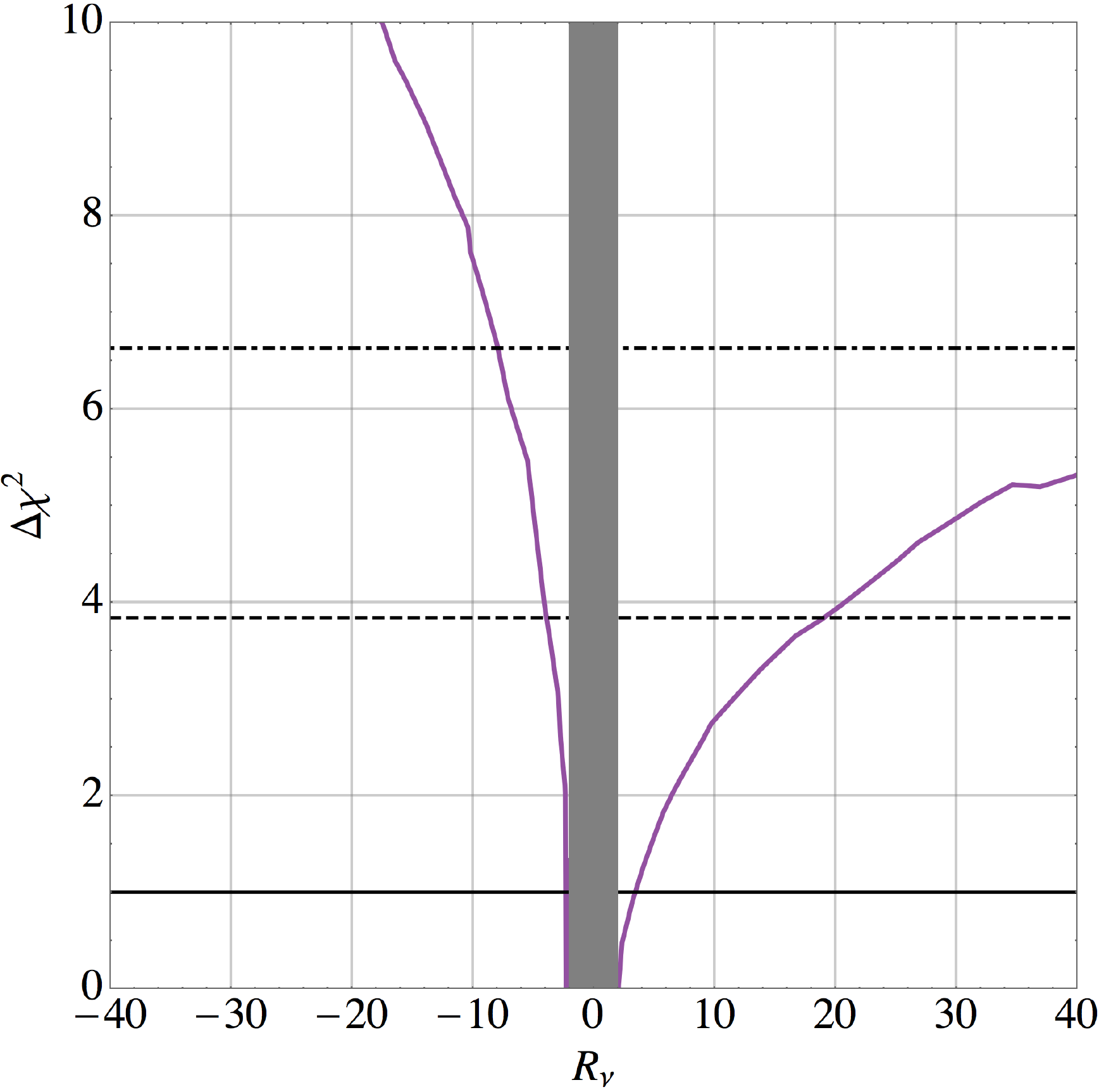}}
\subfigure[]{\includegraphics[width=3in]{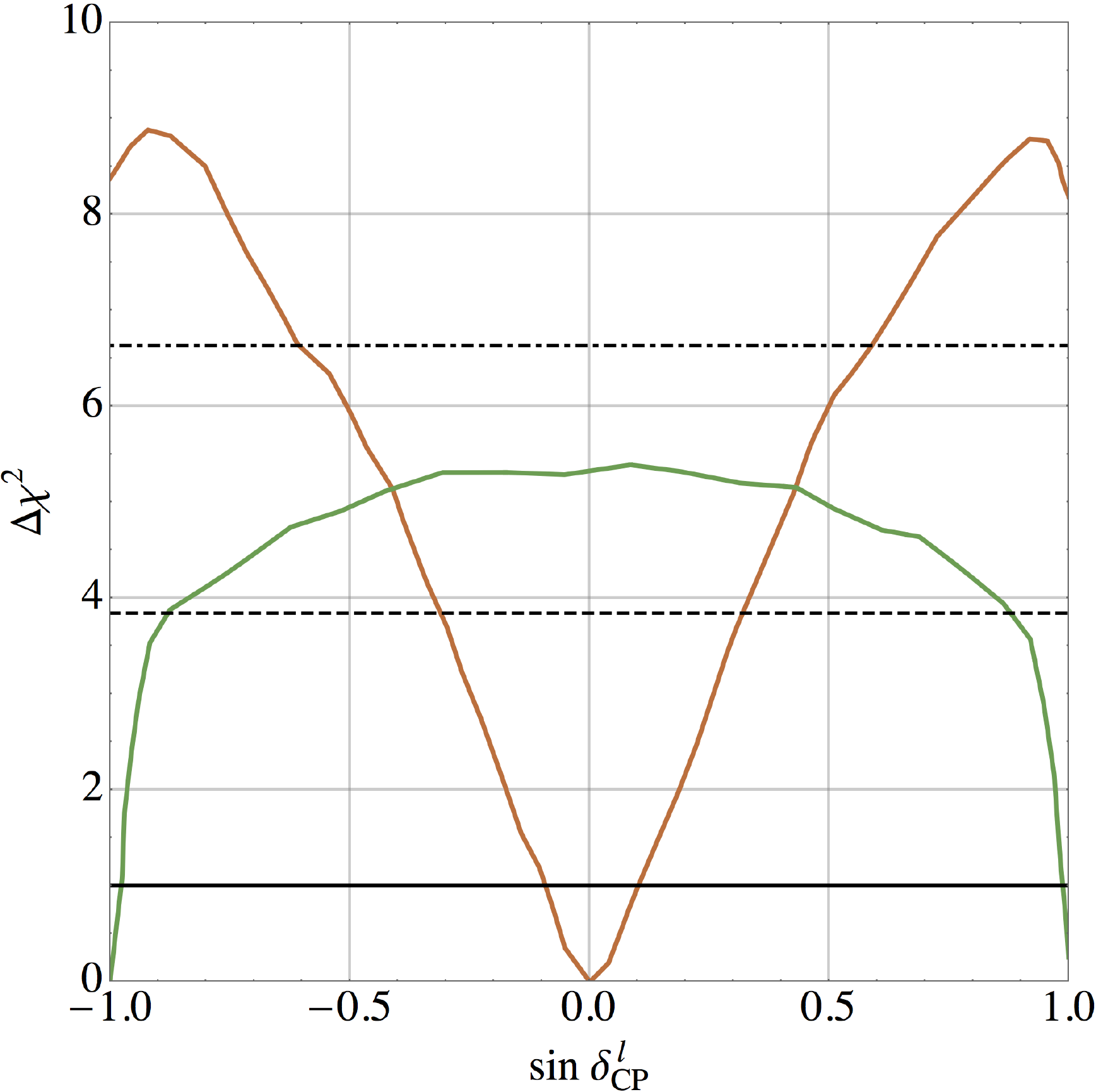}}
\end{center}
\caption{(a) The one-dimensional $\Delta \chi^2$ as a function of $R_\nu$ for a scan over the type I seesaw parameter ranges in Eqs.~\eqref{ranges2}, \eqref{ranges3} and \eqref{ranges4}. The dark gray band covers values of $R_\nu$ that cannot be generated.
(b) The one-dimensional $\Delta \chi^2$ as a function of $\sin \delta_{CP}^l$ for a similar scan. The orange line corresponds to a $30 < R_\nu < 35$, while the green line corresponds to $-35 < R_\nu < -30$. In both panels, the black lines represent the 68.3\% (solid), 95\% (dashed) and 99\% (dot-dashed) confidence levels.}
\label{fig:TypeIHierarchy}
\end{figure}

\begin{figure}[]
\begin{center}
\subfigure[]{\includegraphics[width=3in]{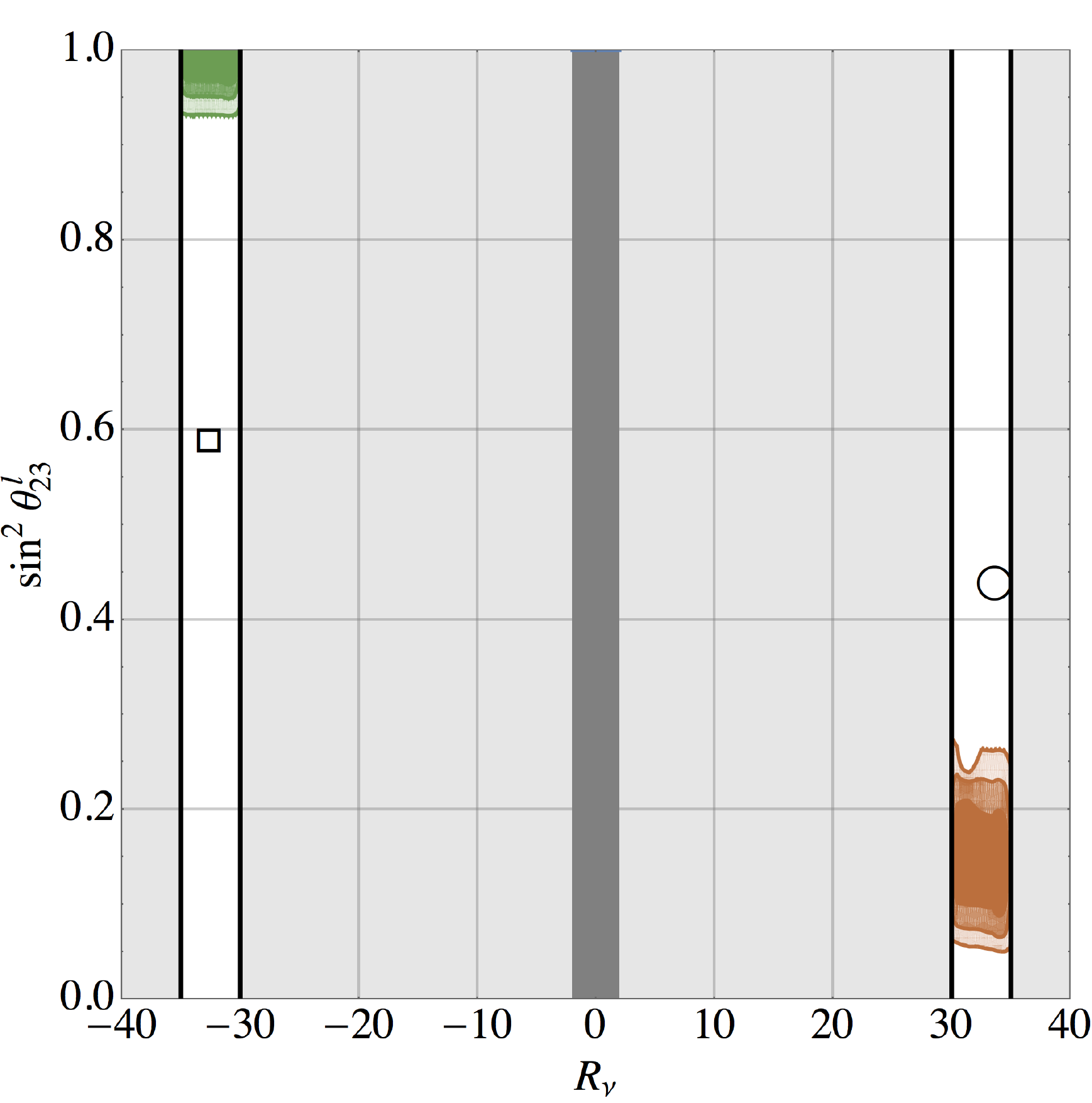}}
\subfigure[]{\includegraphics[width=3in]{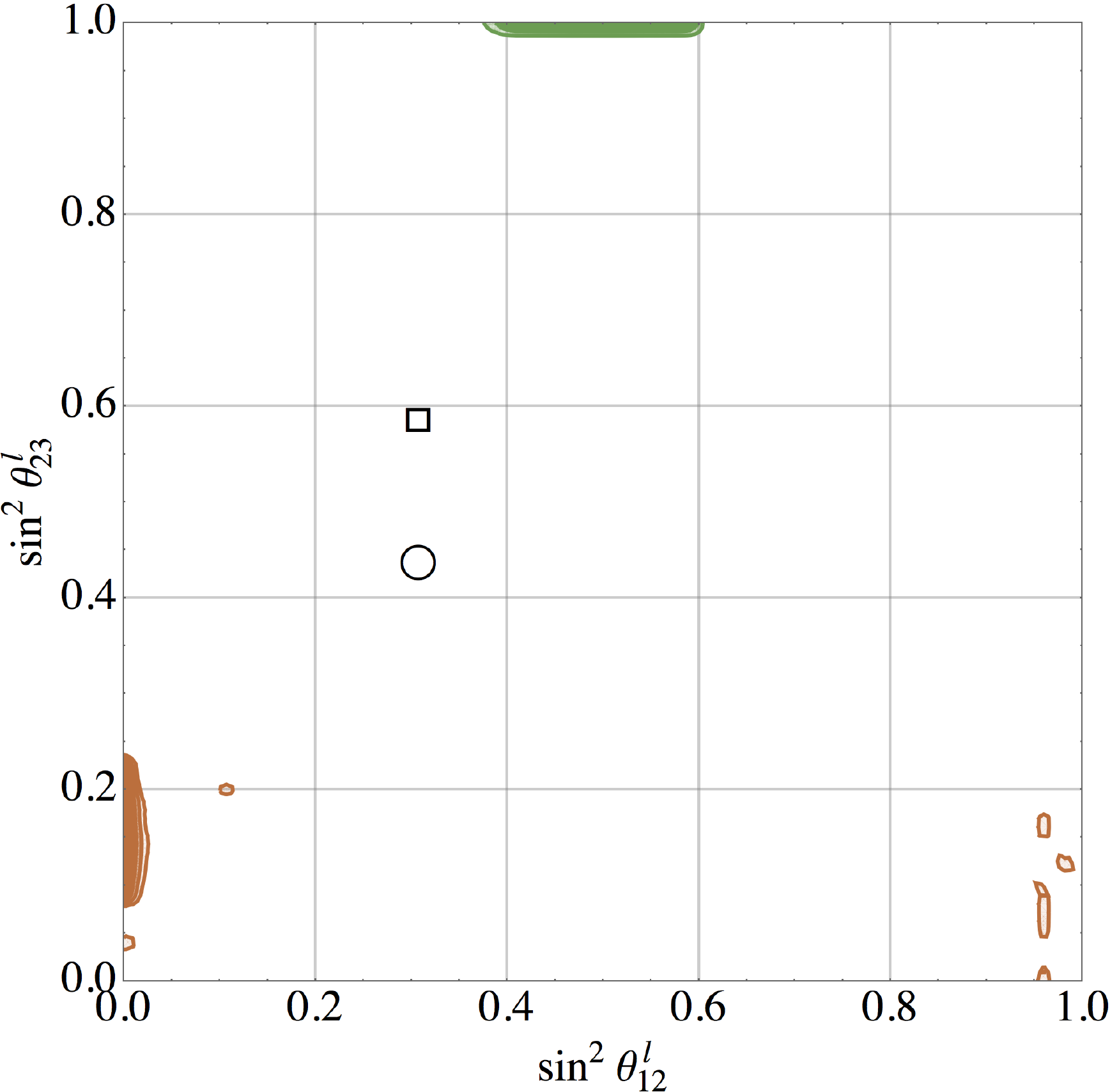}}
\subfigure[]{\includegraphics[width=3in]{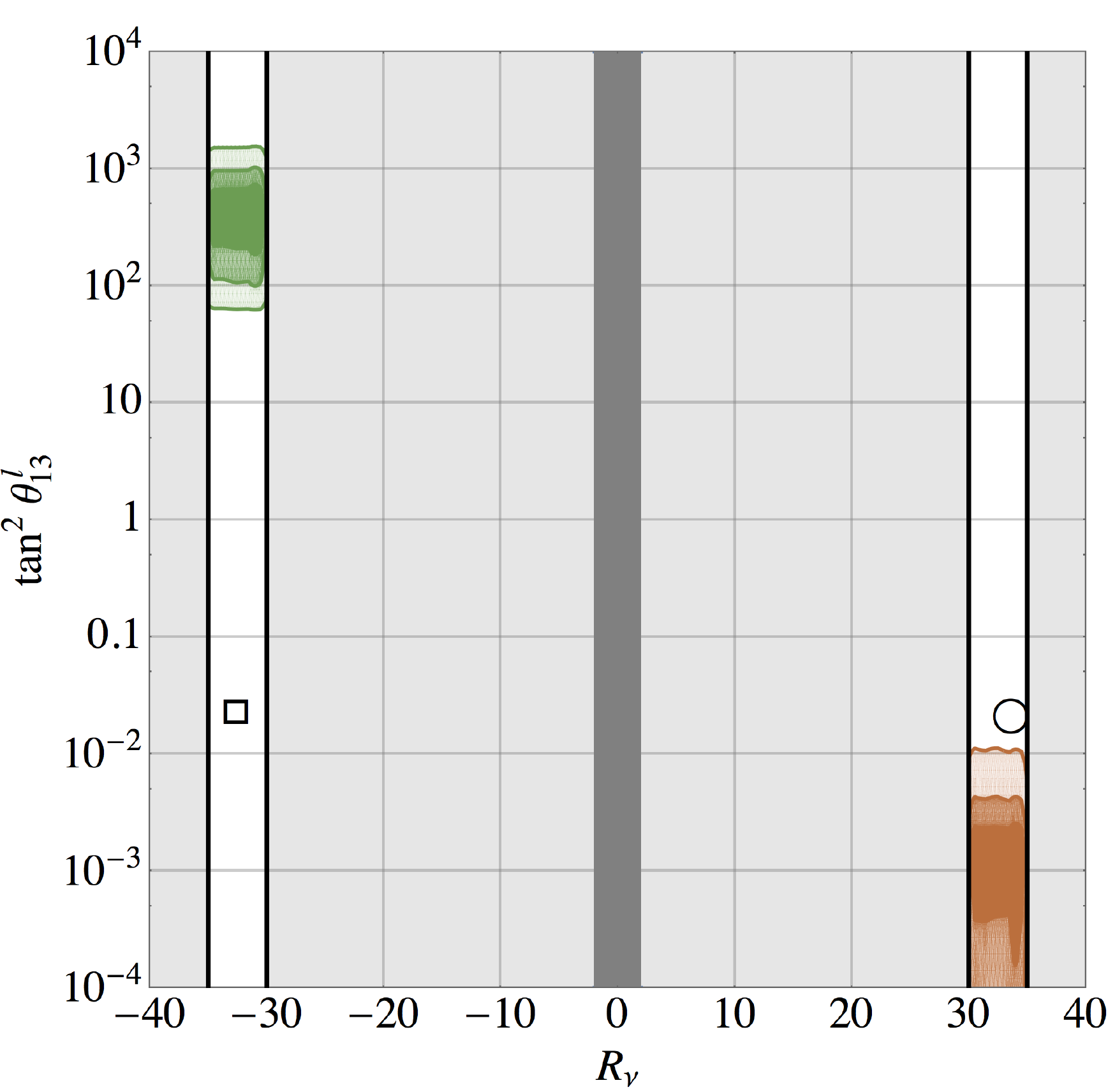}}
\subfigure[]{\includegraphics[width=3in]{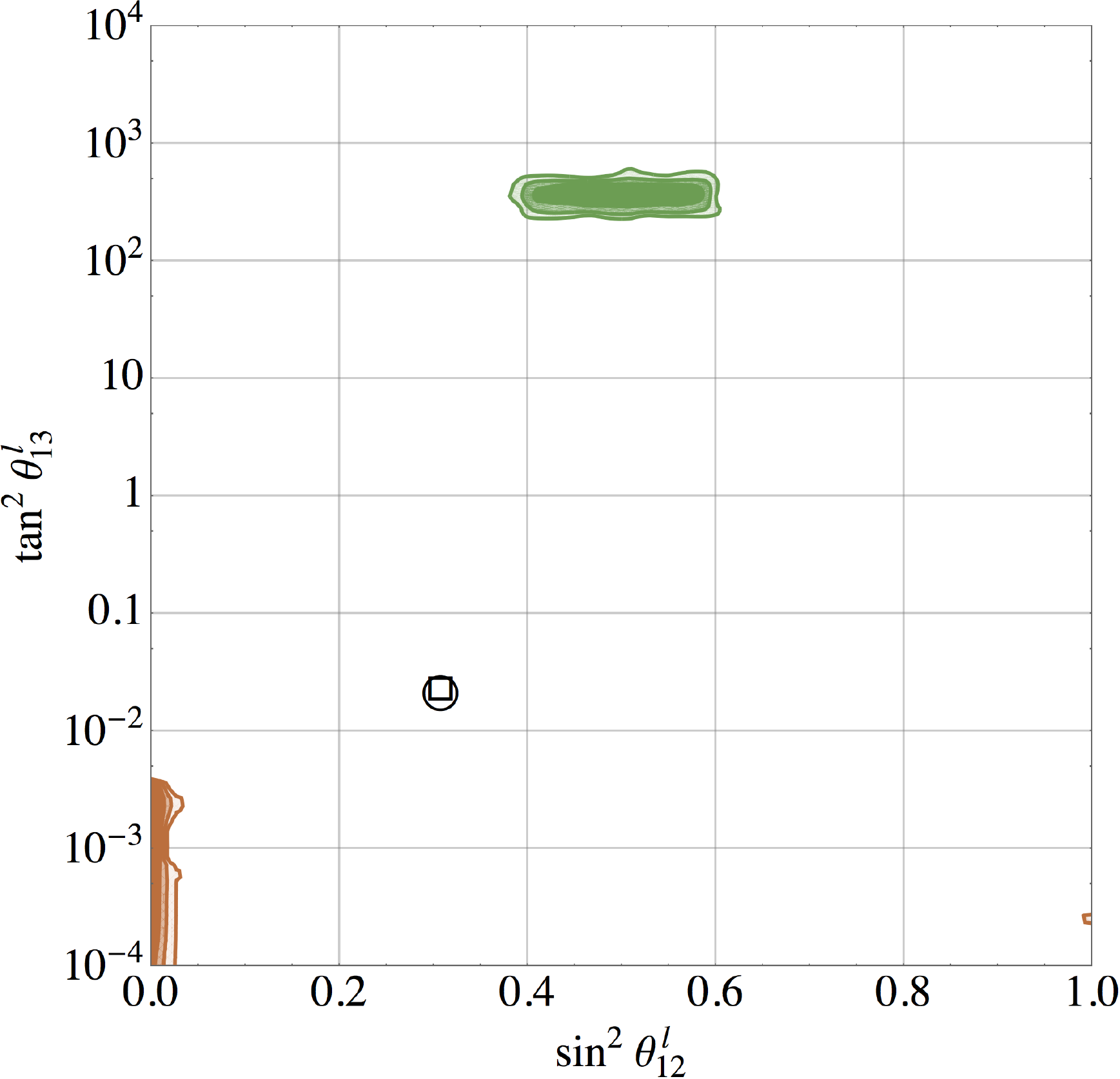}}
\end{center}
\caption{The 68.3\% (dark), 95\% (medium) and 99\% (light) confidence intervals for $R_\nu$ and the leptonic mixing angles produced by a scan over the type I seesaw parameter space given by Eqs.~\eqref{ranges2}, \eqref{ranges3} and \eqref{ranges4}. The NH (IH) is shown in orange (green). The dark gray bands in (a) and (c) cover values of $R_\nu$ that cannot be generated, while light gray bands mask values of $R_\nu$ excluded by our analysis. The circle and square represent the NH and IH solutions in Table~\ref{table:Neutrinos}, respectively.}
\label{fig:TypeI}
\end{figure}

We consider first the type I seesaw formalism of Sec.~\ref{Sec-leptons}, and scan over the neutrino parameters $\xi_\nu$, $\veps_\nu$ and $\overline{\veps}_\nu$, in addition to the parameters in Eqs.~\eqref{ranges2} and \eqref{ranges3}. These parameters are separately varied over the perturbative ranges
\be
\label{ranges4}
|\xi_\nu|, \, |\veps_\nu|, \, |\overline{\veps}_\nu| \in [0,0.1].
\ee
The results of this scan are illustrated in Figs.~\ref{fig:TypeIHierarchy} and \ref{fig:TypeI}. Fig.~\ref{fig:TypeIHierarchy}(a) shows $\Delta \chi^2$ as a function of $R_\nu$ while in Fig.~\ref{fig:TypeIHierarchy}(b), $\Delta \chi^2$ is plotted as a function of $\sin \delta_{CP}^l$ for $30 < R_\nu < 35$ (orange) and for $-35 < R_\nu < -30$ (green). Fig.~\ref{fig:TypeI} shows confidence intervals in two-dimensional slices of the space of observables, where the orange regions are for the NH, while the green regions are for the IH. The circle and square Fig.~\ref{fig:TypeI} represent the NH and IH solutions in Table~\ref{table:Neutrinos}, respectively.

From these plots, we infer that the type I seesaw in Flavorspin is unlikely to simultaneously accommodate for the observed values for the mass differences and leptonic mixing angles. The NH is a somewhat better fit than the IH in the Flavorspin framework for a type I seesaw scenario. In particular, the range $30<R_\nu<35$ is contained in the 95\% CI, while the range $-35<R_\nu<-30$ is excluded at $>99\%$ CL.  Moreover, the NH prefers small values ($\lesssim 5^\circ$) of $\theta_{13}^l$, while the IH prefers large values ($\sim 85^\circ$) thereof. 

On the other hand, neither case can easily accommodate the observed value of $\sin^2 \theta_{12}^{l}$ at 99\% CL. The NH predicts a small $\theta_{12}^l$, $\sin^2 \theta_{12}^l \lesssim 0.05$ at $>$99\% CL, while the IH implies $0.4 \lesssim \sin^2 \theta_{12}^l \lesssim 0.6$ at 99\% CL. The third angle, $\sin^2\theta_{23}^l$ is similarly pooly fit; the NH prefers $\sin^2 \theta_{23}^l \in [0.1, 0.25]$ at 95\% CL, while the IH prefers $\sin^2 \theta_{23}^l \gtrsim 0.95$ at 99\% CL. Finally, Fig.~\ref{fig:TypeIHierarchy}(b) indicates that, while the NH prefers minimal $CP$ violation ($|\sin \delta_{CP}^l| \lesssim 0.3$) at 95\% CL, the IH prefers strictly near-maximal $CP$ violation ($|\sin \delta_{CP}^l| \gtrsim 0.9$) at 95\% CL, with every possible value allowed at 99\% CL.


\subsubsection*{Type II Neutrino Seesaw}


\begin{figure}[]
\begin{center}
\subfigure[]{\includegraphics[width=3in]{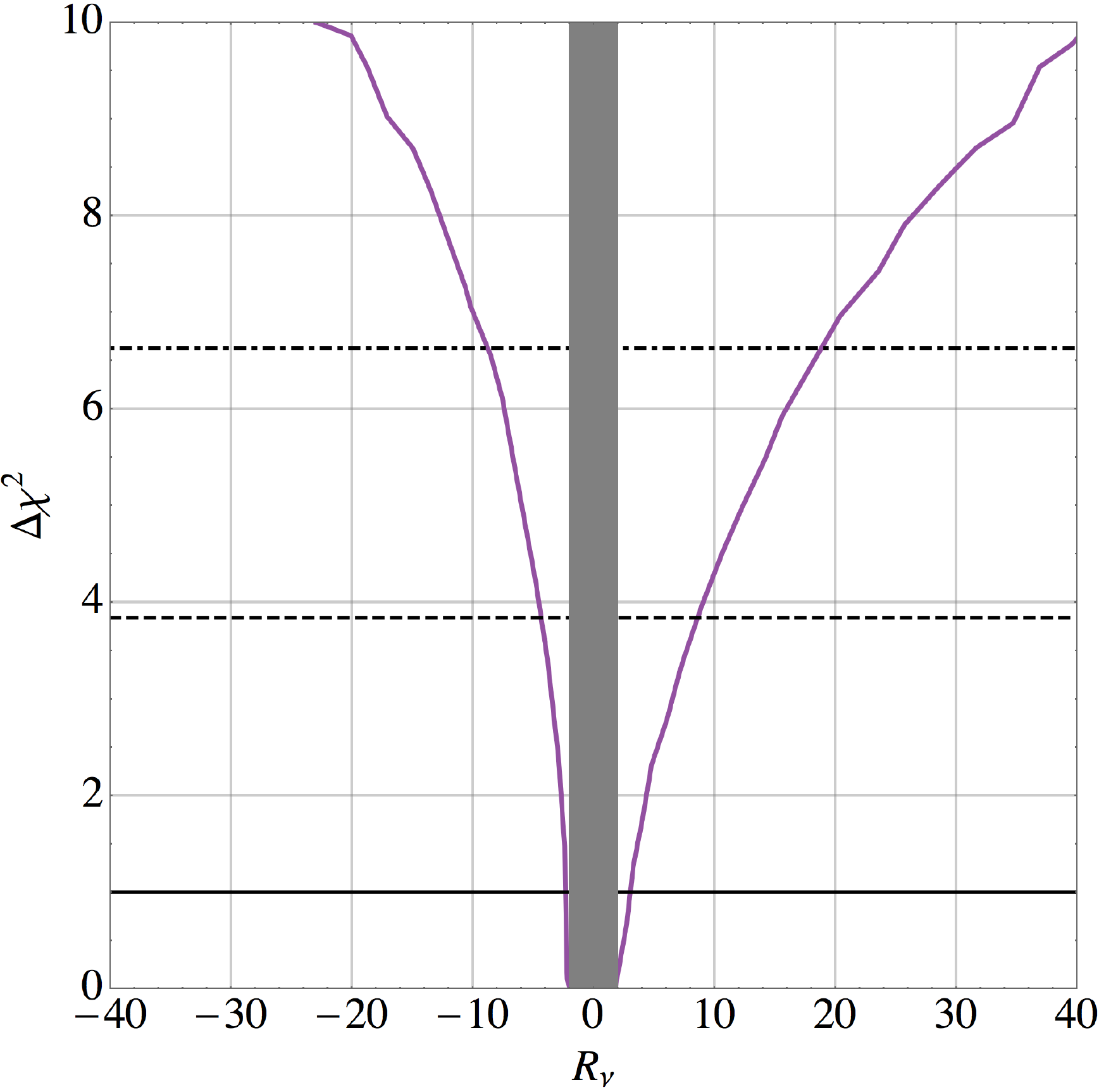}}
\subfigure[]{\includegraphics[width=3in]{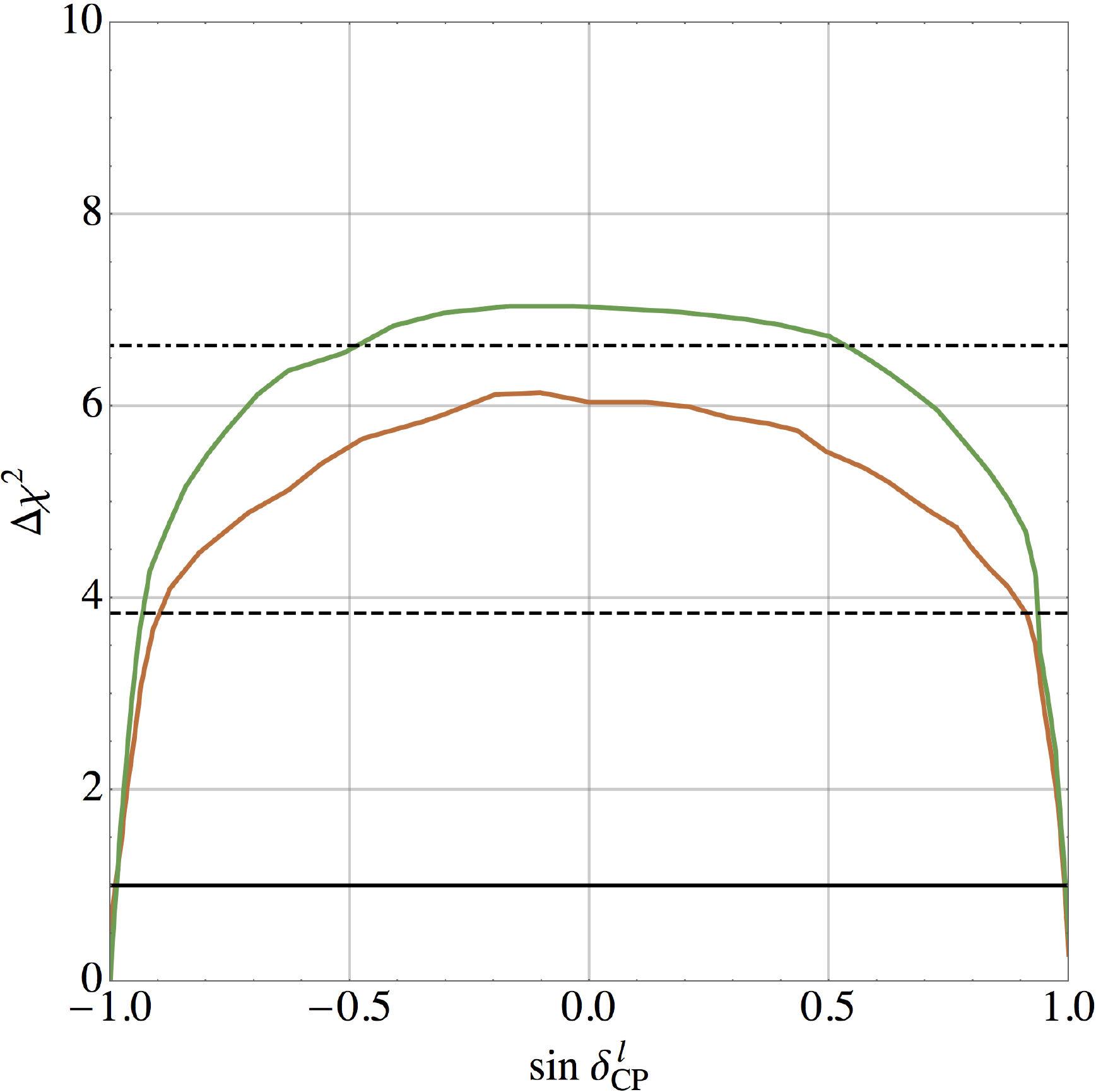}}
\end{center}
\caption{(a) The one-dimensional $\Delta \chi^2$ as a function of $R_\nu$ for a scan over the type II seesaw parameter ranges in Eqs.~\eqref{ranges2}, \eqref{ranges3} and \eqref{ranges5}. The dark gray band covers values of $R_\nu$ that cannot be generated.
(b) The one-dimensional $\Delta \chi^2$ as a function of $\sin \delta_{CP}^l$ for a similar scan. The orange line corresponds to a $30 < R_\nu < 35$, while the green line corresponds to $-35 < R_\nu < -30$. See text for details. In both panels, the black lines represent the 68.3\% (solid), 95\% (dashed) and 99\% (dot-dashed) confidence levels.}
\label{fig:TypeIIHierarchy}
\end{figure}

\begin{figure}[]
\begin{center}
\subfigure[]{\includegraphics[width=3in]{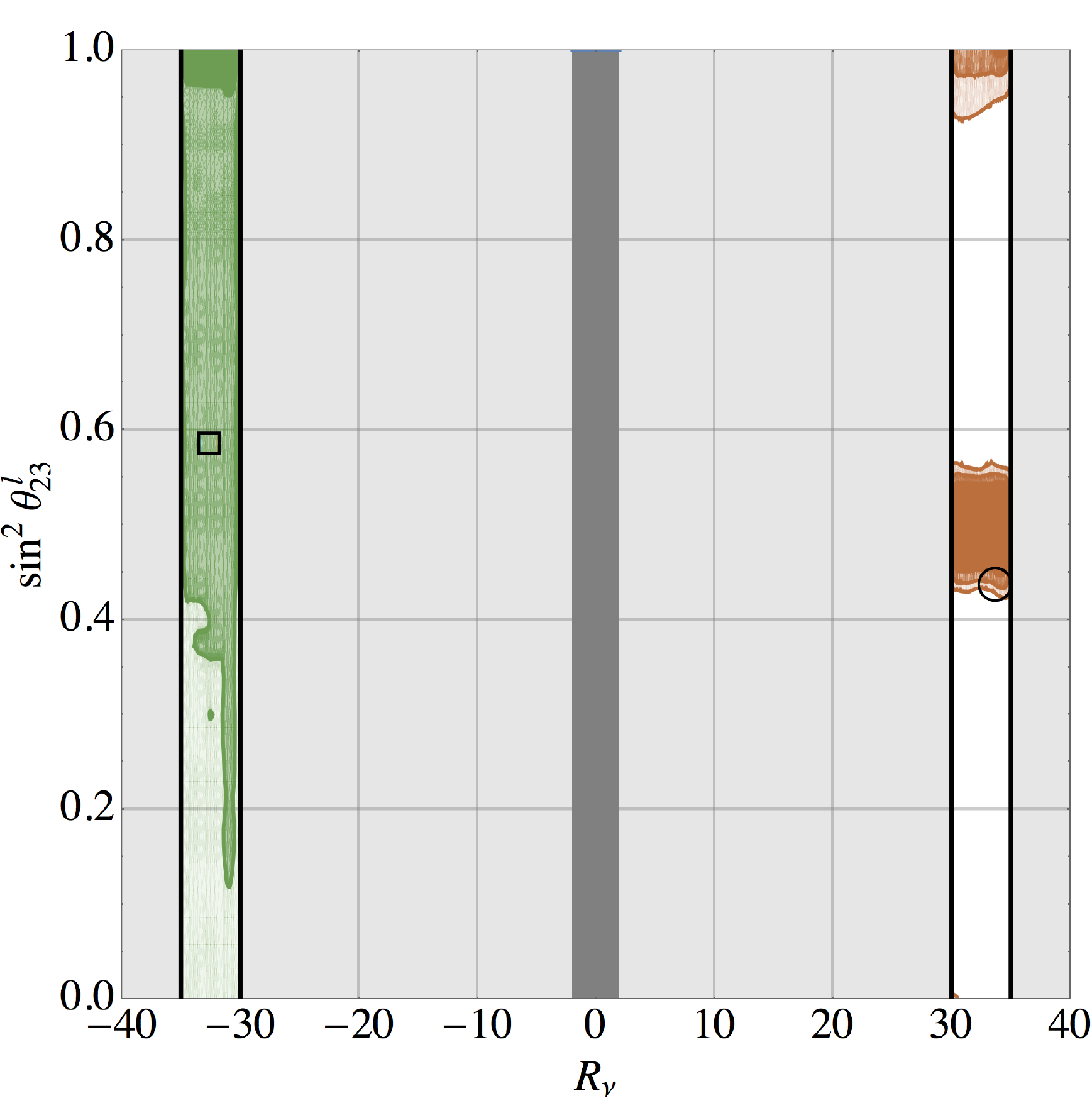}}
\subfigure[]{\includegraphics[width=3in]{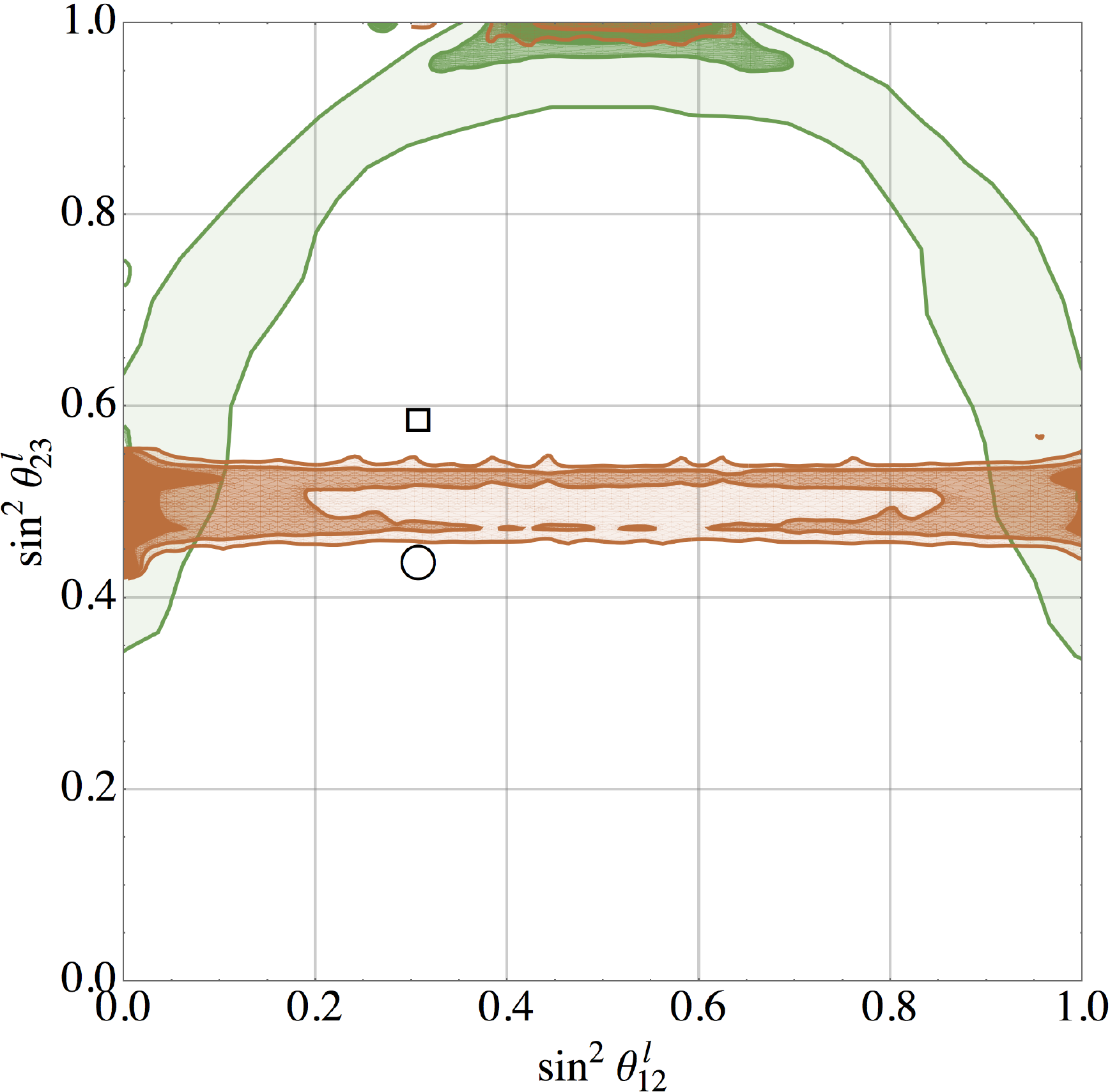}}
\subfigure[]{\includegraphics[width=3in]{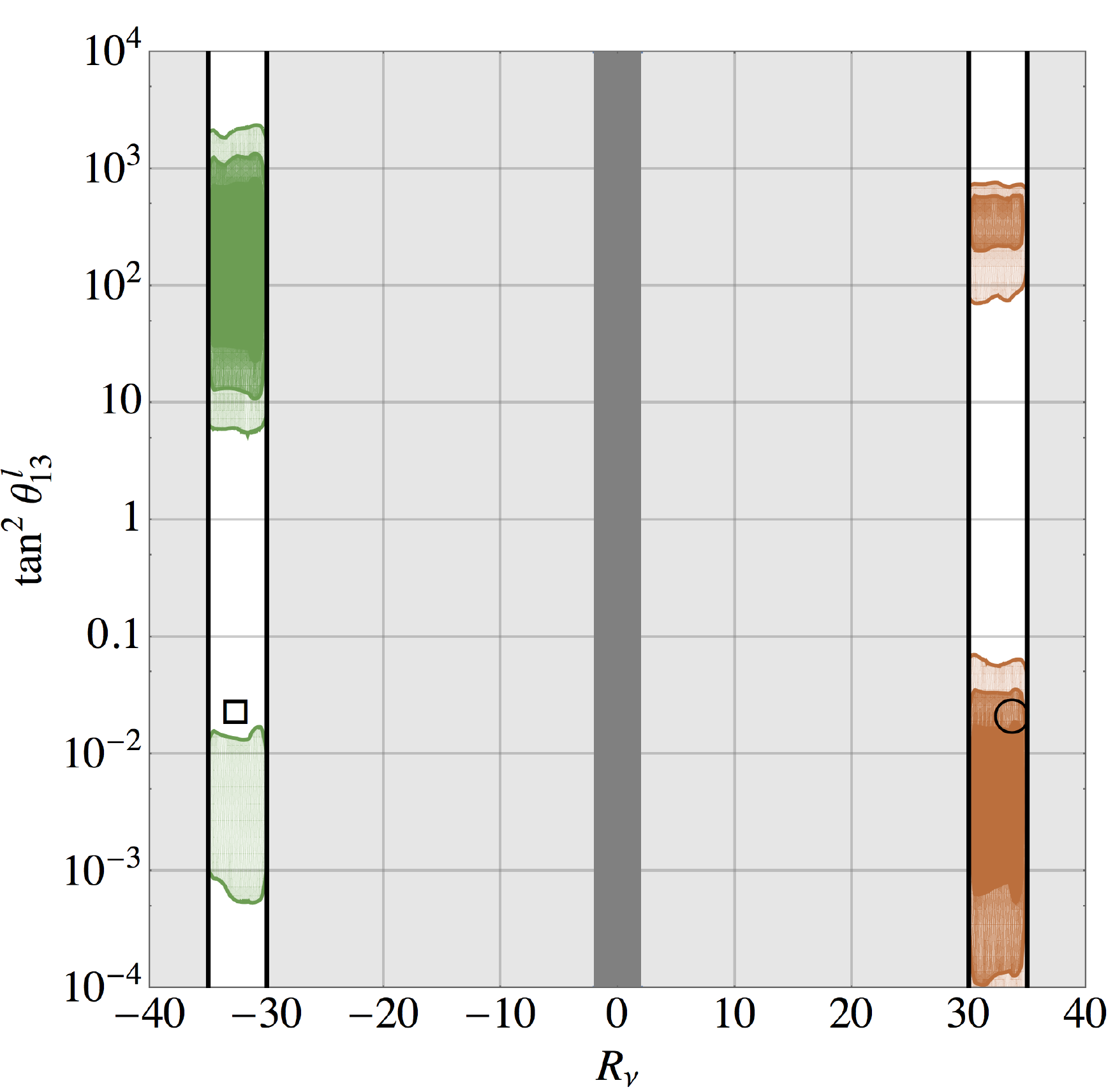}}
\subfigure[]{\includegraphics[width=3in]{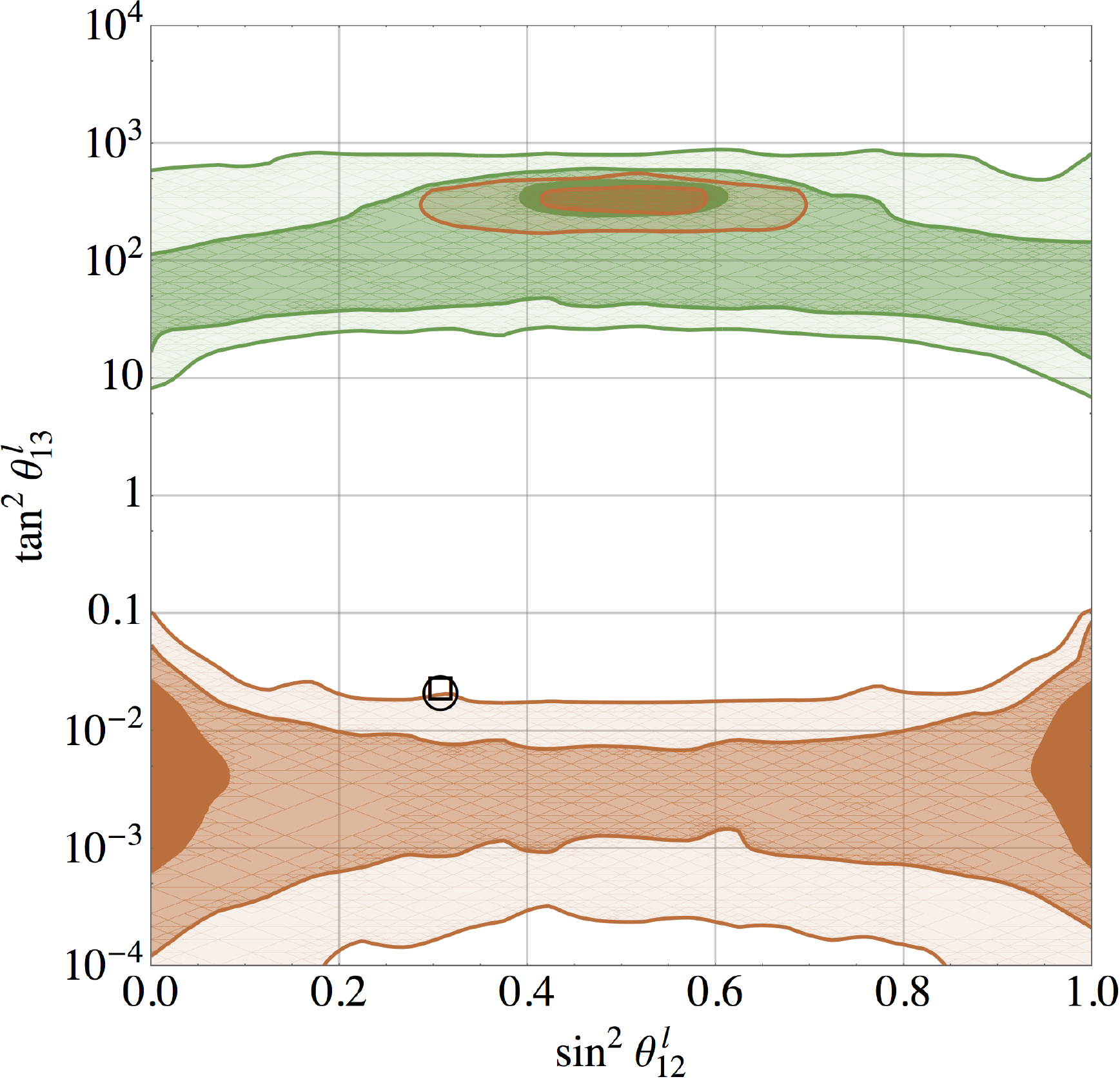}}
\end{center}
\caption{The 68.3\% (dark), 95\% (medium) and 99\% (light) confidence intervals for $R_\nu$ and the leptonic mixing angles produced by a scan over the type II seesaw parameter space given by Eqs.~\eqref{ranges2}, \eqref{ranges3} and \eqref{ranges5}. The NH (IH) is shown in orange (green). The dark gray bands in (a) and (c) cover values of $R_\nu$ that cannot be generated, while light gray bands mask values of $R_\nu$ excluded by our analysis. The circle and square represent the NH and IH solutions in Table~\ref{table:Neutrinos}, respectively.}
\label{fig:TypeII}
\end{figure}

We scan over the parameters $\eta_{33}$, $\eta_5$ and $\eta_1$ of Eq.~\eqref{etastypeI2} in addition to the parameters in Eqs.~\eqref{ranges2} and \eqref{ranges3}. These parameters are separately varied over the ranges
\be
\label{ranges5}
|\eta_{33}|, \, |\eta_5|, \, |\eta_1| \in [0,\,1].
\ee
Note that these ranges are not perturbative.

The results are shown in Figs.~\ref{fig:TypeIIHierarchy} and \ref{fig:TypeII}. Fig.~\ref{fig:TypeIIHierarchy}(a) shows $\Delta \chi^2$ as a function of $R_\nu$. The conclusion of this exploration is that it is hard to reproduce the hierarchy of mass differences in the type II seesaw. More specifically, the observed region  $30<|R_\nu|<35$ does not occur at 99\% CL away from the most likely value for this parameter, irrespective of the hierarchy.

Fig.~\ref{fig:TypeII} shows confidence intervals in two-dimensional slices of the space of observables. In these figures, the orange regions contain NH points, while the green regions contain IH points. From the figures, things are more promising regarding the mixing angles. More specifically, the NH contains all possible values of $\sin^2 \theta_{12}^l$ in the 95\% CI. The IH prefers $\theta_{12}^l \sim 45^\circ$, though it can also accommodate any value at 99\% CL. Both hierarchies allow for $\theta_{13}^l$ to either be small ($\lesssim 10^\circ$) or large ($\gtrsim 80^\circ$), but have exceedingly low probability to produce an intermediate value; the NH prefers small values and the IH prefers large values, both at $>$95\% CL. This framework struggles only to simultaneously accommodate the large value of $\theta_{12}^l$ and the relatively large $\theta_{13}^l$, though the tension is not as severe here as it is for the type I seesaw. 

Regarding $\theta_{23}^l$, the 95\% CI for the IH contains $\sin^2 \theta_{23}^l \gtrsim 0.3$, and the 99\% CI covers the entire allowable range. In the NH, the situation is more predictive, as the 95\% CI covers the regions $0.45 \lesssim \sin^2 \theta_{12}^l \lesssim 0.55$ and $\sin^2\theta_{12}^l \gtrsim 0.98$. While the NH solution (circle) in Fig.~\ref{fig:TypeII}(a) lies inside the 95\% CI, this framework generically has no preference for either octant of $\theta_{23}^l$. From Fig.~\ref{fig:TypeIIHierarchy}(b), we see that both hierarchies have a strong preference for near-maximal $CP$ violation ($|\sin \delta_{CP}^l| \gtrsim 0.9$) at 95\% CL. In fact, the IH prefers $|\sin \delta_{CP}^l| \gtrsim 0.5$ at 99\% CL, though every possible value is allowed at 99\% CL for the NH.


\subsubsection*{General Neutrino Mass Matrix}


\begin{figure}[]
\begin{center}
\subfigure[]{\includegraphics[width=3in]{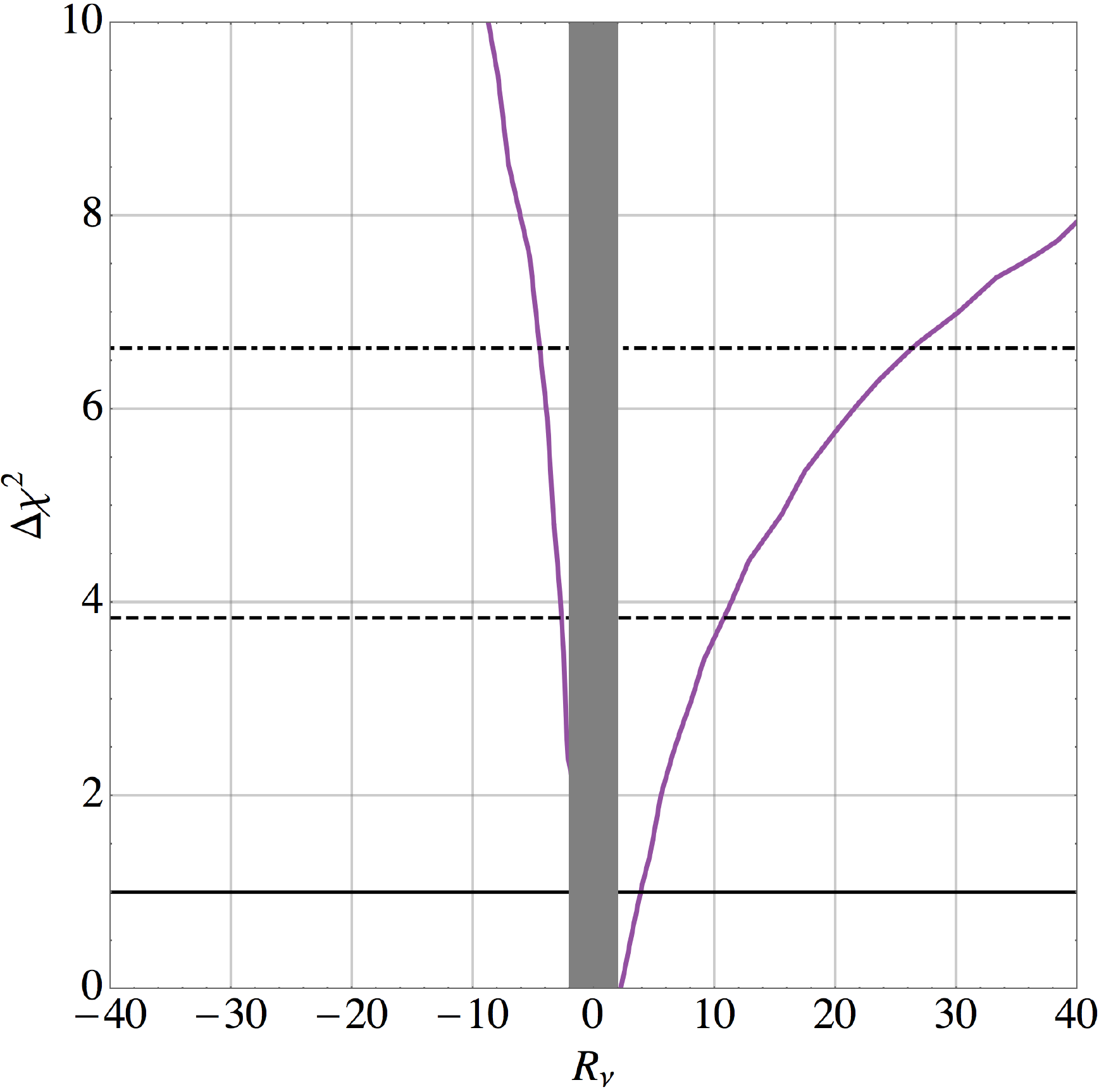}}
\subfigure[]{\includegraphics[width=3in]{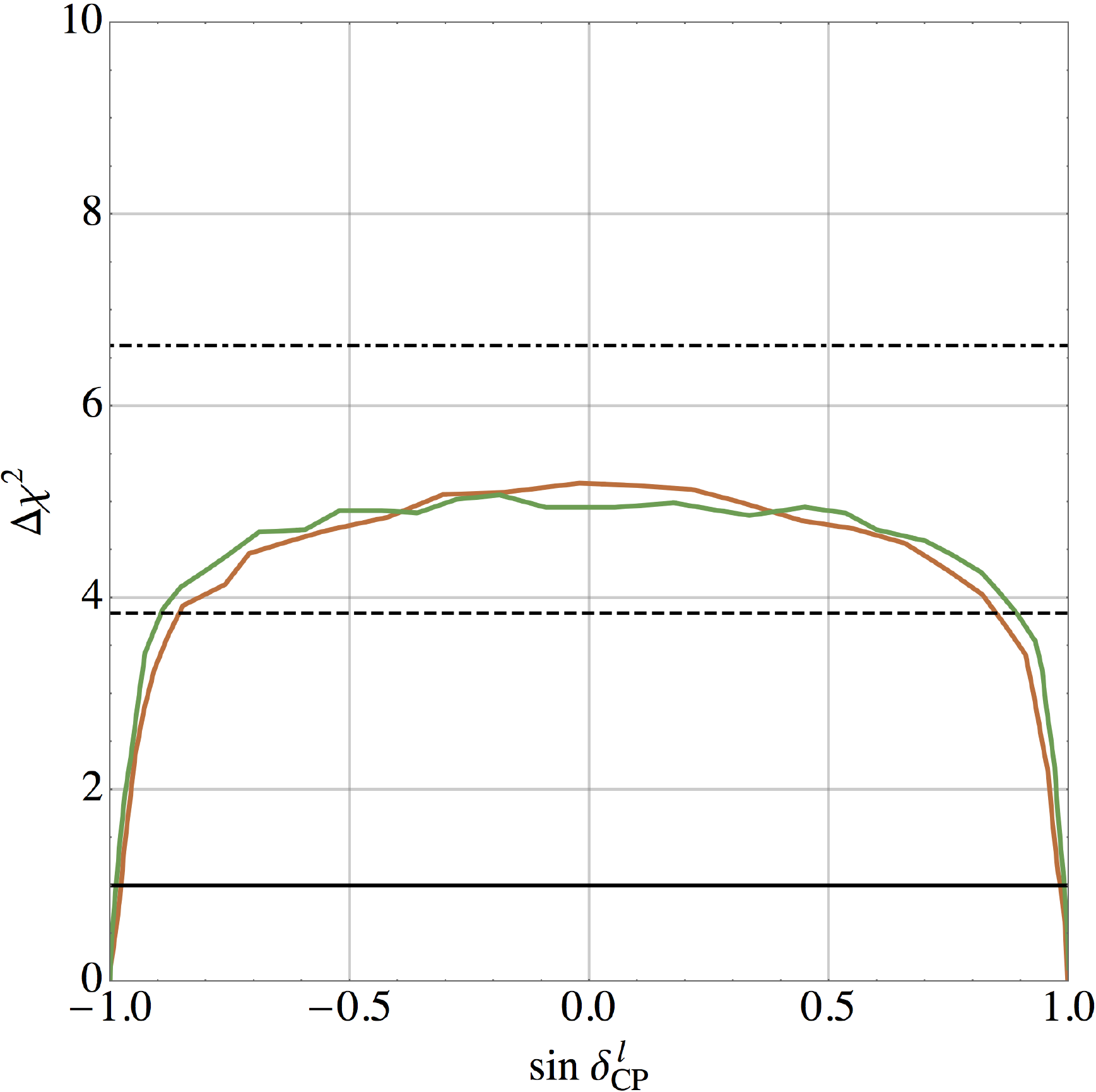}}
\end{center}
\caption{(a) The one-dimensional $\Delta \chi^2$ as a function of $R_\nu$ for a scan over the general neutrino mass matrix parameter ranges in Eqs.~\eqref{ranges2}, \eqref{ranges3} and \eqref{ranges6}. The dark gray band covers values of $R_\nu$ that cannot be generated.
(b) The one-dimensional $\Delta \chi^2$ as a function of $\sin \delta_{CP}^l$ for a similar scan. The orange line corresponds to a $30 < R_\nu$, while the green line corresponds to $-35 < R_\nu < -30$. See text for details. In both panels, the black lines represent the 68.3\% (solid), 95\% (dashed) and 99\% (dot-dashed) confidence levels.}
\label{fig:GeneralHierarchy}
\end{figure}

\begin{figure}[]
\begin{center}
\subfigure[]{\includegraphics[width=3in]{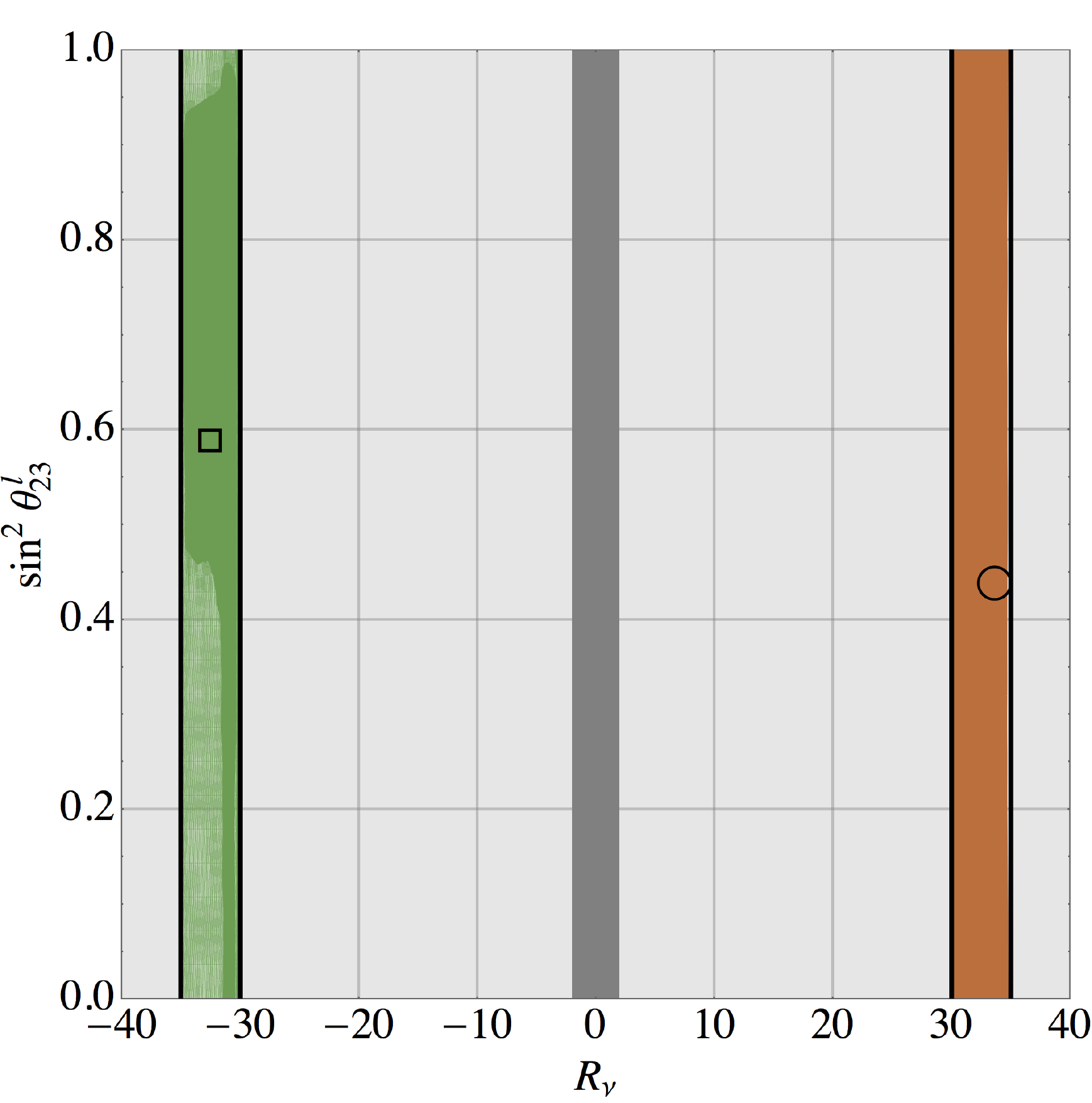}}
\subfigure[]{\includegraphics[width=3in]{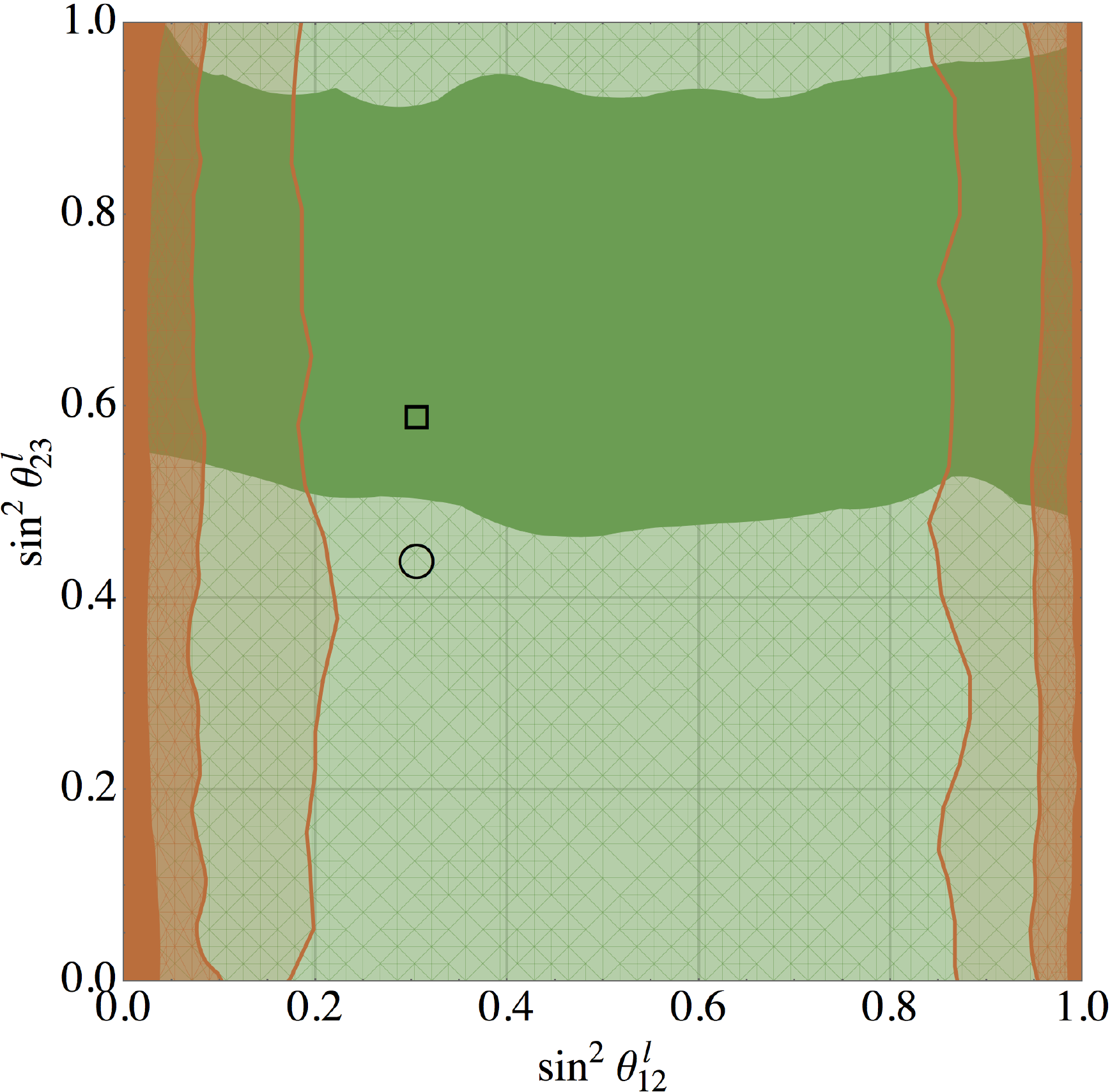}}
\subfigure[]{\includegraphics[width=3in]{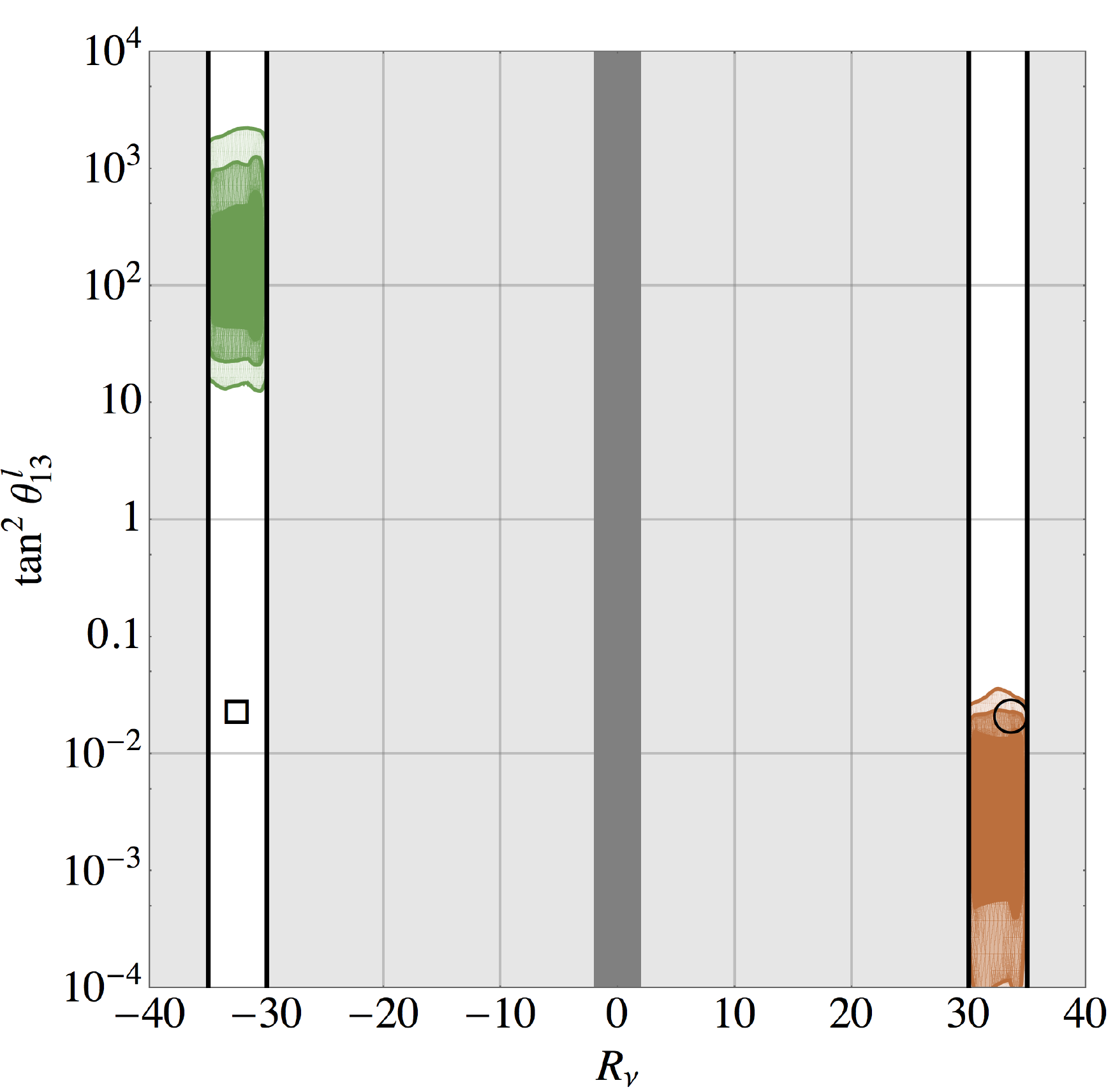}}
\subfigure[]{\includegraphics[width=3in]{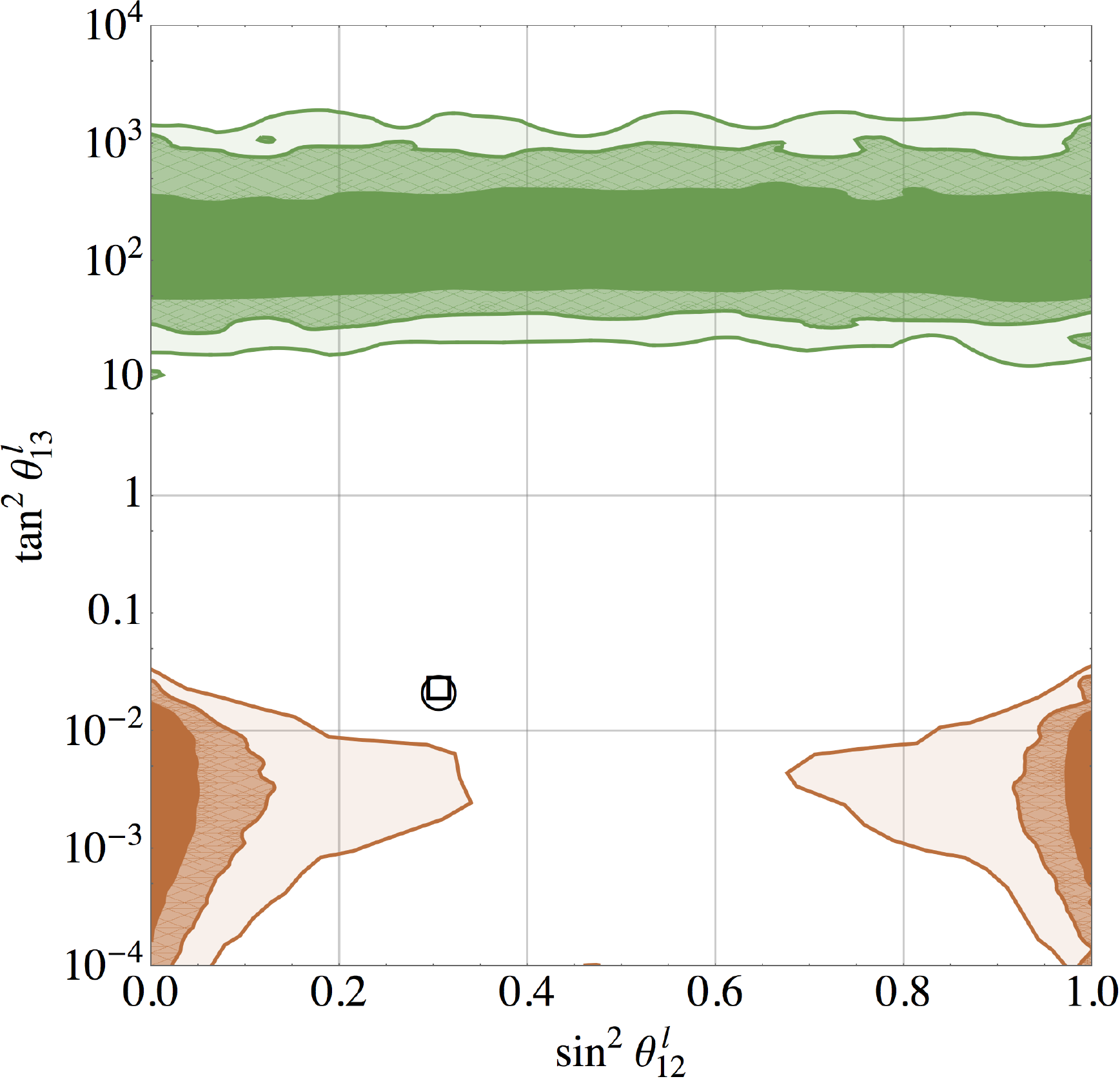}}
\end{center}
\caption{The 68.3\% (dark), 95\% (medium) and 99\% (light) confidence intervals for $R_\nu$ and the leptonic mixing angles produced by a scan over the general neutrino mass matrix parameter space given by Eqs.~\eqref{ranges2}, \eqref{ranges3} and \eqref{ranges6}. The NH (IH) is shown in orange (green). The dark gray bands in (a) and (c) cover values of $R_\nu$ that cannot be generated, while light gray bands mask values of $R_\nu$ excluded by our analysis. The circle and square represent the NH and IH solutions in Table~\ref{table:Neutrinos}, respectively.}
\label{fig:General}
\end{figure}

For completeness, we also explored the general neutrino mass matrix introduced in Eq.~\eqref{Mnu}. We briefly describe the results in this case following the methods previously described. The neutrino-specific parameter ranges over which we scan are
\be
\label{ranges6}
|\eta_{33}|, \, |\eta_5|, \, |\eta_1|, \, |\eta_{35}|, \, |\eta_{55}|, \, |\eta_{335}|, \, |\eta_{353}| \in [0,\,1],
\ee
while the other relevant parameters are scanned over the ranges in Eqs.~\eqref{ranges2} and \eqref{ranges3}. The results are presented Figs.~\ref{fig:GeneralHierarchy} and \ref{fig:General}; the interpretation of those figures being the same as before. As with the type II seesaw, neither $30 < R_\nu < 35$ nor $-35 < R_\nu < -30$ are included in the 99\% CI in Fig.~\ref{fig:GeneralHierarchy}(a), although the IH is significantly less likely than NH. Moreover, in this case, the NH prefers $\theta_{12}^l$ at the extremes of the range: $\theta_{12}^l \lesssim 20^\circ$ or $\theta_{12}^l \gtrsim 70^\circ$ at 95\% CL. The IH, on the other hand, is not $\theta_{12}^l$-predictive; it contains all possible values of $\sin \theta_{12}^l$ within the 68.3\% CI.

The NH prefers small values ($\lesssim 10^\circ$) of $\theta_{13}^l$ at 95\% CL, while the IH prefers large values ($\gtrsim 80^\circ$) at 95\% CL. While the NH (IH) may produce large (small) values of $\theta_{13}^l$, these values lay outside the 99\% CI, and thus do not appear in Fig.~\ref{fig:General}. Both hierarchies contain every possible value of $\sin^2 \theta_{23}^l$ within the 95\% CI. Therefore, there is nothing particularly special about the region around $\theta_{23}^l = 45^\circ$. This disagrees with our theoretical prejudice that $\theta_{23}^l \sim 45^\circ$ indicates that something phenomenologically interesting is happening in the lepton sector, as we saw for the type II seesaw, above. Both hierarchies in this scenario prefer near-maximal $CP$ violation: the 95\% CI consists of $|\sin \delta_{CP}^l| \gtrsim 0.8$, though every possible value is allowed at 99\% CL.


\section{Flavor-Changing Neutral Currents}
\label{nofcncs}
\setcounter{equation}{0}


A full exploration of the phenomenology of the scenario we are proposing is deferred to a later study. However, it is relatively easy to show that the problem of FCNCs is alleviated substantially in Flavorspin models if the coefficients of higher-dimensional operators have a structure
\be
c_\alpha(Y_3,\, \veps Y_5,\, \bveps\mbb{I}),
\ee
where $|\veps|, \,|\bveps| \ll 1$, mimicking the premise that was assumed before for the Yukawa couplings. To see this, note that higher-dimensional operators are comprised of gauge- and Lorentz-invariant combinations of SM matter fields ($Q_L, \, U_R, \, D_R, \, L_L, \, E_R$), Higgs bosons ($H$), field strength tensors ($G_{\mu\nu}^a$, $W^a_{\mu\nu}$, $B_{\mu\nu}$) and (covariant) derivatives ($D_\mu$). After electroweak symmetry is broken, these operators are decomposed in terms of the low-energy degrees of freedom of the SM. Our analysis of nonrenormalizable operators specifically focuses on fermion bilinears of the form\footnote{In this section, we suppress the superscript on $c$ that appeared in, for instance, Eq.~\eqref{first-approx}.}
\be
\mcl{Q}_\alpha \sim c_{\alpha,ij} F_i F^\prime_j, \, c_{\alpha,ij} F_i \overline{F}^\prime_j, \, c_{\alpha,ij} \overline{F}_i \overline{F}^\prime_j, \quad X^{(\prime)} = U_{L,R}, \, D_{L,R}, \, E_{L,R}, \, \nu_L.
\ee
These bilinears are, by construction, singlets of $\mcl{G}_{fl}$, but are not necessarily singlets of the Lorentz or SM gauge groups. Operators may contain any number of bilinears, each with its own $c_\alpha$; operator substructures unrelated to flavor are not relevant here.

We express the flavor-charged coefficients $c_\alpha$ in terms of the spurions $Y_3$ and $Y_5$. For definiteness, we have
\be
\label{definec}
c_{\alpha} = i Y_3 + (1-\xi_\alpha) Y_3^2 + \veps_\alpha Y_5 + \overline{\veps}_\alpha \mbb{I} + \dots ,
\ee
where $|\veps_\alpha|, \, |\overline{\veps}_\alpha|, \, |\xi_\alpha| \ll 1$. Fermion bilinears may be divided into three classes based on their flavor structure.

\begin{enumerate}
\item $\mcl{Q}_1 \sim (\overline{F}_i c_{\alpha,ij} F_j^\prime) $, $F^{(\prime)} = U_{L,R}, \, D_{L,R}, \, E_{L,R},$.

Operators containing bilinears of this class are contributors to the most commonly searched for FCNC processes. In particular, they yield contributions to reactions such as $b \to s \gamma$ \cite{Antonelli:2009ws}, $\mu \to e \gamma$ \cite{Lindner:2016bgg} and meson-antimeson oscillations \cite{Abulencia:2006ze}.
\item $\mcl{Q}_{2} \sim (F_i c_{\alpha,ij} F_j^\prime), \, (\overline{F}_i c_{\beta,ij} \overline{F}_j^\prime)$, $F^{(\prime)} = U_{L,R}, \, D_{L,R}, \, E_{L,R}$.

  Bilinears in this class necessarily violate $B-L$, even though the full operator need not.
  
\item $\mcl{Q}_{3} \sim (\overline{\nu}_i c_{\alpha,ij} F_j), \, (\overline{F}_i c_{\beta,ij} \nu_j)$, $F = U_{L,R}, \, D_{L,R}, \, E_{L,R}, \nu_L$.

  These are bilinears, in any Lorentz or gauge configuration, in which at least one fermion is a neutrino. 
\end{enumerate}

All the bilinears above have been expressed in the flavor basis. After EWSB, whether or not these operators lead to FCNCs is determined upon rotation into the physical basis. The rotation matrices for the charged fermions were discussed in Sec.~\ref{flavorspin}, and that of the neutrinos was discussed in Sec.~\ref{Sec-leptons}; we apply these matrices in each of these cases.

In case 1, after rotation into the mass basis, the matrix $c_\alpha$ transforms to:
\be
\label{classI}
c_\alpha \to V_X c_\alpha V^{\dagger}_{X^\prime},
\ee
where $V_X$ is the matrix that diagonalizes the Yukawa matrix $Y_X$, as in Eqs.~\eqref{defineVX}. In the $\{\veps_X, \veps_\alpha\} \to 0$ limit,  $V_{X}, V_{X^\prime} \to V^0$, as in Eq.~\eqref{V0}. The coefficient $c_\alpha$ is a diagonalized by this rotation, regardless of $\xi_\alpha$. Therefore, this class of bilinears yields no FCNCs at leading order in the small parameters $\veps_\alpha$ and $\bveps_\alpha$. Any FCNCs that do arise -- and that contribute to the above flavor-changing processes -- must be correspondingly suppressed.

As an illustration, consider the bilinear $\overline{D_L} c_{\alpha} D_L$. When the down-type quarks are rotated into their mass basis, the flavor matrix $c_\alpha$ of the bilinear becomes
\be
\label{classIexample}
V_d c_\alpha V_d^\dagger \sim \left( \ba{ccc} 0 & 0 & 0 \\ 0 & \xi_\alpha & (1 + i)(\veps_d - \veps_\alpha) \\ 0 & (1-i) (\veps_d - \veps_\alpha) & \xi_\alpha - 2 \ea \right) + \mcl{O}( \{ \veps_X, \, \xi_X \}^2 )
\ee
where we have simplified this expression by assuming $y_{22} = y_{23} = 1$, $y_{11} = y_{12} = y_{13} = \overline{\veps}_d = \overline{\veps}_\alpha = 0$, and that the remaining Flavorspin parameters are real-valued. The off-diagonal piece of this matrix is proportional to $\veps_{X}$, so the contributions of this bilinear to flavor-changing processes like $b_L \to s_L + \ldots$ are suppressed by the (assumed) smallness of $\veps_X$ relative to the flavor-conserving contributions. While both quarks are left-handed in this example, we emphasize that the same conclusion applies if one or both were right-handed.

In case 2, the matrices $c_\alpha, \, c_\beta$ become, after rotation into the mass basis,
\bea
\label{classII}
c_\alpha \to V^*_X c_\alpha V^\dagger_{X^\prime}, \quad c_\beta \to V_X c_\beta V^T_{X^\prime}.
\eea
Even in the limit $\{\veps_X, \veps_\alpha\} \to 0$, $c_\alpha$ and $c_\beta$ are not a diagonal matrices. Bilinears of this class can then potentially induce large FCNCs. As stressed above, however, these bilinears are exotic, with the flavor coefficients connecting fermions in a $B-L$-violating fashion, pointing to an effective vertex that arises from an underlying $B-L$-violating interaction. In Sec.~\ref{discussion}, we will argue that these vertices can be naturally suppressed by a Froggatt-Nielsen-like mechanism.\footnote{The suppression of the contributions to FCNCs from bilinears that violate $B-L$ also applies to bilinears in class 1 of the form $\overline{E}_{i} c_{\alpha, ij} U_{j}$ or $\overline{E}_{i} c_{\alpha, ij} D_{j}$, even though these bilinears do not produce large FCNCs.}

Finally, for bilinears in class 3, the matrices $c_\alpha$ and $c_\beta$ become:
\be
\label{classIII}
c_\alpha \to V_\nu c_\alpha V^\dagger_X, \quad c_\beta \to V_X c_\beta V^\dagger_\nu.
\ee
The key point here is that in the seesaw scenarios, $V_\nu$, at leading order, Eq.~\eqref{defineVnu}, is different from the matrices that diagonalize the charged-lepton Yukawa matrices. In the limit $\{\veps_X, \veps_\alpha\} \to 0$, $c_\alpha$ and $c_\beta$ can have large off-diagonal components. Therefore, higher-dimensional operators containing neutrinos will produce relatively large amplitudes for FCNC processes.

The kinds of processes to be expected from this third class  of bilinears include rare $\tau$, meson, Higgs and $W$ decays, the flavor-violating structure of which cannot be easily probed at experiments due to the final-state neutrinos. Some of these operators, however, give rise to potentially large nonstandard interactions (NSI) \cite{Ohlsson:2012kf,Miranda:2015dra} for neutrinos. Current measurements of NSI parameters \cite{Coelho:2012bp,Esmaili:2013fva,Fukasawa:2015jaa,Sousa:2015bxa,Liao:2016hsa} are consistent with Flavorspin at the TeV scale. While gauge invariance ensures that operators containing bilinears of this class are accompanied by operators containing charged leptons, bounds on neutrino NSI from charged-lepton flavor change can be partially evaded, due to the differences between the matrices that rotate the charged leptons and the neutrinos into their respective mass bases. Over the next decade or so, a host of experiments \cite{Friedland:2012tq,Choubey:2014iia,Fukasawa:2015jaa,An:2015jdp,Coloma:2015kiu,deGouvea:2015ndi,Choubey:2015xha} will attempt to measure nonzero NSI; these will serve as a critical test of the framework we have introduced. 


\section{Discussion}
\label{discussion}
\setcounter{equation}{0}


In this paper, we proposed a framework to attack the Flavor Puzzle based on the pinciple of decomposition of the SM Yukawas into fundamental spurions. Within this framework, we fully implemented the simplest possible case, in which the flavor structure of the SM is derived from a single horizontal $SU(2)$ flavor symmetry. With respect to Flavorspin, all fermions transform as triplets of flavor $SU(2)$. In addition, we imposed some restrictions on the parameter space, in particular demanding the perturbativity of the set of parameters $\{\veps_X,\, \bveps_X,\,\xi_X, \eta_X\}$.

Phenomenologically desirable highlights that follow from the perturbative Flavorspin scenario include:
\begin{itemize}
\item Naturally small masses for the first and second generations of charged fermions.
\item Naturally small mixing angles in the quark sector, with the Cabibbo angle predicted to be about 100 times larger than $\theta_{13}^q$, see Eq.~\eqref{Ratio_12_13_Quark}.
\item Large $CP$ violation likely in the quark sector, see Fig.~\ref{fig:QuarkAngles}(b).
\item A milder predicted mass hierarchy for Majorana neutrinos.
\item At least one large angle predicted in the lepton sector for the case of Majorana neutrinos.
\item Large $CP$ violation likely in the lepton sector, see Figs.~\ref{fig:TypeIHierarchy}(b), \ref{fig:TypeIIHierarchy}(b) and \ref{fig:GeneralHierarchy}(b).
\item When Flavorspin is extended to nonrenormalizable operators, it naturally suppresses the most common FCNCs involving only charged fermions, while allowing for large FCNCs if neutrinos are involved, see Sec.~\ref{nofcncs}. These could potentially been seen at long-baseline neutrino experiments.
\end{itemize}
Moreover, other features of our setup can be considered aesthetically pleasing. In particular, the quark and leptonic flavor structures both emerge from the same set of fundamental spurions. Quark and lepton flavor are unified in this sense.

Nonetheless, it should be stated that Flavorspin with perturbative $\{\veps_X,\, \bveps_X,\,\xi_X, \eta_X\}$ appears to be somewhat restrictive. In particular,  starting from a good quark fit, it does not do an entirely good job in describing flavor in the leptonic sector. The detailed results are found in Sec.~\ref{results}, where we calculated confidence intervals for the fermion masses and mixing parameters for given ranges of the Flavorspin parameters for several parametrizations of the neutrino mass matrix in terms of fundamental spurions. For the most promising Majorana possibilities, we find that although Flavorspin invariably predicts large mixing angles and $CP$ violation in the lepton sector, it struggles to reproduce the neutrino mass hierarchy. Some tension is also observed between the relatively large values of $\theta_{12}^l$ and $\theta_{13}^l$. The type II scenario yields the best fit, all things considered.

Finally, we comment on the perturbativity of $\veps_X$ and $\bveps_X$. In the previous sections, this was taken as an assumption. However, a Froggatt-Nielsen-like principle could provide partial justification for it. The $U(1)$ Froggatt-Nielsen symmetry would be $B-L$, under which the spurions may also be formally charged. Thus, the formal global symmetry of our model would be thus enlarged to be:
\be
\mcl{G}_g = \Gfl \times U(1)_{B-L} \,.
\ee
Specifically, suppose that in this setup a formal charge of 1 under $B-L$ is assigned to $Y_5$, while $Y_3$ is taken to be $B-L$-neutral. That is, introducing the notation $\mbf{r}_q$ where $\mbf{r}$ is the $\Gfl$ representation and $q$ the $B-L$ charge, we would have
\be
Y_3 \sim \mbf{3}_0\,,\quad\quad Y_5\sim \mbf{5}_{-1} \,,
\ee
Now, the coefficients $\veps_X$, $\veps_\Delta$ for the corresponding operators are introduced with formal charges
\be
\veps_X, \veps_\Delta \sim \mbf{0}_{-1}\ \, .
\ee
The final form of the Yukawas, analogous to Eq.~\eqref{final-form}, necessary to render the Yukawa operator invariant under $\mcl{G}_g$, now under $\Gfl \times B-L$, is  given by:
\be
Y_X \equiv Y_{X}(Y_3,\,\veps_X Y_5) + \bveps_X\mbb{I} \,,\quad \quad Y_\Delta \equiv Y_\Delta\big( (\veps_\Delta^*)^2Y_3,\, \veps_\Delta^* Y_5\big) + \bveps_\Delta\mbb{I} \,, \label{final-form-Y}
\ee
where $|\veps_X|, \,|\veps_\Delta| \ll 1$. More generally, we take the following rule to be valid both for renormalizable and nonrenormalizable operators:
\be
c_{\alpha,\,ij}^{(d)} \equiv  \left\{ \begin{array}{cl}
  \left[\tilde{c}_{\alpha,\,ij}^{(d)} \left( Y_3,\, \veps_\alpha^{(d)}Y_5\right) +  \mbb{I} \right] & \quad \trm{for }  q=0 \\
  \big(\veps_\alpha^{(d)}\big)^{q} \, \left[ \tilde{c}_{\alpha,\,ij}^{(d)} \left( Y_3,\, \big(\veps_\alpha^{(d)}\big)^{-1}Y_5 \right) +  \mbb{I} \right] & \quad \trm{for } q \geq 1
\end{array},
\right.
\ee
where
\be
q = B-L \trm{ charge of } \mcl{O}^{(d)}_\alpha ,
\ee
and where the dimensionless $\veps_\alpha$  parameter is assumed to be parametrically small. The consequence is that the contribution to flavor coming from the $Y_5$ spurion is suppressed with respect to $Y_3$ for $B-L$-conserving operators and vice versa for $B-L$-violating ones, {\`a} la Froggatt-Nielsen. If we assume a mild hierarchy between $Y_3$ and $Y_5$, $|Y_3| > |Y_5|$, then their contributions can be strongly hierarchical in the $B-L$-conserving case and roughly equivalent in the latter, as we found in this study. Moreover, this mechanism can also be used to suppress the flavor singlet fermionic bilinears of case 2 mentioned in Sec.~\ref{nofcncs}. Irrespective of its Lorentz or gauge properties, to each flavor singlet combination corresponds a $c_\alpha$, that would include at least two powers of $\veps_\alpha$ if $B-L$-violating. Thus, by using the familiar $B-L$ as a Froggatt-Nielsen symmetry, this extension would make explicit the difference in flavor structure between $B-L$-conserving and $B-L$-violating operators. One may also consider gauging $U(1)_{L_\mu-L_\tau}$ as part of the flavor group; this has been studied in, for instance, Ref.~\cite{Heeck:2011wj}.

Some other extensions are of potential interest. Although Flavorspin provides a simple explanation for several patterns observed in the spectrum and mixings in the SM, it clearly is not a complete theory of flavor. The parameters must be fit to the data, and it would be interesting to explore whether promoting the Yukawas to true fields and optimizing a scalar potential is beneficial in this case. On the more phenomenological side, since the flavor structure of higher-dimensional operators is determined, deviations from SM branching ratios will be correlated. A full exploration of these is beyond the scope of this paper. Finally, we stress that we have only explored the simplest decomposition of the Yukawas into fundamental spurions, i.e. a sum of two spurions charged under a vectorial $SU(2)$ symmetry. This is of course not the only possibility.


\begin{acknowledgements}
We thank Andr\'e de Gouv\^ea for many illuminating conversations regarding this work and for reviewing this manuscript. J.M.B. thanks Kevin Kelly for useful conversations. This work is supported in part by DOE grant \#DE-SC0010143.
\end{acknowledgements}


\bibliographystyle{apsrev-title}
\bibliography{FS_Bib}{}

\begin{thebibliography}{55}
\expandafter\ifx\csname natexlab\endcsname\relax\def\natexlab#1{#1}\fi
\expandafter\ifx\csname bibnamefont\endcsname\relax
  \def\bibnamefont#1{#1}\fi
\expandafter\ifx\csname bibfnamefont\endcsname\relax
  \def\bibfnamefont#1{#1}\fi
\expandafter\ifx\csname citenamefont\endcsname\relax
  \def\citenamefont#1{#1}\fi
\expandafter\ifx\csname url\endcsname\relax
  \def\url#1{\texttt{#1}}\fi
\expandafter\ifx\csname urlprefix\endcsname\relax\def\urlprefix{URL }\fi
\providecommand{\bibinfo}[2]{#2}
\providecommand{\eprint}[2][]{\url{#2}}

\bibitem[{\citenamefont{Froggatt and Nielsen}(1979)}]{Froggatt:1978nt}
\bibinfo{author}{\bibfnamefont{C.~D.} \bibnamefont{Froggatt}} \bibnamefont{and}
  \bibinfo{author}{\bibfnamefont{H.~B.} \bibnamefont{Nielsen}}, ``{Hierarchy of
  quark masses, Cabibbo angles and $CP$ violation},'' \bibinfo{journal}{Nucl.
  Phys.} \textbf{\bibinfo{volume}{B147}}, \bibinfo{pages}{277}
  (\bibinfo{year}{1979}).

\bibitem[{\citenamefont{Glashow et~al.}(1970)\citenamefont{Glashow, Iliopoulos,
  and Maiani}}]{Glashow:1970gm}
\bibinfo{author}{\bibfnamefont{S.~L.} \bibnamefont{Glashow}},
  \bibinfo{author}{\bibfnamefont{J.}~\bibnamefont{Iliopoulos}},
  \bibnamefont{and} \bibinfo{author}{\bibfnamefont{L.}~\bibnamefont{Maiani}},
  ``{Weak Interactions with lepton-hadron symmetry},'' \bibinfo{journal}{Phys.
  Rev.} \textbf{\bibinfo{volume}{D2}}, \bibinfo{pages}{1285}
  (\bibinfo{year}{1970}).

\bibitem[{\citenamefont{Chivukula and Georgi}(1987)}]{Chivukula:1987py}
\bibinfo{author}{\bibfnamefont{R.~S.} \bibnamefont{Chivukula}}
  \bibnamefont{and} \bibinfo{author}{\bibfnamefont{H.}~\bibnamefont{Georgi}},
  ``{Composite technicolor standard model},'' \bibinfo{journal}{Phys. Lett.}
  \textbf{\bibinfo{volume}{B188}}, \bibinfo{pages}{99} (\bibinfo{year}{1987}).

\bibitem[{\citenamefont{D'Ambrosio et~al.}(2002)\citenamefont{D'Ambrosio,
  Giudice, Isidori, and Strumia}}]{D'Ambrosio:2002ex}
\bibinfo{author}{\bibfnamefont{G.}~\bibnamefont{D'Ambrosio}},
  \bibinfo{author}{\bibfnamefont{G.~F.} \bibnamefont{Giudice}},
  \bibinfo{author}{\bibfnamefont{G.}~\bibnamefont{Isidori}}, \bibnamefont{and}
  \bibinfo{author}{\bibfnamefont{A.}~\bibnamefont{Strumia}}, ``{Minimal flavor
  violation: an effective field theory approach},'' \bibinfo{journal}{Nucl.
  Phys.} \textbf{\bibinfo{volume}{B645}}, \bibinfo{pages}{155}
  (\bibinfo{year}{2002}), \eprint{hep-ph/0207036}.

\bibitem[{\citenamefont{Buras}(2003)}]{Buras:2003jf}
\bibinfo{author}{\bibfnamefont{A.~J.} \bibnamefont{Buras}}, ``{Minimal flavor
  violation},'' \bibinfo{journal}{Acta Phys. Polon.}
  \textbf{\bibinfo{volume}{B34}}, \bibinfo{pages}{5615} (\bibinfo{year}{2003}),
  \eprint{hep-ph/0310208}.

\bibitem[{\citenamefont{Cirigliano et~al.}(2005)\citenamefont{Cirigliano,
  Grinstein, Isidori, and Wise}}]{Cirigliano:2005ck}
\bibinfo{author}{\bibfnamefont{V.}~\bibnamefont{Cirigliano}},
  \bibinfo{author}{\bibfnamefont{B.}~\bibnamefont{Grinstein}},
  \bibinfo{author}{\bibfnamefont{G.}~\bibnamefont{Isidori}}, \bibnamefont{and}
  \bibinfo{author}{\bibfnamefont{M.~B.} \bibnamefont{Wise}}, ``{Minimal flavor
  violation in the lepton sector},'' \bibinfo{journal}{Nucl. Phys.}
  \textbf{\bibinfo{volume}{B728}}, \bibinfo{pages}{121} (\bibinfo{year}{2005}),
  \eprint{hep-ph/0507001}.

\bibitem[{\citenamefont{Agashe et~al.}(2005)\citenamefont{Agashe, Papucci,
  Perez, and Pirjol}}]{Agashe:2005hk}
\bibinfo{author}{\bibfnamefont{K.}~\bibnamefont{Agashe}},
  \bibinfo{author}{\bibfnamefont{M.}~\bibnamefont{Papucci}},
  \bibinfo{author}{\bibfnamefont{G.}~\bibnamefont{Perez}}, \bibnamefont{and}
  \bibinfo{author}{\bibfnamefont{D.}~\bibnamefont{Pirjol}}, ``{Next to minimal
  flavor violation},''  (\bibinfo{year}{2005}), \eprint{hep-ph/0509117}.

\bibitem[{\citenamefont{Kagan et~al.}(2009)\citenamefont{Kagan, Perez,
  Volansky, and Zupan}}]{Kagan:2009bn}
\bibinfo{author}{\bibfnamefont{A.~L.} \bibnamefont{Kagan}},
  \bibinfo{author}{\bibfnamefont{G.}~\bibnamefont{Perez}},
  \bibinfo{author}{\bibfnamefont{T.}~\bibnamefont{Volansky}}, \bibnamefont{and}
  \bibinfo{author}{\bibfnamefont{J.}~\bibnamefont{Zupan}}, ``{General minimal
  flavor violation},'' \bibinfo{journal}{Phys. Rev.}
  \textbf{\bibinfo{volume}{D80}}, \bibinfo{pages}{076002}
  (\bibinfo{year}{2009}), \eprint{0903.1794}.

\bibitem[{\citenamefont{Gavela et~al.}(2009)\citenamefont{Gavela, Hambye,
  Hernandez, and Hernandez}}]{Gavela:2009cd}
\bibinfo{author}{\bibfnamefont{M.~B.} \bibnamefont{Gavela}},
  \bibinfo{author}{\bibfnamefont{T.}~\bibnamefont{Hambye}},
  \bibinfo{author}{\bibfnamefont{D.}~\bibnamefont{Hernandez}},
  \bibnamefont{and}
  \bibinfo{author}{\bibfnamefont{P.}~\bibnamefont{Hernandez}}, ``{Minimal
  flavour seesaw models},'' \bibinfo{journal}{JHEP}
  \textbf{\bibinfo{volume}{09}}, \bibinfo{pages}{038} (\bibinfo{year}{2009}),
  \eprint{0906.1461}.

\bibitem[{\citenamefont{Alonso et~al.}(2011{\natexlab{a}})\citenamefont{Alonso,
  Gavela, Merlo, and Rigolin}}]{Alonso:2011yg}
\bibinfo{author}{\bibfnamefont{R.}~\bibnamefont{Alonso}},
  \bibinfo{author}{\bibfnamefont{M.~B.} \bibnamefont{Gavela}},
  \bibinfo{author}{\bibfnamefont{L.}~\bibnamefont{Merlo}}, \bibnamefont{and}
  \bibinfo{author}{\bibfnamefont{S.}~\bibnamefont{Rigolin}}, ``{On the scalar
  potential of minimal flavour violation},'' \bibinfo{journal}{JHEP}
  \textbf{\bibinfo{volume}{07}}, \bibinfo{pages}{012}
  (\bibinfo{year}{2011}{\natexlab{a}}), \eprint{1103.2915}.

\bibitem[{\citenamefont{Alonso et~al.}(2012)\citenamefont{Alonso, Gavela,
  Hernandez, and Merlo}}]{Alonso:2012fy}
\bibinfo{author}{\bibfnamefont{R.}~\bibnamefont{Alonso}},
  \bibinfo{author}{\bibfnamefont{M.~B.} \bibnamefont{Gavela}},
  \bibinfo{author}{\bibfnamefont{D.}~\bibnamefont{Hernandez}},
  \bibnamefont{and} \bibinfo{author}{\bibfnamefont{L.}~\bibnamefont{Merlo}},
  ``{On the potential of leptonic minimal flavour violation},''
  \bibinfo{journal}{Phys. Lett.} \textbf{\bibinfo{volume}{B715}},
  \bibinfo{pages}{194} (\bibinfo{year}{2012}), \eprint{1206.3167}.

\bibitem[{\citenamefont{Cabibbo and Maiani}(1970)}]{Cabibbo:1970rza}
\bibinfo{author}{\bibfnamefont{N.}~\bibnamefont{Cabibbo}} \bibnamefont{and}
  \bibinfo{author}{\bibfnamefont{L.}~\bibnamefont{Maiani}}, in
  \emph{\bibinfo{booktitle}{Evolution of particle physics: A volume dedicated
  to Edoardo Amaldi in his sixtieth birthday}}, edited by
  \bibinfo{editor}{\bibfnamefont{M.}~\bibnamefont{Conversi}}
  (\bibinfo{year}{1970}), pp. \bibinfo{pages}{50--80}.

\bibitem[{\citenamefont{Alonso et~al.}(2011{\natexlab{b}})\citenamefont{Alonso,
  Isidori, Merlo, Munoz, and Nardi}}]{Alonso:2011jd}
\bibinfo{author}{\bibfnamefont{R.}~\bibnamefont{Alonso}},
  \bibinfo{author}{\bibfnamefont{G.}~\bibnamefont{Isidori}},
  \bibinfo{author}{\bibfnamefont{L.}~\bibnamefont{Merlo}},
  \bibinfo{author}{\bibfnamefont{L.~A.} \bibnamefont{Munoz}}, \bibnamefont{and}
  \bibinfo{author}{\bibfnamefont{E.}~\bibnamefont{Nardi}}, ``{Minimal flavour
  violation extensions of the seesaw},'' \bibinfo{journal}{JHEP}
  \textbf{\bibinfo{volume}{06}}, \bibinfo{pages}{037}
  (\bibinfo{year}{2011}{\natexlab{b}}), \eprint{1103.5461}.

\bibitem[{\citenamefont{Alonso et~al.}(2013{\natexlab{a}})\citenamefont{Alonso,
  Gavela, Hern\'andez, Merlo, and Rigolin}}]{Alonso:2013mca}
\bibinfo{author}{\bibfnamefont{R.}~\bibnamefont{Alonso}},
  \bibinfo{author}{\bibfnamefont{M.~B.} \bibnamefont{Gavela}},
  \bibinfo{author}{\bibfnamefont{D.}~\bibnamefont{Hern\'andez}},
  \bibinfo{author}{\bibfnamefont{L.}~\bibnamefont{Merlo}}, \bibnamefont{and}
  \bibinfo{author}{\bibfnamefont{S.}~\bibnamefont{Rigolin}}, ``{Leptonic
  dynamical yukawa couplings},'' \bibinfo{journal}{JHEP}
  \textbf{\bibinfo{volume}{08}}, \bibinfo{pages}{069}
  (\bibinfo{year}{2013}{\natexlab{a}}), \eprint{1306.5922}.

\bibitem[{\citenamefont{Alonso et~al.}(2013{\natexlab{b}})\citenamefont{Alonso,
  Gavela, Isidori, and Maiani}}]{Alonso:2013nca}
\bibinfo{author}{\bibfnamefont{R.}~\bibnamefont{Alonso}},
  \bibinfo{author}{\bibfnamefont{M.~B.} \bibnamefont{Gavela}},
  \bibinfo{author}{\bibfnamefont{G.}~\bibnamefont{Isidori}}, \bibnamefont{and}
  \bibinfo{author}{\bibfnamefont{L.}~\bibnamefont{Maiani}}, ``{Neutrino mixing
  and masses from a minimum principle},'' \bibinfo{journal}{JHEP}
  \textbf{\bibinfo{volume}{11}}, \bibinfo{pages}{187}
  (\bibinfo{year}{2013}{\natexlab{b}}), \eprint{1306.5927}.

\bibitem[{\citenamefont{Terazawa et~al.}(1977)\citenamefont{Terazawa, Akama,
  and Chikashige}}]{Terazawa:1976xx}
\bibinfo{author}{\bibfnamefont{H.}~\bibnamefont{Terazawa}},
  \bibinfo{author}{\bibfnamefont{K.}~\bibnamefont{Akama}}, \bibnamefont{and}
  \bibinfo{author}{\bibfnamefont{Y.}~\bibnamefont{Chikashige}}, ``{Unified
  model of the Nambu-Jona-Lasinio type for all elementary particle forces},''
  \bibinfo{journal}{Phys. Rev.} \textbf{\bibinfo{volume}{D15}},
  \bibinfo{pages}{480} (\bibinfo{year}{1977}).

\bibitem[{\citenamefont{Terazawa}(1977)}]{Terazawa:1977ee}
\bibinfo{author}{\bibfnamefont{H.}~\bibnamefont{Terazawa}}, ``{A gauge model
  for muon-number changing processes},'' \bibinfo{journal}{Prog. Theor. Phys.}
  \textbf{\bibinfo{volume}{57}}, \bibinfo{pages}{1808} (\bibinfo{year}{1977}).

\bibitem[{\citenamefont{Maehara and Yanagida}(1978)}]{Maehara:1978ts}
\bibinfo{author}{\bibfnamefont{T.}~\bibnamefont{Maehara}} \bibnamefont{and}
  \bibinfo{author}{\bibfnamefont{T.}~\bibnamefont{Yanagida}}, ``{ $CP$
  violation and off-diagonal neutral currents},'' \bibinfo{journal}{Prog.
  Theor. Phys.} \textbf{\bibinfo{volume}{60}}, \bibinfo{pages}{822}
  (\bibinfo{year}{1978}).

\bibitem[{\citenamefont{Wilczek and Zee}(1979)}]{Wilczek:1978xi}
\bibinfo{author}{\bibfnamefont{F.}~\bibnamefont{Wilczek}} \bibnamefont{and}
  \bibinfo{author}{\bibfnamefont{A.}~\bibnamefont{Zee}}, ``{Horizontal
  interaction and weak mixing angles},'' \bibinfo{journal}{Phys. Rev. Lett.}
  \textbf{\bibinfo{volume}{42}}, \bibinfo{pages}{421} (\bibinfo{year}{1979}).

\bibitem[{\citenamefont{Yanagida}(1979)}]{Yanagida:1979gs}
\bibinfo{author}{\bibfnamefont{T.}~\bibnamefont{Yanagida}}, ``{Horizontal
  symmetry and mass of the top quark},'' \bibinfo{journal}{Phys. Rev.}
  \textbf{\bibinfo{volume}{D20}}, \bibinfo{pages}{2986} (\bibinfo{year}{1979}).

\bibitem[{\citenamefont{Chikashige et~al.}(1980)\citenamefont{Chikashige,
  Gelmini, Peccei, and Roncadelli}}]{Chikashige:1980ht}
\bibinfo{author}{\bibfnamefont{Y.}~\bibnamefont{Chikashige}},
  \bibinfo{author}{\bibfnamefont{G.}~\bibnamefont{Gelmini}},
  \bibinfo{author}{\bibfnamefont{R.~D.} \bibnamefont{Peccei}},
  \bibnamefont{and}
  \bibinfo{author}{\bibfnamefont{M.}~\bibnamefont{Roncadelli}}, ``{Horizontal
  symmetries, dynamical symmetry breaking and neutrino masses},''
  \bibinfo{journal}{Phys. Lett.} \textbf{\bibinfo{volume}{B94}},
  \bibinfo{pages}{499} (\bibinfo{year}{1980}).

\bibitem[{\citenamefont{Yanagida}(1980)}]{Yanagida:1980wd}
\bibinfo{author}{\bibfnamefont{T.}~\bibnamefont{Yanagida}}, ``{Origin of
  horizontal symmetry and $SU(5) \times SU(2)_F$ unification},''
  \bibinfo{journal}{Prog. Theor. Phys.} \textbf{\bibinfo{volume}{63}},
  \bibinfo{pages}{354} (\bibinfo{year}{1980}).

\bibitem[{\citenamefont{Terazawa}(2011)}]{Terazawa:2011ci}
\bibinfo{author}{\bibfnamefont{H.}~\bibnamefont{Terazawa}}, in
  \emph{\bibinfo{booktitle}{{Journal of Modern Physics, Vol.5, Nov.5,
  205-208(2014)}}} (\bibinfo{year}{2011}), \eprint{1109.3705},
  \urlprefix\url{http://inspirehep.net/record/927777/files/arXiv:1109.3705.pdf}.

\bibitem[{\citenamefont{Aulakh and Khosa}(2014)}]{Aulakh:2013kha}
\bibinfo{author}{\bibfnamefont{C.~S.} \bibnamefont{Aulakh}} \bibnamefont{and}
  \bibinfo{author}{\bibfnamefont{C.~K.} \bibnamefont{Khosa}}, ``{SO(10) grand
  unified theories with dynamical Yukawa couplings},'' \bibinfo{journal}{Phys.
  Rev.} \textbf{\bibinfo{volume}{D90}}, \bibinfo{pages}{045008}
  (\bibinfo{year}{2014}), \eprint{1308.5665}.

\bibitem[{\citenamefont{Aulakh}(2015)}]{Aulakh:2014wsa}
\bibinfo{author}{\bibfnamefont{C.~S.} \bibnamefont{Aulakh}}, ``{Bajc-Melfo
  vacua enable Yukawon ultraminimal grand unified theories},''
  \bibinfo{journal}{Phys. Rev.} \textbf{\bibinfo{volume}{D91}},
  \bibinfo{pages}{055012} (\bibinfo{year}{2015}), \eprint{1402.3979}.

\bibitem[{\citenamefont{Terazawa and Yasue}(2016)}]{Terazawa:2015bsa}
\bibinfo{author}{\bibfnamefont{H.}~\bibnamefont{Terazawa}} \bibnamefont{and}
  \bibinfo{author}{\bibfnamefont{M.}~\bibnamefont{Yasue}}, ``{Excited gauge and
  higgs bosons in the unified composite model},'' \bibinfo{journal}{Nonlin.
  Phenom. Complex Syst.} \textbf{\bibinfo{volume}{19}}, \bibinfo{pages}{1}
  (\bibinfo{year}{2016}), \eprint{1508.00172}.

\bibitem[{\citenamefont{Buchmuller and Wyler}(1986)}]{Buchmuller:1985jz}
\bibinfo{author}{\bibfnamefont{W.}~\bibnamefont{Buchmuller}} \bibnamefont{and}
  \bibinfo{author}{\bibfnamefont{D.}~\bibnamefont{Wyler}}, ``{Effective
  lagrangian analysis of new interactions and flavor conservation},''
  \bibinfo{journal}{Nucl. Phys.} \textbf{\bibinfo{volume}{B268}},
  \bibinfo{pages}{621} (\bibinfo{year}{1986}).

\bibitem[{\citenamefont{Manohar}(1997)}]{Manohar:1996cq}
\bibinfo{author}{\bibfnamefont{A.~V.} \bibnamefont{Manohar}}, ``{Effective
  field theories},'' \bibinfo{journal}{Lect. Notes Phys.}
  \textbf{\bibinfo{volume}{479}}, \bibinfo{pages}{311} (\bibinfo{year}{1997}),
  \eprint{hep-ph/9606222}.

\bibitem[{\citenamefont{Burgess}(2007)}]{Burgess:2007pt}
\bibinfo{author}{\bibfnamefont{C.~P.} \bibnamefont{Burgess}}, ``{Introduction
  to effective field theory},'' \bibinfo{journal}{Ann. Rev. Nucl. Part. Sci.}
  \textbf{\bibinfo{volume}{57}}, \bibinfo{pages}{329} (\bibinfo{year}{2007}),
  \eprint{hep-th/0701053}.

\bibitem[{\citenamefont{Olive et~al.}(2014)}]{Agashe:2014kda}
\bibinfo{author}{\bibfnamefont{K.~A.} \bibnamefont{Olive}} \bibnamefont{et~al.}
  (\bibinfo{collaboration}{Particle Data Group}), ``{Review of particle
  physics},'' \bibinfo{journal}{Chin. Phys.} \textbf{\bibinfo{volume}{C38}},
  \bibinfo{pages}{090001} (\bibinfo{year}{2014}).

\bibitem[{\citenamefont{Antusch and Maurer}(2013)}]{Antusch:2013jca}
\bibinfo{author}{\bibfnamefont{S.}~\bibnamefont{Antusch}} \bibnamefont{and}
  \bibinfo{author}{\bibfnamefont{V.}~\bibnamefont{Maurer}}, ``{Running quark
  and lepton parameters at various scales},'' \bibinfo{journal}{JHEP}
  \textbf{\bibinfo{volume}{11}}, \bibinfo{pages}{115} (\bibinfo{year}{2013}),
  \eprint{1306.6879}.

\bibitem[{\citenamefont{Esteban et~al.}(2016)\citenamefont{Esteban,
  Gonzalez-Garcia, Maltoni, Martinez-Soler, and Schwetz}}]{Esteban:2016qun}
\bibinfo{author}{\bibfnamefont{I.}~\bibnamefont{Esteban}},
  \bibinfo{author}{\bibfnamefont{M.~C.} \bibnamefont{Gonzalez-Garcia}},
  \bibinfo{author}{\bibfnamefont{M.}~\bibnamefont{Maltoni}},
  \bibinfo{author}{\bibfnamefont{I.}~\bibnamefont{Martinez-Soler}},
  \bibnamefont{and} \bibinfo{author}{\bibfnamefont{T.}~\bibnamefont{Schwetz}},
  ``{Updated fit to three neutrino mixing: exploring the accelerator-reactor
  complementarity},''  (\bibinfo{year}{2016}), \eprint{1611.01514}.

\bibitem[{\citenamefont{Antusch et~al.}(2005)\citenamefont{Antusch, Kersten,
  Lindner, Ratz, and Schmidt}}]{Antusch:2005gp}
\bibinfo{author}{\bibfnamefont{S.}~\bibnamefont{Antusch}},
  \bibinfo{author}{\bibfnamefont{J.}~\bibnamefont{Kersten}},
  \bibinfo{author}{\bibfnamefont{M.}~\bibnamefont{Lindner}},
  \bibinfo{author}{\bibfnamefont{M.}~\bibnamefont{Ratz}}, \bibnamefont{and}
  \bibinfo{author}{\bibfnamefont{M.~A.} \bibnamefont{Schmidt}}, ``{Running
  neutrino mass parameters in see-saw scenarios},'' \bibinfo{journal}{JHEP}
  \textbf{\bibinfo{volume}{03}}, \bibinfo{pages}{024} (\bibinfo{year}{2005}),
  \eprint{hep-ph/0501272}.

\bibitem[{\citenamefont{Mei}(2005)}]{Mei:2005qp}
\bibinfo{author}{\bibfnamefont{J.-w.} \bibnamefont{Mei}}, ``{Running neutrino
  masses, leptonic mixing angles and $CP$-violating phases: From $M(Z)$ to
  $\Lambda(GUT)$},'' \bibinfo{journal}{Phys. Rev.}
  \textbf{\bibinfo{volume}{D71}}, \bibinfo{pages}{073012}
  (\bibinfo{year}{2005}), \eprint{hep-ph/0502015}.

\bibitem[{\citenamefont{Ellis et~al.}(2005)\citenamefont{Ellis, Hektor,
  Kadastik, Kannike, and Raidal}}]{Ellis:2005dr}
\bibinfo{author}{\bibfnamefont{J.~R.} \bibnamefont{Ellis}},
  \bibinfo{author}{\bibfnamefont{A.}~\bibnamefont{Hektor}},
  \bibinfo{author}{\bibfnamefont{M.}~\bibnamefont{Kadastik}},
  \bibinfo{author}{\bibfnamefont{K.}~\bibnamefont{Kannike}}, \bibnamefont{and}
  \bibinfo{author}{\bibfnamefont{M.}~\bibnamefont{Raidal}}, ``{Running of
  low-energy neutrino masses, mixing angles and $CP$ violation},''
  \bibinfo{journal}{Phys. Lett.} \textbf{\bibinfo{volume}{B631}},
  \bibinfo{pages}{32} (\bibinfo{year}{2005}), \eprint{hep-ph/0506122}.

\bibitem[{\citenamefont{Xing et~al.}(2008)\citenamefont{Xing, Zhang, and
  Zhou}}]{Xing:2007fb}
\bibinfo{author}{\bibfnamefont{Z.-z.} \bibnamefont{Xing}},
  \bibinfo{author}{\bibfnamefont{H.}~\bibnamefont{Zhang}}, \bibnamefont{and}
  \bibinfo{author}{\bibfnamefont{S.}~\bibnamefont{Zhou}}, ``{Updated values of
  running quark and lepton masses},'' \bibinfo{journal}{Phys. Rev.}
  \textbf{\bibinfo{volume}{D77}}, \bibinfo{pages}{113016}
  (\bibinfo{year}{2008}), \eprint{0712.1419}.

\bibitem[{\citenamefont{Lin et~al.}(2010)\citenamefont{Lin, Merlo, and
  Paris}}]{Lin:2009sq}
\bibinfo{author}{\bibfnamefont{Y.}~\bibnamefont{Lin}},
  \bibinfo{author}{\bibfnamefont{L.}~\bibnamefont{Merlo}}, \bibnamefont{and}
  \bibinfo{author}{\bibfnamefont{A.}~\bibnamefont{Paris}}, ``{Running effects
  on lepton mixing angles in flavour models with type I seesaw},''
  \bibinfo{journal}{Nucl. Phys.} \textbf{\bibinfo{volume}{B835}},
  \bibinfo{pages}{238} (\bibinfo{year}{2010}), \eprint{0911.3037}.

\bibitem[{\citenamefont{Ohlsson and Zhou}(2014)}]{Ohlsson:2013xva}
\bibinfo{author}{\bibfnamefont{T.}~\bibnamefont{Ohlsson}} \bibnamefont{and}
  \bibinfo{author}{\bibfnamefont{S.}~\bibnamefont{Zhou}}, ``{Renormalization
  group running of neutrino parameters},'' \bibinfo{journal}{Nature Commun.}
  \textbf{\bibinfo{volume}{5}}, \bibinfo{pages}{5153} (\bibinfo{year}{2014}),
  \eprint{1311.3846}.

\bibitem[{\citenamefont{Antonelli et~al.}(2010)}]{Antonelli:2009ws}
\bibinfo{author}{\bibfnamefont{M.}~\bibnamefont{Antonelli}}
  \bibnamefont{et~al.}, ``{Flavor physics in the quark sector},''
  \bibinfo{journal}{Phys. Rept.} \textbf{\bibinfo{volume}{494}},
  \bibinfo{pages}{197} (\bibinfo{year}{2010}), \eprint{0907.5386}.

\bibitem[{\citenamefont{Lindner et~al.}(2016)\citenamefont{Lindner, Platscher,
  and Queiroz}}]{Lindner:2016bgg}
\bibinfo{author}{\bibfnamefont{M.}~\bibnamefont{Lindner}},
  \bibinfo{author}{\bibfnamefont{M.}~\bibnamefont{Platscher}},
  \bibnamefont{and} \bibinfo{author}{\bibfnamefont{F.~S.}
  \bibnamefont{Queiroz}}, ``{A call for new physics: the muon anomalous
  magnetic moment and lepton flavor violation},''  (\bibinfo{year}{2016}),
  \eprint{1610.06587}.

\bibitem[{\citenamefont{Abulencia et~al.}(2006)}]{Abulencia:2006ze}
\bibinfo{author}{\bibfnamefont{A.}~\bibnamefont{Abulencia}}
  \bibnamefont{et~al.} (\bibinfo{collaboration}{CDF}), ``{Observation of $B^0_s
  - \bar{B}^0_s$ oscillations},'' \bibinfo{journal}{Phys. Rev. Lett.}
  \textbf{\bibinfo{volume}{97}}, \bibinfo{pages}{242003}
  (\bibinfo{year}{2006}), \eprint{hep-ex/0609040}.

\bibitem[{\citenamefont{Ohlsson}(2013)}]{Ohlsson:2012kf}
\bibinfo{author}{\bibfnamefont{T.}~\bibnamefont{Ohlsson}}, ``{Status of
  non-standard neutrino interactions},'' \bibinfo{journal}{Rept. Prog. Phys.}
  \textbf{\bibinfo{volume}{76}}, \bibinfo{pages}{044201}
  (\bibinfo{year}{2013}), \eprint{1209.2710}.

\bibitem[{\citenamefont{Miranda and Nunokawa}(2015)}]{Miranda:2015dra}
\bibinfo{author}{\bibfnamefont{O.~G.} \bibnamefont{Miranda}} \bibnamefont{and}
  \bibinfo{author}{\bibfnamefont{H.}~\bibnamefont{Nunokawa}}, ``{Non standard
  neutrino interactions: current status and future prospects},''
  \bibinfo{journal}{New J. Phys.} \textbf{\bibinfo{volume}{17}},
  \bibinfo{pages}{095002} (\bibinfo{year}{2015}), \eprint{1505.06254}.

\bibitem[{\citenamefont{Coelho et~al.}(2012)\citenamefont{Coelho, Kafka, Mann,
  Schneps, and Altinok}}]{Coelho:2012bp}
\bibinfo{author}{\bibfnamefont{J.~A.~B.} \bibnamefont{Coelho}},
  \bibinfo{author}{\bibfnamefont{T.}~\bibnamefont{Kafka}},
  \bibinfo{author}{\bibfnamefont{W.~A.} \bibnamefont{Mann}},
  \bibinfo{author}{\bibfnamefont{J.}~\bibnamefont{Schneps}}, \bibnamefont{and}
  \bibinfo{author}{\bibfnamefont{O.}~\bibnamefont{Altinok}}, ``{Constraints for
  non-standard interaction $\epsilon_{e \tau}V_e$ from $\nu_e$ appearance in
  MINOS and T2K},'' \bibinfo{journal}{Phys. Rev.}
  \textbf{\bibinfo{volume}{D86}}, \bibinfo{pages}{113015}
  (\bibinfo{year}{2012}), \eprint{1209.3757}.

\bibitem[{\citenamefont{Esmaili and Smirnov}(2013)}]{Esmaili:2013fva}
\bibinfo{author}{\bibfnamefont{A.}~\bibnamefont{Esmaili}} \bibnamefont{and}
  \bibinfo{author}{\bibfnamefont{A.~{\relax Yu}.} \bibnamefont{Smirnov}},
  ``{Probing non-standard interaction of neutrinos with IceCube and
  DeepCore},'' \bibinfo{journal}{JHEP} \textbf{\bibinfo{volume}{06}},
  \bibinfo{pages}{026} (\bibinfo{year}{2013}), \eprint{1304.1042}.

\bibitem[{\citenamefont{Fukasawa and Yasuda}(2015)}]{Fukasawa:2015jaa}
\bibinfo{author}{\bibfnamefont{S.}~\bibnamefont{Fukasawa}} \bibnamefont{and}
  \bibinfo{author}{\bibfnamefont{O.}~\bibnamefont{Yasuda}}, ``{Constraints on
  the nonstandard interaction in propagation from atmospheric neutrinos},''
  \bibinfo{journal}{Adv. High Energy Phys.} \textbf{\bibinfo{volume}{2015}},
  \bibinfo{pages}{820941} (\bibinfo{year}{2015}), \eprint{1503.08056}.

\bibitem[{\citenamefont{Sousa}(2015)}]{Sousa:2015bxa}
\bibinfo{author}{\bibfnamefont{A.~B.} \bibnamefont{Sousa}}
  (\bibinfo{collaboration}{MINOS+, MINOS}), ``{First MINOS+ data and new
  results from MINOS},'' \bibinfo{journal}{AIP Conf. Proc.}
  \textbf{\bibinfo{volume}{1666}}, \bibinfo{pages}{110004}
  (\bibinfo{year}{2015}), \eprint{1502.07715}.

\bibitem[{\citenamefont{Liao et~al.}(2016)\citenamefont{Liao, Marfatia, and
  Whisnant}}]{Liao:2016hsa}
\bibinfo{author}{\bibfnamefont{J.}~\bibnamefont{Liao}},
  \bibinfo{author}{\bibfnamefont{D.}~\bibnamefont{Marfatia}}, \bibnamefont{and}
  \bibinfo{author}{\bibfnamefont{K.}~\bibnamefont{Whisnant}}, ``{Degeneracies
  in long-baseline neutrino experiments from nonstandard interactions},''
  \bibinfo{journal}{Phys. Rev.} \textbf{\bibinfo{volume}{D93}},
  \bibinfo{pages}{093016} (\bibinfo{year}{2016}), \eprint{1601.00927}.

\bibitem[{\citenamefont{Friedland and Shoemaker}(2012)}]{Friedland:2012tq}
\bibinfo{author}{\bibfnamefont{A.}~\bibnamefont{Friedland}} \bibnamefont{and}
  \bibinfo{author}{\bibfnamefont{I.~M.} \bibnamefont{Shoemaker}}, ``{Searching
  for novel neutrino interactions at NOvA and beyond in light of large
  $\theta_{13}$},''  (\bibinfo{year}{2012}), \eprint{1207.6642}.

\bibitem[{\citenamefont{Choubey and Ohlsson}(2014)}]{Choubey:2014iia}
\bibinfo{author}{\bibfnamefont{S.}~\bibnamefont{Choubey}} \bibnamefont{and}
  \bibinfo{author}{\bibfnamefont{T.}~\bibnamefont{Ohlsson}}, ``{Bounds on
  non-standard neutrino interactions Using PINGU},'' \bibinfo{journal}{Phys.
  Lett.} \textbf{\bibinfo{volume}{B739}}, \bibinfo{pages}{357}
  (\bibinfo{year}{2014}), \eprint{1410.0410}.

\bibitem[{\citenamefont{An et~al.}(2016)}]{An:2015jdp}
\bibinfo{author}{\bibfnamefont{F.}~\bibnamefont{An}} \bibnamefont{et~al.}
  (\bibinfo{collaboration}{JUNO}), ``{Neutrino physics with JUNO},''
  \bibinfo{journal}{J. Phys.} \textbf{\bibinfo{volume}{G43}},
  \bibinfo{pages}{030401} (\bibinfo{year}{2016}), \eprint{1507.05613}.

\bibitem[{\citenamefont{Coloma}(2016)}]{Coloma:2015kiu}
\bibinfo{author}{\bibfnamefont{P.}~\bibnamefont{Coloma}}, ``{Non-standard
  interactions in propagation at the Deep Underground Neutrino Experiment},''
  \bibinfo{journal}{JHEP} \textbf{\bibinfo{volume}{03}}, \bibinfo{pages}{016}
  (\bibinfo{year}{2016}), \eprint{1511.06357}.

\bibitem[{\citenamefont{de~Gouv\^ea and Kelly}(2016)}]{deGouvea:2015ndi}
\bibinfo{author}{\bibfnamefont{A.}~\bibnamefont{de~Gouv\^ea}} \bibnamefont{and}
  \bibinfo{author}{\bibfnamefont{K.~J.} \bibnamefont{Kelly}}, ``{Non-standard
  neutrino interactions at DUNE},'' \bibinfo{journal}{Nucl. Phys.}
  \textbf{\bibinfo{volume}{B908}}, \bibinfo{pages}{318} (\bibinfo{year}{2016}),
  \eprint{1511.05562}.

\bibitem[{\citenamefont{Choubey et~al.}(2015)\citenamefont{Choubey, Ghosh,
  Ohlsson, and Tiwari}}]{Choubey:2015xha}
\bibinfo{author}{\bibfnamefont{S.}~\bibnamefont{Choubey}},
  \bibinfo{author}{\bibfnamefont{A.}~\bibnamefont{Ghosh}},
  \bibinfo{author}{\bibfnamefont{T.}~\bibnamefont{Ohlsson}}, \bibnamefont{and}
  \bibinfo{author}{\bibfnamefont{D.}~\bibnamefont{Tiwari}}, ``{Neutrino physics
  with non-standard interactions at INO},'' \bibinfo{journal}{JHEP}
  \textbf{\bibinfo{volume}{12}}, \bibinfo{pages}{126} (\bibinfo{year}{2015}),
  \eprint{1507.02211}.

\bibitem[{\citenamefont{Heeck and Rodejohann}(2011)}]{Heeck:2011wj}
\bibinfo{author}{\bibfnamefont{J.}~\bibnamefont{Heeck}} \bibnamefont{and}
  \bibinfo{author}{\bibfnamefont{W.}~\bibnamefont{Rodejohann}}, ``{Gauged
  $L_\mu - L_\tau$ symmetry at the electroweak scale},''
  \bibinfo{journal}{Phys. Rev.} \textbf{\bibinfo{volume}{D84}},
  \bibinfo{pages}{075007} (\bibinfo{year}{2011}), \eprint{1107.5238}.

\end{thebibliography}

\end{document}